\documentclass[english]{iopart}

\usepackage{graphicx}
\usepackage{url}
\usepackage{iopams} 

\expandafter\let\csname equation*\endcsname\relax
\expandafter\let\csname endequation*\endcsname\relax

\usepackage{amssymb}
\usepackage[latin1]{inputenc}
\usepackage{amsmath}
\usepackage{epsfig}
\newcounter{fig}
\begin{document}

\title[Selected differential operators: a canonical decomposition]
{\Large Canonical decomposition of linear differential operators
  with selected differential Galois groups}
\vskip .3cm 

\author{S. Boukraa$||$, S. Hassani$^\S$, 
J-M. Maillard$^\pounds$, J-A. Weil$^\ddag$}
\address{$||$  \ LPTHIRM and IAESB,
 Universit\'e de Blida, Algeria}
\address{\S  Centre de Recherche Nucl\'eaire d'Alger, 
2 Bd. Frantz Fanon, BP 399, 16000 Alger, Algeria}
\address{$^\pounds$ LPTMC, UMR 7600 CNRS, 
Universit\'e de Paris 6, Sorbonne Universit\'es, 
Tour 23, 5\`eme \'etage, case 121, 
 4 Place Jussieu, 75252 Paris Cedex 05, France} 
\address{$^\ddag$ XLIM, Universit\'e de Limoges, 123 Avenue
Albert Thomas, 87060 Limoges Cedex, France}

\begin{abstract}

We first revisit an order-six linear differential operator, already
introduced in a previous paper, having a solution
which is a diagonal of a rational function of three variables. This 
linear differential operator is such that its exterior square has 
a rational solution, indicating that it has a selected 
differential Galois group, and is actually homomorphic to its adjoint. We obtain 
the two corresponding intertwiners giving this homomorphism 
to the adjoint. We show that these intertwiners are also homomorphic to their
adjoint and have a simple decomposition, already underlined in a previous 
paper, in terms of order-two self-adjoint operators. From these results,
we deduce a new form of decomposition of operators for this selected order-six linear 
differential operator in terms of three order-two self-adjoint 
operators. We generalize this decomposition to decomposition in terms of three
self-adjoint operators of arbitrary orders, provided the three 
orders have the same parity. We then generalize the 
previous decomposition to 
decompositions in terms of an arbitrary number of self-adjoint operators 
of the same parity order. This yields an infinite family of 
linear differential operators homomorphic to their adjoint, and, thus, 
with a selected differential Galois group.
We show that the equivalence of such operators, 
with selected differential Galois groups, is compatible 
with these canonical decompositions.
The rational solutions of the symmetric, or exterior, squares of 
these selected operators are, noticeably,
 seen to depend only on the rightmost self-adjoint 
operator in the decomposition. These results, and tools, are 
applied on operators of large orders. For instance, 
it is seen that a large set of (quite massive) operators, associated
with reflexive 4-polytopes defining  Calabi-Yau 3-folds, 
obtained recently by P. Lairez,
correspond to a particular form of the decomposition detailed in this paper.
All the results of this paper can be seen as providing an 
algebraic characterization of linear differential 
operators with selected symplectic or orthogonal differential Galois groups.

\end{abstract}

\vskip .1cm

\vskip .1cm


\noindent {\bf AMS Classification scheme numbers}: 34M55, 
47E05, 81Qxx, 32G34, 34Lxx, 34Mxx, 14Kxx 

\vskip .2cm


\section{Introduction}
\label{intro}

The $\, n$-fold integrals occurring in theoretical physics 
(lattice statistical mechanics, enumerative combinatorics, ...)
are, quite systematically\footnote[5]{In an ``experimental
mathematics'' approach.}, seen to be highly selected. For instance, the corresponding 
series expansions are {\em globally bounded}~\cite{Christol}, 
the linear differential 
operators, that annihilate them, are not only 
{\em Fuchsian}, but {\em globally nilpotent}~\cite{bo-bo-ha-ma-we-ze-09}. This is  
sometimes encapsulated in the wording ``modularity'', 
well-defined in a Calabi-Yau framework~\cite{Short,Big,Bogner}, but a work-in-progress
concept\footnote[1]{The mix between analytic, arithmetic, algebraic-geometry,
differential geometry, differential algebra, ... properties
being often a source of confusion in the literature.} outside this framework. It 
corresponds to two different kinds of 
``special properties'': firstly, properties of algebraic geometry, or of 
arithmetic character~\cite{Short,Big} (occurrence of miscellaneous series 
with {\em integer coefficients} 
like the nome, or the Yukawa couplings, 
emergence of {\em modular forms}~\cite{CalabiYauIsing}, 
algebraic varieties of Kodeira dimension zero~\cite{Wu}, ...),
and, secondly, properties of differential geometry character (the associated 
linear differential operators have 
{\em selected}  (or {\em special}~\cite{L12L21}) 
{\em differential Galois groups}~\cite{KatzInventiones,Katz,vdP}),
and this can be rephrased as 
{\em differential algebra properties}~\cite{bridged,unabridged}:
these operators are {\em homomorphic to their adjoint}, the symmetric, or 
exterior, powers of these operators, or of equivalent operators,
have  {\em rational solutions}. We have addressed these two different kinds of 
``special properties'' in two recent sets of 
papers. In a first set 
of papers~\cite{Short,Big}, we have shown that the $\, n$-fold 
integrals $\, \chi^{(n)}$, associated with the $\,n$-particle contribution
to the magnetic susceptibility of the Ising 
model~\cite{bo-gu-ha-je-ma-ni-ze-08}, as well as various other 
$\,n$-fold integrals of the ``Ising class''~\cite{Isingclass}, 
or $\, n$-fold integrals from
enumerative combinatorics~\cite{GoodGuttmann}, like lattice Green
 functions, are actually
{\em diagonals of rational functions}\footnote[3]{Diagonals of rational 
functions can be seen as the simplest generalization of algebraic functions
to transcendental (holonomic) functions~\cite{Short,Big}.}. As a consequence, 
they are  solutions of linear differential 
equations ``{\em Derived From Geometry}'', and
their power series expansions are {\em globally bounded}~\cite{Christol}, which 
means that, after just one rescaling of the expansion variable, they can be 
cast into series expansions with \emph{integer coefficients}.
In a second set of papers~\cite{bridged,unabridged}, we revisited
 miscellaneous linear differential operators, mostly associated with lattice
Green functions in arbitrary dimensions~\cite{GoodGuttmann,Guttmann}, but 
also {\em Calabi-Yau operators}~\cite{Guillera,TablesCalabi}, and order-seven 
operators corresponding to {\em exceptional differential 
Galois groups}~\cite{BognerGood,DettReit}. 
We have shown that the fact that these irreducible operators
have special differential Galois groups\footnote[8]{In the regular case 
the differential Galois group forms the Zariski closure
of the monodromy group (Schlesinger~\cite{Schlesinger}).}, 
can be simply understood, in a 
differential algebra viewpoint, from the fact that they are {\em homomorphic 
to their} (formal) {\em adjoints}\footnote[9]{The adjoint of a linear differential  operator
is the (formal) adjoint defined as in~\cite{bridged} 
(see equations (3) and (4) in~\cite{bridged}). As far as formal calculations
in Maple (DEtools)  are concerned, there is a command ``adjoint''
which can be used, see also~\cite{Homomorphisms}
 the command ``Homomorphisms''.}, and 
this can also be seen on the fact that the
{\em symmetric squares, or the exterior squares}, of these operators,
or of equivalent operators, have a {\em rational solution}.
Furthermore, in the examples displayed in~\cite{bridged,unabridged},
we saw that this homomorphism to the adjoint property always 
corresponded to a decomposition~\cite{bridged,unabridged}
 of the order-$2p \,$ linear 
differential operator $\, M_{2p}^{(n, \, 2p\, -n)}$, as (see equations 
(60), (83), (90),(91) in~\cite{unabridged})
\begin{eqnarray}
\label{genr1}
 \hspace{-0.95in}&& \quad \quad \quad\quad   \qquad 
 M_{2p}^{(n, \, 2p\, -n)} \, \,\, = \, \, \,  \, \, 
L_{2p \, -n} \cdot \, a(x) \cdot \, L_n 
\, \,\, + \, \, \, {{ \lambda} \over { a(x)}}, 
\end{eqnarray}
or, introducing\footnote[1]{Do note that 
the $\, {\tilde M}_{2p}^{(n, \, 2p\, -n)}$ operators (\ref{genr2}) are 
such that the functions, annihilated by $\, L_n$, are
automatically {\em eigenfunctions} of $\,{\tilde M}_{2p}^{(n, \, 2p\, -n)}$
with {\em eigenvalue} 
$\, \lambda$.} $ \,\,  {\tilde M}_{2p}^{(n, \, 2p\, -n)} 
\, = \, \, a(x) \cdot \, M_{2p}^{(n, \, 2p\, -n)}$,  as
\begin{eqnarray}
\label{genr2}
 \hspace{-0.95in}&& \quad \quad \quad \quad  \qquad 
 {\tilde M}_{2p}^{(n, \, 2p\, -n)} \, \,\, = \, \, \,  \, \, 
a(x) \cdot \, L_{2p \, -n} \cdot \, a(x) \cdot \, L_n 
\, \,\, + \, \, \,  \lambda, 
\end{eqnarray}
where the $\, L_m$'s  are {\em self-adjoint operators} 
of order $\, m$.
Such operators are, naturally, homomorphic to their adjoint, 
with intertwiners corresponding to these
decompositions (\ref{genr1}) and (\ref{genr2}):
\begin{eqnarray}
\label{genr1inter}
 \hspace{-0.95in}&& \quad \quad   \quad 
   L_n \cdot \,  a(x) \cdot \,  M_{2p}^{(n, \, 2p\, -n)} 
\, \,\, = \, \, \,  \, \,  
 adjoint(M_{2p}^{(n, \, 2p\, -n)}) \cdot \, a(x) \cdot \, L_n,  
\nonumber \\
\hspace{-0.95in}&& \quad \quad  \quad 
   M_{2p}^{(n, \, 2p\, -n)}  \cdot \,a(x)  \cdot \, L_{2p \, -n} 
\, \,\, = \, \, \,  \, \,  
L_{2p \, -n} \cdot \, a(x) \cdot \,  adjoint(M_{2p}^{(n, \, 2p\, -n)}).
\end{eqnarray}
In other words, these decompositions (\ref{genr1}), or (\ref{genr2}),
are closely related to 
the left, or right, intertwiners of the operator with its adjoint.
These decompositions have been seen
in all the quite large number of non-trivial
 lattice statistical physics examples,
 or enumerative combinatorics examples
in~\cite{bridged,unabridged}. Note that such decompositions 
enable to understand why certain
 differential Galois groups, appearing in lattice Green,
are included in {\em orthogonal groups} $\, O(n, \, \mathbb{C})$,
instead of symplectic groups  $\, Sp(n, \, \mathbb{C})$, that one might expect 
at first sight for an even order
 operator\footnote[9]{The intertwiners are of {\em odd orders}.}.

On all the examples of (minimal order) linear differential operators 
we have encountered in lattice statistical physics, and beyond, in 
enumerative combinatorics (see for 
instance~\cite{bo-bo-ha-ma-we-ze-09,Short,Big,CalabiYauIsing,Isingclass,High,bernie2010,Khi6,ze-bo-ha-ma-04,ze-bo-ha-ma-05c,
Renorm,mccoy3}), 
we have verified\footnote[5]{Except on the order twelve and order-21
operators occurring~\cite{High,Khi6} with $\, \chi^{(5)}$ 
and  $\, \chi^{(6)}$, because of their sizes.}
that they were actually homomorphic to their adjoint.

Since many {\em Derived From Geometry} $\, n$-fold 
integrals (``{\em Periods}''~\cite{Gri70}) occurring in physics, are seen 
to be {\em diagonals of rational functions}~\cite{Short,Big}, we also addressed
in~\cite{bridged,unabridged} several examples of (minimal order) 
operators annihilating diagonals of rational functions (not necessarily 
emerging from physics), and remarked, again,
that their irreducible factors\footnote[2]{The associated Hodge mixed structure
explains, to some extent, why the linear differential operators
annihilating diagonals of rational functions 
(like the $\, \chi^{(n)}$'s) have a large number of factors.} 
were, systematically, {\em homomorphic 
to their adjoint}.
This yields to envisage the conjecture\footnote[3]{This conjecture 
will be ruled out below in section (\ref{Specul}). }
that  all the {\em irreducible factors} of the {\em minimal order}
 linear differential 
operator annihilating a {\em diagonal of a rational function}, 
{\em should be homomorphic to their adjoint} (possibly 
on an algebraic extension).

Again a decomposition like (\ref{genr1}), or (\ref{genr2}), has been seen
in all these diagonals of rational functions  examples
in~\cite{bridged,unabridged}, except an order-six operator\footnote[8]{This 
operator is obtained from
a creative telescopic code (we thank A. Bostan for this calculation).
Following our traditional 
methods~\cite{ze-bo-ha-ma-04,High,bernie2010,Khi6,ze-bo-ha-ma-05c} 
to find the linear ODEs annihilating
a given series, one needs 234 coefficients to find an order-nine operator 
annihilating the series (the polynomial coefficients being of degree at most 22), 
or one finds the minimal order-six operator
with 390 coefficients (the polynomial coefficients 
being of degree at most 55).} $\, {\cal L}_6$
that was too large (see section (\ref{revisit}) below)
to quickly check whether it is {\em homomorphic to its adjoint}. 
This order-six linear differential
operator annihilates the diagonal of the (three variables) 
rational function
\begin{eqnarray}
\label{generic}
\hspace{-0.95in}&& \qquad  \qquad 
R(x,\, y, \, z)\, \, 
\,  = \, \, \,  \, 
 {{1} \over { 
1 \, -3\, x \, -5\, y \, -7\, z \,+x\, y \,+2\,y\,z^2\, +3\,x^2\,z^2}},
\end{eqnarray}
whose series expansion reads\footnote[2]{Use the maple 
command mtaylor(F, [x,y,z], terms), to get the series 
in three variables, then take the diagonal. Another method, 
in Mathematica is to install the risc package 
Riscergosum~\cite{Mathematica}, and in HolonomicFunctions
use the command FindCreativeTelescoping.}:
\begin{eqnarray}
\label{genericdiag}
\hspace{-0.95in}&&  \, \quad 
Diag(R(x,\, y, \, z))
\, \, = \, \, \,  \,\, \, 1 \,\, \,  +616 \, x \, \, 
 \, +947175 \, x^2 \, \, + \,1812651820 \, x^3 
\,\, \, + \, \,  \cdots 
\end{eqnarray}
This expansion of the rational function (\ref{generic}) 
can also be obtained from an expansion using {\em multinomial coefficients}
(see \ref{analyL6}). 

All these results are a strong incentive to accumulate other examples of 
minimal order operators annihilating
{\em diagonals of rational functions}, and analyze all their irreducible 
factors to confirm, or discard, the previous conjecture that
these factors are necessarily homomorphic to their adjoints, and 
see whether this homomorphism to the adjoint property is always 
associated to decompositions like (\ref{genr1}) or (\ref{genr2}).

Let us try to address this conjecture revisiting the order-six operator $\, {\cal L}_6$
in~\cite{bridged,unabridged}, in order  to see if $\, {\cal L}_6$ is also of the
form (\ref{genr1}) or (\ref{genr2}).

\vskip .1cm 

\section{Revisiting the order-six operator $\, {\cal L}_6$}
\label{revisit}

A first sketchy analysis of this operator $\, {\cal L}_6$
was performed in~\cite{bridged,unabridged}, which we recall 
now. We saw, for instance, that this operator is 
{\em not} MUM\footnote[1]{MUM means maximally unipotent 
monodromy~\cite{CalabiYauIsing,GoodGuttmann,Almkvist1}.}: it 
has four solution-series analytic at
the origin $\, x \, = \, 0$, one, among them, being not 
globally bounded~\cite{Christol}, and two being 
log-dependent formal series solutions.

Even though the order-six operator $\, {\cal L}_6$, which annihilates the 
diagonal of the rational function (\ref{generic}), was quite large, we 
were able to check that its {\em exterior square} is of generic order 15.
Switching to the associated differential {\em theta-system}~\cite{forthcoming}, 
we have been able to see that $\, {\cal L}_6$ (seen as a differential 
system) is  actually  
{\em homomorphic to its adjoint}. Furthermore, one actually finds 
that the {\em exterior square} of the {\em associated differential system}
 has a {\em rational solution} (but 
not its symmetric square). The differential Galois group thus corresponds 
to a {\em symplectic structure}. 

Since this order-six operator $\, {\cal L}_6$ has this symplectic structure, 
one can expect that its order-15 {\em exterior square} 
has a {\em rational solution}.
Actually, after some formal calculations work, we have first been able
to find this rational solution $\, R(x)$ which can be written
 as $\, R(x) \, = \, \, p_{10}/p_{12}/x$,
where $\, p_{10}$ and $\, p_{12}$ are two 
polynomials\footnote[9]{Everywhere in this paper $\, p_n$ will denote a polynomial 
of degree $\, n$ in $\, x$, with integer coefficients.} of degree
ten and twelve, with integer coefficients given in  \ref{polynomial}.

We can also consider the order-six linear
differential operator  $\, {\cal L}_6$ 
in~\cite{bridged,unabridged}, seen as a linear differential
 operator with {\em polynomial coefficients}.
 The head polynomial $\, h_6$ of the 
order-six operator $\, {\cal L}_6$, such that
${\cal L}_6 \,\, = \, \, \,h_6 \cdot D_x^6 \, + \, \, \cdots$,
reads $\, h_6 \, = \, \,  x^2 \cdot \, p_{12} \cdot \, p_{43}$, 
where  $\, p_{12}$
is the previous degree-twelve polynomial, and where $\, p_{43}$ 
is a polynomial with integer coefficients 
of degree $\, 43$ in $\, x$, given in \ref{polynomial}. 
As usual (see~\cite{High,bernie2010,Khi6,ze-bo-ha-ma-05c}), 
the roots of polynomial  $\, p_{43}$ 
corresponds to {\em apparent singularities} of the order-six 
operator  $\, {\cal L}_6$.
It is worth noting that, remarkably, the roots of the 
degree twelve polynomial $\, p_{12}$ {\em do not correspond to apparent 
singularities} but, actually, to {\em true singularities} 
of the order-six operator $\, {\cal L}_6$.

\vskip .1cm

\subsection{Homomorphisms of $\, {\cal L}_6$ with its adjoint}
\label{afirst}

Let us now focus on the fundamental relation,
underlined in~\cite{bridged,unabridged}, between
a linear differential operator and its adjoint,
seeking for an homomorphism between $\, {\cal L}_6$ and its adjoint,
and the associated intertwiners.
In a second step we will also consider the homomorphism of the previous intertwiners
with their adjoints, and so on. We will see that
finding this ``tower'' of intertwiners eventually
yields a simple decomposition of the order-six operator $\, {\cal L}_6$.

\vskip .1cm

\subsubsection{Homomorphism of $\, {\cal L}_6$ with its adjoint: 
   the $\, {\cal L}_4$ intertwiner \newline}
\label{P4int}

After some large formal calculations, performed using the DEtools Maple command
``Homomorphisms($\, {\cal L}_6$, adjoint(${\cal L}_6$))'',  we obtained 
an intertwiner, that we will denote $\, {\cal L}_4$, such that
\begin{eqnarray}
\label{defP4}
\hspace{-0.9in}&& \quad  \quad \quad \quad \quad  \quad 
adjoint({\cal L}_4)  \cdot \, {\cal L}_6  \,  \, \, \, = \, \, \, \,\, \, 
  adjoint({\cal L}_6) \cdot \, {\cal L}_4. 
\end{eqnarray}
The intertwiner $\, {\cal L}_4$ is a quite large {\em order-four} linear differential
operator. The coefficients of $\, D_x^n$, appearing in the 
 operator $\, p_{43}^2 \cdot \, {\cal L}_4$, are 
(quite large\footnote[8]{The polynomial 
coefficient of $\, D_x^n$ is of degree $\, 38 \, +n$, the head polynomial
being, up to an integer factor,
 the product of $\, x^2$ and of a polynomial $\, p_{28}$ of degree 28.}) 
polynomials with integer coefficients. This intertwiner $\, {\cal L}_4$ 
is {\em not conjugated to its adjoint}, 
which {\em excludes decompositions of $\, {\cal L}_6$ of the form}
 (\ref{genr1}) or (\ref{genr2}).

Remarkably the order-four intertwiner $\, {\cal L}_4$ is such that
its {\em exterior square} has the {\em same rational function solution} 
 $\, R(x) \, = \, \, p_{10}/p_{12}/x \, $ as $\, {\cal L}_6$. We explain 
this result later on in the paper (see Remark 1 in section (\ref{decomp})).
Since $\, {\cal L}_4$
has this symplectic structure, it 
is natural to seek for a decomposition of $\, {\cal L}_4$
of the form (\ref{genr1}), or (\ref{genr2}), by looking at the
homomorphisms of $\, {\cal L}_4$ with its adjoint\footnote[2]{Namely 
performing the Maple DEtools command``Homomorphisms(${\cal L}_4$, adjoint(${\cal L}_4$))''
and then, ``Homomorphisms(adjoint(${\cal L}_4$), $\, {\cal L}_4$)''.}.
Performing these calculations, we, indeed, obtained a decomposition of 
this form for $\, {\cal L}_4$, namely
\begin{eqnarray}
\label{decompP4}
\hspace{-0.9in}&& \quad  \,  \quad  \quad  \quad \quad \quad  \quad  
{\cal L}_4  \,  \, \, = \, \, \, \,\, 
 (N \cdot \, P \, \, + \, 1) \cdot \, r(x), 
\end{eqnarray}
where  $\, N $ and $\,P $ are two order-two {\em self-adjoint operators}, and
where $\, r(x)$ is a rational function. The operators  $\, N $ and $\,P $, 
and the rational function  $\, r(x)$, are given in \ref{decompL4app}.

\subsubsection{Decomposition of $\, {\cal L}_6$ \newline}
\label{decompcalL6}

Let us now perform the {\em euclidean right division} of  
$\, {\cal L}_6$ by  $\, {\cal L}_4$:
\begin{eqnarray}
\label{rightdivL6L4}
\hspace{-0.9in}&&  \quad  \quad  \quad \quad \quad    \quad \quad   \quad \quad 
{\cal L}_6 \, \, \, = \, \, \, \, M \cdot \, {\cal L}_4  \,\, + \, \,{\cal L}_2.  
\end{eqnarray}
The two operators $\, M$ and $\, {\cal L}_2$ are two order-two operators. One 
remarks, from direct calculations, that the order-two operator
$\, M$ is {\em exactly self-adjoint}. The exact expression of the order-two operator
$\, M$ is given in \ref{decompL4app} (see equation (\ref{defM})). 

One also remarks that  the order-two operator $\, {\cal L}_2$ is {\em exactly equal to 
the product}
\begin{eqnarray}
\label{defL2Pr}
\hspace{-0.9in}&& \quad  \quad \quad \quad \quad \quad  \quad \qquad 
 {\cal L}_2\,\, \, = \, \,\, \,  P \cdot r(x), 
\end{eqnarray}
 where $\, P$ is the self-adjoint operator introduced in (\ref{decompP4}).
Using (\ref{defP4}) and (\ref{decompP4}), and the fact that $\, M$ is self-adjoint, we 
note that  $\, {\cal L}_2\, = \, \, P \cdot r(x)$ can be seen as 
an intertwiner of the homomorphism of $\, {\cal L}_4$ with its adjoint:
\begin{eqnarray}
\label{defL2}
\hspace{-0.9in}&& \quad  \quad \quad \quad \quad
adjoint({\cal L}_2)  \cdot \, {\cal L}_4  \,  \, \, \, = \, \, \, \,\, 
  adjoint({\cal L}_4) \cdot \, {\cal L}_2. 
\end{eqnarray}
The fact that $\, P$ is self-adjoint, and that $\, {\cal L}_2\, = \, \, P \cdot r(x)$, 
corresponds to a last intertwining relation of $\, {\cal L}_2$ with its adjoint:
\begin{eqnarray}
\label{defL0}
\hspace{-0.9in}&& \quad  \quad \quad \quad \quad 
adjoint({\cal L}_0)  \cdot \, {\cal L}_2  \,  \, \, \, = \, \, \, \,\, 
  adjoint({\cal L}_2) \cdot \, {\cal L}_0, \quad \quad \quad
 {\cal L}_0 \, \, = \, \, \, r(x).
\end{eqnarray}

\vskip .1cm
\vskip .1cm 

{\bf Decomposition of $\, {\cal L}_6$:} From  (\ref{decompP4}) and (\ref{rightdivL6L4})
 one immediately deduces a {\em very
simple decomposition} for $\, {\cal L}_6$, {\em generalizing the 
decompositions} (\ref{genr1}) or (\ref{genr2}) of~\cite{bridged,unabridged}:
\begin{eqnarray}
\label{decompL6}
\hspace{-0.95in}&& \quad    \quad   \qquad \quad \quad  
{\cal L}_6 \, \,\,  \, = \, \, \,\, 
M \cdot \, (N \cdot \, P \, + \, 1) \cdot r(x)  \,\,\, + \,  P \cdot r(x)
\nonumber \\
\hspace{-0.95in}&& \quad    \quad   \qquad \quad \quad  \quad  
\, \,\,  \, \, \, = \, \, \,\, 
 (M \cdot \, N \cdot \, P \, + \, M \, + \, P) \cdot r(x). 
\end{eqnarray}

\vskip .1cm

The other intertwining relation between 
 $\, {\cal L}_6$  and its adjoint reads
\begin{eqnarray}
\label{defP4dual}
\hspace{-0.9in}&& \quad  \quad \quad \quad \quad \quad \quad 
 {\cal L}_6  \cdot \, {\cal M}_4  \,  \, \, \, = \, \, \, \,\, 
 adjoint({\cal M}_4)  \cdot \, adjoint({\cal L}_6), 
\end{eqnarray}
where the {\em order-four} intertwiner $\, {\cal M}_4$ can be simply expressed 
in terms of the two previous self-adjoint order-two operators $\, M$ and $\, N$:
\begin{eqnarray}
\label{calM4}
\hspace{-0.9in}&& \quad  \quad \quad \quad \quad \quad 
{\cal M}_4 \, \,= \,\, \,  \, 
{{1} \over {r(x)}} \cdot \, (N \cdot \, M \, + \, 1).
\end{eqnarray}

\vskip .1cm

\subsection{Similar decompositions\newline}
\label{Similar}

This order-six operator, $\, {\cal L}_6$, associated 
with the diagonal of a rational function~\cite{Short,Big},
shows that there exist operators, with selected 
differential Galois groups, with decompositions 
that {\em do not reduce} to the decompositions of~\cite{bridged,unabridged}, namely 
(\ref{genr1}) or (\ref{genr2}). Let us now show two other examples also generalizing 
decompositions (\ref{genr1}) and (\ref{genr2}).

\subsubsection{A simple order-three operator \newline}
\label{decompcalL3}

In fact, a much simpler example, 
corresponding to decomposition 
(\ref{decompL6}), can easily be found. Let us consider an order-two operator
 ($W(x)$ denotes its Wronskian)
\begin{eqnarray}
\label{formU1deuxun} 
\hspace{-0.95in}&&   \quad \quad \quad \quad  \quad 
 {\cal L}_2 \,\, = \, \, \,\,   
a_2(x) \cdot \, \Bigl(D_x^2 \,\, \,  
- \,  \, {{1} \over {W(x)}} \cdot \, {{d W(x)} \over {dx}} \cdot \, D_x\Bigr) 
 \,\, \,  + \, a_0(x), 
\end{eqnarray}
and let us consider an order-three linear differential 
operator $\, {\tilde {\cal L}}_3$, equivalent\footnote[5]{In the sense of
the equivalence of linear differential operators~\cite{vdP}.
}to the symmetric square 
of operator $\, {\cal L}_2$ given by\footnote[1]{Just perform the right division by $\, D_x$ 
of the LCLM of $\,Sym^2({\cal L}_{2})$ and $\, D_x$.} (\ref{formU1deuxun}):
\begin{eqnarray}
\label{equivL5}
\hspace{-0.95in}&& \quad  \quad \qquad \qquad 
  {\cal I}_1 \cdot \, Sym^2({\cal L}_{2}) \,\, \,  \,  = \,\, \, \,\, {\tilde {\cal L}}_3  \cdot \, D_x.
\end{eqnarray}
where $\, {\cal I}_1$ denotes an order-one intertwiner.
It is clear that this order-three operator $\, {\tilde {\cal L}}_3$ has, by construction, 
a selected differential Galois group, since it must reduce to the differential
Galois group of the ``underlying'' order-two operator $\, {\cal L}_{2}$, 
namely $\, SL(2, \,  \mathbb{C})$,
which is known to be, up to a 2-to-1 homomorphism,
 isomorphic to the {\em orthogonal group} $\, SO(3, \,  \mathbb{C})$.
One easily finds that the symmetric square of this 
order-three operator $\, {\tilde {\cal L}}_3$
has a rational solution, which is nothing but $\, W(x)^2$, the square of 
the Wronskian of $\, {\tilde {\cal L}}_2$. 
Let us introduce the order-two intertwiner $\, {\tilde {\cal L}}_2$ corresponding 
to the homomorphism of $\, {\tilde {\cal L}}_3$ with its adjoint:
\begin{eqnarray}
\label{tildecalL3adj}
\hspace{-0.9in}&& \quad  \quad \quad \quad \quad \quad  \quad 
adjoint({\tilde {\cal L}}_2)  \cdot \, {\tilde {\cal L}}_3   \,  \, \, \, = \, \, \, \,\, 
  adjoint({\tilde {\cal L}}_3) \cdot \, {\tilde {\cal L}}_2. 
\end{eqnarray}
Let us perform the {\em euclidean right division} of  $\, {\tilde {\cal L}}_3$ 
by  $\, {\tilde {\cal L}}_2$:
\begin{eqnarray}
\label{rightdivL3L2}
\hspace{-0.9in}&&  \quad  \quad  \quad \quad \quad    \quad \quad   \quad \quad 
{\tilde {\cal L}}_3 \, \, \, = \, \, \, \, 
M \cdot \, {\tilde {\cal L}}_2 \, \, + \, \, {\tilde {\cal L}}_1,  
\end{eqnarray}
The order-one operator $\, M$ is found to be {\em self-adjoint}. 
Let us perform, again, the euclidean right division of  $\, {\tilde {\cal L}}_2$ 
by $\, {\tilde {\cal L}}_1$ (namely the rest of the previous euclidean 
right division (\ref{rightdivL3L2})):
\begin{eqnarray}
\label{rightdivL2L1}
\hspace{-0.9in}&&  \quad  \quad  \quad \quad \quad    \quad \quad   \quad \quad 
{\tilde {\cal L}}_2 \, \, \, = \, \, \, \, 
N \cdot \, {\tilde {\cal L}}_1 \,\,  + \, \, {\tilde {\cal L}}_0.  
\end{eqnarray}
The order-one operator $\, N$ is found to be {\em self-adjoint}. 
One also finds that 
$\, {\tilde {\cal L}}_1  \, = \, \, P \cdot \, {\tilde {\cal L}}_0$,
where $\, {\tilde {\cal L}}_0$ is a function $\, r(x)$, and where 
$\, P$ is found to be {\em self-adjoint}. One thus deduces 
a decomposition of  $\, {\tilde {\cal L}}_3$ 
also of the form (\ref{decompL6}) 
\begin{eqnarray}
\label{decompL3}
\hspace{-0.95in}&& \quad    \quad   \qquad \quad \quad 
 {\tilde {\cal L}}_3 \, \,\,  \, = \, \, \,\, 
 (M \cdot \, N \cdot \, P \, + \, M \, + \, P) \cdot r(x), 
\end{eqnarray}
but where the {\em self-adjoint} operators $ \, M$, $\, N$ and $\, P$ 
are, this time, of {\em order one}.

\subsubsection{Similar decompositions for simple order-$n$ operators \newline}
\label{decompcalL5}

In a similar way, one considers, for 
$\, n \ge 5$ {\em odd} ($\, n \, = \,5, \, 7 \, \cdots  $)
 an order-$n$ linear differential 
operator $\,{\tilde {\cal L}}_n$, equivalent to the symmetric $(n-1)$-th power
of operator $\, {\cal L}_2$, given by (\ref{formU1deuxun}):
\begin{eqnarray}
\label{equivL5-30}
\hspace{-0.95in}&& \quad  \quad  \quad  \quad  \quad  \quad 
  {\cal I}_1^{(n-1)} \cdot \,Sym^{n-1}({\cal L}_{2}) \,\, \,  \,  = \,\, \, \,\, 
{\tilde {\cal L}}_n  \cdot \, D_x,
\end{eqnarray}
where $\, {\cal I}_1^{(n-1)}$ denotes an order-one intertwiner.
Again, one expects the differential Galois group of 
$\, {\tilde {\cal L}}_n$
to correspond to the differential Galois group 
of the underlying order-two operator $\, {\cal L}_2$, namely 
$\, SL(2, \,  \mathbb{C})$
or $\, PSL(2, \,  \mathbb{C}) \, \simeq \, \, SO(3, \,  \mathbb{C})$.
Performing the same calculations as in the previous 
section (\ref{decompcalL3}), 
one thus deduces a decomposition of  $\, {\tilde {\cal L}}_n$,  
also of the form (\ref{decompL6}) 
\begin{eqnarray}
\label{decompL5L7}
\hspace{-0.95in}&& \quad    \quad   \qquad \quad \quad 
 {\tilde {\cal L}}_n \, \,\,  \, = \, \, \,\, 
 (M \cdot \, N \cdot \, P \, + \, M \, + \, P) \cdot r(x), 
\end{eqnarray}
where the {\em self-adjoint} operators $ \, M$, $\, N$  
are of {\em order one}, but where the {\em self-adjoint} operator $\, P$ 
is of {\em odd order $\,\, n-2$}. The symmetric square of $\, {\tilde {\cal L}}_n$
does not have a rational solution, but has a {\em drop of order} : its order 
is less than the order
$\, n \cdot \, (n+1)/2$ one expects generically for 
an order-$n$ operator. In contrast the symmetric  square of the {\em adjoint}
of operator $\, {\tilde {\cal L}}_n$ has a rational solution which is the 
same as the {\em rational solution} of the symmetric  square of operator $\, M$,
namely the inverse of the head coefficient of the self-adjoint operator $\, M$.

\vskip .1cm 

{\bf Remark:} For $\, n$ {\em even} the order-$n$ linear differential 
operator $\, {\tilde {\cal L}}_n$, equivalent to the symmetric $\, (n-1)$-th power
of operator $\, {\cal L}_2$ (see (\ref{equivL5-30})), gives decompositions of the form 
$\, {\tilde {\cal L}}_n \, = \, \, (M \cdot \, N \, + \, 1) \cdot r(x)$, 
corresponding to {\em symplectic Galois groups}, 
where  $\, M$ is of order two and  $\, N$ are of {\em even order}
 $\, n-2$. This corresponds
 to the fact that the differential Galois group 
of the order-two operator $\, {\cal L}_2$, namely 
$\, SL(2, \, \mathbb{C})$, is {\em also\footnote[1]{$\, SL(2, \, \mathbb{C})$ 
is isomorphic to $Sp(2, \, \mathbb{C})$, to $Spin(3, \, \mathbb{C})$, 
and isomorphic, up to a 2-to-1 homomorphism,
to $\, SO(3, \, \mathbb{C}) \simeq \, PSL(2,\, \mathbb{C})$.} a symplectic group} 
 $\, SL(2, \, \mathbb{C})\, \simeq \, \, Sp(2, \, \mathbb{C})$.
The {\em exterior square} of $\, {\tilde {\cal L}}_n$ has, for $\, n=4$, 
a solution which is $\, W(x)^3$, but for $\, n$ even, $\, n>4$,  
this exterior square has no rational solution, it has a drop of order: its 
order is less than the order $\, n \cdot \, (n-1)/2$ one expects generically for 
an order-$n$ operator. In contrast the exterior square of the {\em adjoint}
of operator $\, {\tilde {\cal L}}_n$ has a rational solution which is the 
same as the rational solution of the exterior square of operator $\, M$,
namely the inverse of the head coefficient of the self-adjoint operator $\, M$.
To be symplectic or orthogonal is a property of the representation. It is 
not an intrinsic property of the group.

\vskip .1cm 

\subsection{Terminology: to be or not to be a selected differential Galois group}
\label{terminology}

A simple generalization of section (\ref{decompcalL3}) amounts to introducing 
 an order-three operator
 ($W(x)$ denotes its Wronskian)
\begin{eqnarray}
\label{formU1deuxuntrois} 
\hspace{-0.95in}&&   \quad  
 {\cal L}_3 \,\,\, = \, \, \,\,   
a_3(x) \cdot \, \Bigl(D_x^3 \,\, \,  
- \,  \, {{1} \over {W(x)}} \cdot \, {{d W(x)} \over {dx}} \cdot \, D_x^2\Bigr) 
 \,\, \,  + \, a_1(x) \cdot \, D_x \,\, + \, a_0(x), 
\end{eqnarray}
and considering, for instance, an order-six linear differential 
operator, equivalent to the symmetric square 
of operator $\, {\cal L}_3$, given by (\ref{formU1deuxun}):
\begin{eqnarray}
\label{equivL6symL3}
\hspace{-0.95in}&& \quad  \quad \qquad \qquad 
  {\cal I}_1 \cdot \,  Sym^2({\cal L}_{3}) 
\,\, \,  \,  = \,\, \, \,\,  {\tilde {\cal L}}_6 \cdot \, D_x.
\end{eqnarray}
where $\, {\cal I}_1$ is an order-one intertwiner.
This order-six operator $\, {\tilde {\cal L}}_6$ has, by construction, 
a ``special'' differential Galois group, since it must reduce to the differential
Galois group of the ``underlying''  order-three operator $\, {\cal L}_{3}$, namely 
$\, SL(3, \, C)$. However, the symmetric square, or exterior square,
of this order-six operator {\em does not have a rational solution}, or even a
{\em hyperexponential}~\cite{SingUlm} {\em solution}\footnote[9]{This can be
seen, more clearly, switching to the symmetric square of
companion system  of $\, {\cal L}_{3}$.}. This operator is {\em not homomorphic 
to its adjoint} (even in some algebraic extension). 

We will {\em not} say that such an operator corresponds to 
``Special Geometry''~\cite{L12L21}, even if it is clearly extremely
``special''. By ``Special Geometry'' we do mean ({\em only}) that the operator is 
{\em homomorphic to its adjoint}~\cite{bridged,unabridged}. 

\vskip .1cm 

\vskip .1cm 

\subsection{A first set of generalizations of this result \newline}
\label{decompcalL6}

In this section we will consider {\em self-adjoint} linear differential 
operators, denoted $\, M$, $\, N$, $\, P$, $\, Q$, ... 
not necessarily of the same 
order, but such that their orders have the {\em same parity} (all the 
operators are even order, or all the operators are odd order).
Recalling the result that two operators $\, A$ and $\, B$,
such that their orders have the {\em same parity}, are such 
that\footnote[1]{See also footnote 19 in~\cite{bridged}.}
$\, adjoint(A+B) \, = \, \, adjoint(A) \, + adjoint(B)$, one immediately 
deduces relations like
\begin{eqnarray}
\label{deux}
\hspace{-0.95in}&& \quad \quad \quad  
adjoint(N \cdot \, P \, + \, 1) \, \,  \, = \, \, \,  \, P \cdot \, N \, + \, 1, 
  \\
\label{un}
\hspace{-0.95in}&& \quad  \quad \quad
adjoint(M \cdot \, N \cdot \, P \, + \, M \, + \, P) 
\,  \, \, = \, \, \,  \, P \cdot \, N \cdot \, M \, \, + \, M \, + \, P,
\end{eqnarray}
enabling to deduce a decomposition for the adjoint of operators like 
(\ref{decompP4}) or (\ref{decompL6}) without any new calculations.

The intertwining relations  (\ref{defP4}), (\ref{defL2}),  (\ref{defL0})
form a ``{\em tower of intertwiners}''. Once the decompositions of $\, {\cal L}_6$, 
$\, {\cal L}_4$, $\, {\cal L}_2$ in terms of self-adjoint 
linear differential operators, and of the function $\, r(x)$, 
is known (see (\ref{decompP4}), (\ref{defL2Pr}), (\ref{decompL6})), the ``russian-doll'' 
structure of this ``tower of intertwiners'' becomes 
obvious, corresponding, in fact, to 
{\em simple operator identities}. Actually  the intertwining relation (\ref{defP4}) 
is, because of (\ref{decompL6}), (\ref{un}), (\ref{deux}),  nothing but the identity:
\begin{eqnarray}
\label{identity1} 
\hspace{-0.95in}&& \quad  \quad \quad  
 \Bigl(r(x) \cdot \, (P \cdot \, N \, + \, 1) \Bigr)\cdot \,
 \Bigl((M \cdot \, N \cdot \, P \, + \, M \, + \, P) \cdot \, r(x) \Bigr)
 \nonumber  \\
\hspace{-0.95in}&& \quad \quad \quad   \quad   \quad    \quad  
 \,\, = \, \, \,  \, \, 
  \Bigl(r(x) \cdot \, (P \cdot \, N \cdot \, M \, + \, M \, + \, P)\Bigr) 
 \cdot \,  \Bigl((N \cdot \, P \, + \, 1) \cdot \, r(x) \Bigr),
\end{eqnarray}
Obviously, we also have the identity
\begin{eqnarray}
\label{identity2} 
\hspace{-0.95in}&& \quad  \quad 
\Bigl((M \cdot \, N \cdot \, P \, + \, M \, + \, P) \cdot \, r(x) \Bigr) 
\cdot \, {{1} \over {r(x)}}  \cdot \, 
(N \cdot \, M \, + \, 1)
 \nonumber  \\
\hspace{-0.95in}&& \quad \quad \quad   \quad   \quad \,\, = \,\, \,  
 \Bigl((M \cdot \, N \, + \, 1) \cdot \, {{1} \over {r(x)}}\Bigr) 
 \cdot \, 
\Bigl(r(x) \cdot \,  (P \cdot \, N \cdot \, M \, + \, M \, + \, P)\Bigr),
\end{eqnarray}
which actually corresponds to the other intertwining 
relation (\ref{defP4dual}) between an operator, like $\, {\cal L}_6$
 in (\ref{decompL6}), and its adjoint, the exact 
expression (\ref{calM4}) of the intertwiner $\, {\cal M}_4$ 
 being deduced, without any further calculations, from identity (\ref{identity2}).

\vskip .1cm 

{\bf Remark 1:} As noticed\footnote[5]{See the sentence after equation (8) 
in ~\cite{bridged}.} in a previous paper~\cite{bridged}, the two intertwiners 
$\, {\cal L}_4$  and  $\, {\cal M}_4$ are inverse operators 
modulo $\, {\cal L}_6$. This, in fact, corresponds to the following identity:
\begin{eqnarray}
\label{inversemodulo}
 \hspace{-0.9in}&& \quad  \quad \quad  
\Bigl( {{1} \over {r(x)}} \cdot \, (N \cdot \, M \, + \, 1)\Bigr)  \cdot \,
\Bigl( (N \cdot \, P \, + \, 1) \cdot \, r(x)\Bigr) 
\nonumber \\ 
 \hspace{-0.9in}&& \quad \quad  \quad \quad \quad  \, \, \, = \, \, \, \, \, 
1 \, \, \, \,  + \, \, \, 
 \Bigl({{1} \over {r(x)}} \cdot \, N \Bigr) \cdot \, 
 \Bigl(  (M \cdot \, N \cdot \, P \, + \, M \, + \, P)  \cdot \, r(x) \Bigr), 
\end{eqnarray}
which means that 
$\, \, {\cal M}_4 \cdot \,{\cal L}_4 \, = \, \, 1 \,\, (mod. \,  {\cal L}_6)$.
Of course, we also have a ``dual'' inverse identity for the adjoint of the operator
(namely $\, \, {\cal L}_4 \cdot \, {\cal M}_4 \, = \, \, 1$
$ \,\, (mod. \,  adjoint({\cal L}_6))$):
\begin{eqnarray}
\label{inversemoduloadj}
 \hspace{-0.9in}&& \quad  \quad \quad  
\Bigl( (N \cdot \, P \, + \, 1)  \cdot \,  r(x) \Bigr)  \cdot \,
\Bigl( {{1} \over {r(x)}} \cdot \, (N \cdot \, M \, + \, 1) \Bigr) 
\nonumber \\ 
 \hspace{-0.9in}&& \quad \quad  \quad \quad \quad  \, \, \, = \, \, \, \, \, \, \, 
1 \, \, \,\,  + \, \, \,  \,  
 \Bigl( N \cdot \, {{1} \over {r(x)}} \Bigr) \cdot \, 
 \Bigl( r(x)  \cdot \,  (P \cdot \, N \cdot \, M \, + \, M \, + \, P) \Bigr).
\end{eqnarray}

{\bf Remark 2:} It is easy to generalize  identities (\ref{identity1}), (\ref{identity2})
with more operators (we remove, here, the ``dressing'' by the function $\, r(x)$):
\begin{eqnarray}
\label{identity4}
\hspace{-0.95in}&& \qquad  
(Q \cdot \, P \cdot \, N \, + \, Q \, + \, N) 
\cdot \, (M \cdot \, N \cdot \, P  \cdot \, Q 
\, + \, M \cdot \, Q  \, + \, P \cdot \, Q \, 
+ \, M \cdot \, N \, + \, 1) 
\nonumber \\
\hspace{-0.95in}&&\, \qquad \qquad    \,\, \, = \, \, \,  \, \,   
 (Q \cdot \, P \cdot \, N  \cdot \, M \, + \, Q \cdot \, M  \, + \, Q \cdot \, P \, 
+ \, N \cdot \, M \, + \, 1)
\nonumber \\
\hspace{-0.95in}&& \qquad  \qquad \qquad \qquad   
 \times \, (N \cdot \, P \cdot \, Q \, + \, Q \, + \, N),  
\end{eqnarray}
and:
\begin{eqnarray}
\label{identity3}
\hspace{-0.95in}&& \qquad  
(M \cdot \, N \cdot \, P  \cdot \, Q \, + \, M \cdot \, Q  \, + \, P \cdot \, Q \, 
+ \, M \cdot \, N \, + \, 1) \cdot \, 
 (P \cdot \, N \cdot \, M \, + \, M \, + \, P) 
\nonumber \\
\hspace{-0.9in}&&\, \, \qquad \qquad    \, \, = \, \, \, \,  \,   
 (M \cdot \, N \cdot \, P \, + \, M \, + \, P)
 \nonumber \\
\hspace{-0.95in}&& \qquad  \qquad \qquad  \quad  
\times \,  (Q \cdot \, P \cdot \, N  \cdot \, M \, + \, Q \cdot \, M 
 \, + \, Q \cdot \, P \, + \, N \cdot \, M \, + \, 1). 
\end{eqnarray}
If one assumes that the four operators $\, M$, $\, N$, $\, P$ and $\, Q$
 are {\em self-adjoint} operators of the {\em same parity order},
these identities can be interpreted as intertwining relations 
between an operator and its adjoint, the operator having the new decomposition:
\begin{eqnarray}
\label{identity4decF}
\hspace{-0.9in}&&\quad \quad \quad  \quad 
L \, \, = \, \, \, 
(M \cdot \, N \cdot \, P  \cdot \, Q \, + \, M \cdot \, Q  \, + \, P \cdot \, Q \, 
+ \, M \cdot \, N \, + \, 1) \cdot \, r(x).
\end{eqnarray}
Since these intertwining relations {\em do not require} that
the {\em self-adjoint} operators are of the {\em same order}\footnote[2]{But are
of the same parity order.}, we thus discover, with 
decompositions (\ref{decompL6}) or (\ref{identity4decF}), {\em extremely large families} 
of linear differential operators for which we are sure that their 
{\em differential Galois groups will be special}. In the next section we generalize
the decompositions (\ref{decompP4}), (\ref{decompL6}), (\ref{identity4decF}), 
with, respectively, two, three, four {\em self-adjoint} operators, to an 
{\em arbitrary number} of {\em self-adjoint} operators.

\vskip .1cm 

{\bf Remark 3:} The smaller factors, in the last two identities (\ref{identity4})
and (\ref{identity3}), can, thus, be 
seen as intertwiners. Again, one has two of these intertwiners 
which are {\em inverse operators modulo the operator}. This corresponds to the 
following identity generalizing (\ref{inversemodulo}):
\begin{eqnarray}
\label{inversemodulo2}
\hspace{-0.95in}&&   
\Bigl( {{1} \over {r(x)}}  \cdot \, (P \cdot \, N \cdot \, M \, + \, M \, + \, P) \Bigr)
\cdot \, \Bigl((N \cdot \, P \cdot \, Q \, + \, N \, + \, Q) \cdot \, r(x) \Bigr) 
 \, \,\,\, \, = \, \, \,\, \, -1 
 \\ 
 \hspace{-0.95in}&& \,    \,  
+ \Bigl({{1} \over {r(x)}}  \cdot \, (P \cdot \, N   \, +1) \Bigr)  \cdot \, 
\Bigl( (M \cdot \, N \cdot \, P  \cdot \, Q \, 
+ \, M \cdot \, Q  \, + \, P \cdot \, Q \, + \, M \cdot \, N \, + \, 1)  \cdot \, r(x) \Bigr).
\nonumber
\end{eqnarray}
which actually amounts to saying that two intertwiners are {\em inverse  
operators modulo the operator} $\, L$ given by (\ref{identity4decF}). Of 
course, we also have the ``dual'' 
inverse relation {\em modulo the adjoint of the operator}
 $\, L$ given by (\ref{identity4decF}):
\begin{eqnarray}
\label{inversemodulo2adj}
\hspace{-0.97in}&&   
\Bigl((N \cdot \, P \cdot \, Q \, + \, N \, + \, Q) \cdot \, r(x) \Bigr)\cdot \, 
\Bigl( {{1} \over {r(x)}}  \cdot \, (P \cdot \, N \cdot \, M \, + \, M \, + \, P) \Bigr)
 \, \,\,\, \, = \, \, \,\, \, \, -1 
 \\ 
 \hspace{-0.97in}&& \,     
+ \Bigl((N \cdot \, P   \, +1) \cdot  {{1} \over {r(x)}} \Bigr) \cdot  
\Bigl( r(x) \cdot   (Q \cdot \, P \cdot \, N  \cdot \, M \, 
+ \, Q \cdot \, M  \, + \, Q \cdot \, P \, + \, N \cdot \, M \, + \, 1) \Bigr).
\nonumber
\end{eqnarray}

\vskip .1cm 

\section{Tower of intertwiners and canonical decomposition
 of linear differential operators}
\label{tower}

With the previous identities (\ref{identity1}), (\ref{identity2}), 
one sees that the selected linear differential
operator, and its successive intertwiners (between operators and their adjoints),
have decompositions of a similar form. It is thus tempting to try
to find, systematically, the decompositions of these selected operators
from successive intertwiners of operators with their adjoints, ideally in 
an algorithmic recursion process.

In the next section (and in \ref{towerdualapp} and \ref{inverseapp})  
the operators will be denoted $\, L_{[N]}$, where 
$\, N$ will {\em not denote the order} of the operators, as 
we always do~\cite{Short,Big,bridged,unabridged,L12L21}, but 
an integer associated with the number of successive intertwiners.
Similarly the integer $\, n$
of the operators denoted $\, U_n$ (or $\, V_n$ in \ref{towerdualright})
{\em does not correspond to the order} of these operators.

\subsection{The tower of intertwiners from a simple euclidean right division}
\label{towersimpleright}

Let us consider an order-$q$ linear differential operator $\, L_{[N]}$,
homomorphic to its adjoint. We have shown, in 
previous papers~\cite{bridged,unabridged},
that this means that there exists an intertwiner, we will denote $\, L_{[N-1]}$,
such that 
\begin{eqnarray}
\label{tower1} 
\hspace{-0.8in}&& \quad \quad \quad \quad 
adjoint(L_{[N]}) \cdot  \, L_{[N-1]}
 \,\, \,  \,  = \, \, \,\, \,  \,  adjoint(L_{[N-1]})  \cdot  \, L_{[N]}. 
\end{eqnarray}
In Maple, $\, Homomorphisms(L_{[N]}, \, adjoint(L_{[N]}))$ is the  command
one should use to obtain this intertwiner $\, L_{[N-1]}$.
From the previous intertwining relation (\ref{tower1}), it is natural to compare 
the original operator $\, L_{[N]}$ and this new intertwiner $\, L_{[N-1]}$,
performing an {\em euclidean right division}:
\begin{eqnarray}
\label{euclid1} 
\hspace{-0.8in}&& \quad \quad \quad \quad \quad \quad 
L_{[N]} \,\, \,  \,  = \, \, \,\, \,  \,  U_N \cdot \, L_{[N-1]} \, \,  + \,  L_{[N-2]}, 
\end{eqnarray}
where $\,  U_N$  is the quotient
of the euclidean right division, and $\,  L_{[N-2]}$ 
is the remainder of the euclidean right division. 

Reinjecting the euclidean right division decomposition (\ref{euclid1}) 
in the intertwining relation (\ref{tower1}), one gets
\begin{eqnarray}
\label{tower1bis} 
\hspace{-0.8in}&& \quad \,
adjoint(L_{[N]}) \cdot  \, L_{[N-1]}
 \,\, \,  \,  = \, \, \,\, \,  \,
  adjoint(L_{[N-1]})  \cdot  \, \Bigl(U_N \cdot \,L_{[N-1]} \, + \,  L_{[N-2]}\Bigr), 
\end{eqnarray}
or, equivalently:  
\begin{eqnarray}
\label{tower1ter} 
\hspace{-0.75in}&&  \,\quad \quad 
\Bigl(adjoint(L_{[N]}) \, - \, \, adjoint(L_{[N-1]})  \cdot  \, U_N\Bigr) \cdot  \, L_{[N-1]} 
\nonumber \\
\hspace{-0.75in}&& \quad \quad\quad \quad  \quad \quad  \quad
   \, \,\, \,  \,  = \, \, \,\, \,  \, \, 
  adjoint(L_{[N-1]})  \cdot  \,  L_{[N-2]}. 
\end{eqnarray}
Since it was shown in~\cite{bridged,unabridged} that the intertwining relation
between an {\em irreducible} operator $\, L_{[N-1]}$ and its adjoint is necessarily 
of the form (\ref{tower1}), one 
deduces the equality
\begin{eqnarray}
\label{deduced1}
\hspace{-0.8in}&& \quad \quad \quad 
adjoint(L_{[N]}) \, - \, \, adjoint(L_{[N-1]})  \cdot  \, U_N
 \, \, = \, \, \, \,  adjoint(L_{[N-2]}),
\end{eqnarray}
the previous relation (\ref{tower1ter}), rewriting as:
\begin{eqnarray}
\label{tower2}
\hspace{-0.8in}&& \quad \quad \quad 
adjoint(L_{[N-2]}) \cdot  \, L_{[N-1]} 
 \,\, \,  \,  = \, \, \,\, \,  \,
  adjoint(L_{[N-1]})  \cdot  \,  L_{[N-2]}, 
\end{eqnarray}
which is a new intertwining relation exactly of the {\em same form} as the first 
intertwining relation (\ref{tower1}).

By definition of the euclidean  right division, the two terms 
$\, L_{[N]}$ and $\, U_N \cdot  \, L_{[N-1]}$
in (\ref{euclid1}) are of the same order. Recalling, 
for two operators $\, A$ and $\, B$ of the same order (or even 
orders of the same parity),
the result\footnote[1]{Obvious from
 the definition of the adjoint of an operator, see~\cite{bridged,unabridged}.}
 that $\, adjoint(A \, -B) \,$
$ = \, adjoint(A) \, -adjoint(B)$,  one gets from (\ref{euclid1})
\begin{eqnarray}
\label{adjeuclid1} 
\hspace{-0.9in}&& \, \quad \quad \quad \quad \,  adjoint(L_{[N-2]})\,  \,  = \, \, \,\, \, 
 adjoint(L_{[N]} \, - \, U_N \cdot  \, L_{[N-1]})
\nonumber \\ 
\hspace{-0.9in}&& \, \quad \quad \quad \quad  \quad \,  
 \,  \,  = \, \, \,\, \, 
adjoint(L_{[N]}) \,\, \, - \, \, adjoint(U_N \cdot  \, L_{[N-1]}) 
\nonumber \\ 
\hspace{-0.9in}&& \, \quad \quad \quad \quad  \quad \,\, \,\,  = \, \, \,\,  
adjoint(L_{[N]}) \,\, \, - \, \, adjoint(L_{[N-1]}) \cdot \, adjoint(U_N). 
\end{eqnarray}
Comparing (\ref{adjeuclid1}) with (\ref{deduced1}), one deduces the result that the
operator $\, U_N$ , in the  euclidean division (\ref{euclid1}), is, necessarily, 
{\em exactly self-adjoint}:
\begin{eqnarray}
\label{self1}
\hspace{-0.8in}&& \quad \quad \quad \qquad \quad \quad 
U_N \,\, \,  \,  = \, \, \,\, \, adjoint(U_N).
\end{eqnarray}
It is straightforward to see that one can go on recursively 
\begin{eqnarray}
\label{recurs}
\hspace{-0.95in}&& \quad  
L_{[N]} \,\, \,  \,  = \, \, \,\, \,  \, 
 U_N \cdot \,L_{[N-1]} \, + \,  L_{[N-2]}, 
\quad \quad  \quad \,
L_{[N-1]} \,\, \,  \,  = \, \, \,\, \,  \, 
 U_{N-1} \cdot \,L_{[N-2]} \, + \,  L_{[N-3]}, \quad 
\nonumber \\
\hspace{-0.95in}&& \quad  
L_{[N-2]} \,\, \,  \,  = \, \, \,\, \,  \,  U_{N-2} \cdot \,L_{[N-3]} \, + \,  L_{[N-4]}, 
\quad  \quad \cdots 
\nonumber \\
\hspace{-0.95in}&& \quad  
L_{[N-p]} \,\, \,  \,  = \, \, \,\, \,  \,  U_{N-p} \cdot \,L_{[N-p-1]} \, + \,  L_{[N-p-2]},
\quad  \quad \cdots 
\end{eqnarray}
and 
\begin{eqnarray}
\label{recurs2}
\hspace{-0.95in}&& \quad   \quad  \quad 
adjoint(L_{[N-1]}) \cdot  \, L_{[N]} 
 \,\, \,  \,  = \, \, \,\, \,  \,
  adjoint(L_{[N]})  \cdot  \,  L_{[N-1]},
 \nonumber \\
\hspace{-0.95in}&& \quad  \quad  \quad 
adjoint(L_{[N-2]}) \cdot  \, L_{[N-1]} 
 \,\, \,  \,  = \, \, \,\, \,  \,
  adjoint(L_{[N-1]})  \cdot  \,  L_{[N-2]},
\nonumber \\
\hspace{-0.95in}&& \quad  \quad  \quad 
 adjoint(L_{[N-3]}) \cdot  \, L_{[N-2]} 
 \,\, \,  \,  = \, \, \,\, \,  \,
  adjoint(L_{[N-2]})  \cdot  \,  L_{[N-3]}, 
 \quad  \, \, \, \, \, \, \cdots  \\
\hspace{-0.95in}&& \quad  \quad  \quad 
 adjoint(L_{[N-p-1]}) \cdot  \, L_{[N-p]} 
 \,\, \,  \,  = \, \, \,\, \,  \,
  adjoint(L_{[N-p]})  \cdot  \,  L_{[N-p-1]}, 
 \quad  \, \,  \, \, \, \, \cdots \nonumber 
\end{eqnarray}
thus building a ``{\em tower of intertwiners}''. 
Let us see how this recursion stops. 
In the sequence of intertwining relation (\ref{recurs2}), the orders 
of the intertwiners decrease, and finally reach the moment where
$\,  L_{[N-p-1]}$ is just a function $\, r(x)$, which means 
that the operator $\,  L_{[N-p]}$ is {\em simply conjugated} to its adjoint:
\begin{eqnarray}
\label{rL}
\hspace{-0.95in}&& \quad  \quad  \quad 
 r(x) \cdot \, L_{[N-p]} 
 \,\, \,  \,  = \, \, \,\, \, 
\, adjoint(L_{[N-p]}) \cdot  \, r(x), 
\quad \qquad \quad \quad \hbox{or:}
\end{eqnarray}
\begin{eqnarray}
\label{rL}
\hspace{-0.95in}&&  \quad \quad \quad 
 U_{N-p}\,  \, \,  = \, \,  \, \, L_{[N-p]} \cdot \, {{1} \over {r(x)}}
 \,   \,  = \, \, \, \, 
\,  {{1} \over {r(x)}}  \cdot \, adjoint(L_{[N-p]})
\nonumber \\ 
\hspace{-0.95in}&& \qquad \qquad \qquad \qquad  \,   = \, \,  \, \, \, \,
 adjoint\Bigl(L_{[N-p]} \cdot \, {{1} \over {r(x)}}\Bigr),
\end{eqnarray}
which means that the operator $\, U_{N-p} \,$ is {\em exactly self-adjoint}.

This is in agreement with the last euclidean division in (\ref{recurs}),
i.e.  $\,\, L_{[1]} \,  = \, \, U_{1} \cdot \,L_{[0]} \, \, + \,  L_{[-1]}$, 
with $\, L_{[0]} \,  = \, \, r(x)$ and $\, L_{[-1]} \,  = \, \,0$, and where $\, U_{1}$
is exactly self-adjoint, namely:
\begin{eqnarray}
\label{last}
\hspace{-0.95in}&& \quad  \quad  \quad \quad \quad  \quad \quad  \quad \quad  \quad 
L_{[1]} \,\, \,  \,  = \, \, \,\, \,  \,  U_{1} \cdot \, r(x).
\end{eqnarray}
This operator $\, U_{1}$ appears, in the decomposition of $\, L_{[N]}$, 
as the {\em rightmost}\footnote[5]{Throughout the paper we call ``rightmost operator''
in the decomposition of $\, L_{[N]}$, the rightmost operator 
appearing in the first and largest term
of the decomposition: see the examples (\ref{lastm1}), ... (\ref{last6}) below.} 
operator of this decomposition (see (\ref{lastm1}), ... (\ref{last6}) below). 
This rightmost self-adjoint operator $\, U_1$ plays a {\em selected role 
in the decomposition}. It will be seen, in \ref{comment}, that the
rational solutions of the exterior, or symmetric, squares of the operators 
with special differential Galois groups depend only on $\, U_{1} \cdot \, r(x)$,
and {\em not} on the other $\, U_{N}$'s in the decomposition. It will be seen 
in section (\ref{parity}) that the order of $\, U_1$ is just constrained to have
the {\em same parity} as the orders of the other $\, U_{n}$'s, the {\em even} order 
corresponding to {\em symplectic} differential Galois groups, and the {\em odd} order
corresponding to {\em orthogonal} differential Galois groups.
 
\vskip .1cm 

\subsection{Canonical decomposition \newline}
\label{decomp}

From these sequences of euclidean right-divisions on the 
successive intertwiners (\ref{recurs}), 
together with the initial operators $\, L_{[0]} \,  = \, \, r(x)$ 
and $\, L_{[-1]} \,  = \, \,0$,
one immediately deduces canonical decompositions of the operator $\, L_{[N]}$.
Let us show the first decompositions 
\begin{eqnarray}
\label{lastm2}
\hspace{-0.95in}&& \quad    
L_{[0]} \,\, \,  \,  = \, \, \,\, \, r(x), \qquad \qquad  \qquad 
L_{[1]} \,\, \,  \,  = \, \, \,\, \,   U_{1} \cdot r(x), 
\end{eqnarray}
\begin{eqnarray}
\label{lastm1}
\hspace{-0.95in}&& \quad   
L_{[2]} \,\, \,  \,  = \, \, \, 
U_{2} \cdot \, L_{[1]} \, + \, L_{[0]}
 \, \,  \,  = \, \, \,\, 
 (U_{2} \cdot \, U_{1} \,\, + 1) \cdot r(x), 
\end{eqnarray}
\begin{eqnarray}
\label{last}
\hspace{-0.95in}&& \quad   
L_{[3]} \,\, \, \, = \, \,\, \, \,\,   U_{3} \cdot \, L_{[2]} \, + \, L_{[1]} 
\, \, \, = \, \, \,\, \,
 (U_{3} \cdot \, U_{2} \cdot \, U_{1} \,\, + U_{1} \,+ \, U_{3}) \cdot r(x), 
\end{eqnarray}
\begin{eqnarray}
\hspace{-0.95in}&& \quad    
L_{[4]} \,\, \, \, = \, \, \,\, \, \,  U_{4} \cdot \, L_{[3]} \, + \, L_{[2]}
 \, \, \,\,  = \, \, \,\, \,\, 
 U_{4} \cdot \, (U_{3} \cdot \, L_{[2]} \, + \, L_{[1]}) \, + \, L_{[2]} 
\nonumber \\
\label{last2}
\hspace{-0.95in}&& \quad \quad   \quad    \,  = \, \, \, 
 (U_{4} \cdot \, U_{3} \cdot \, U_{2} \cdot \,  U_{1}
\, + U_{4} \cdot \, U_{1} \, 
+ \, U_{2} \cdot \,  U_{1} \, 
+ \, U_{4} \cdot \,  U_{3} \, + \, \, 1) \cdot r(x),  
\end{eqnarray}
\begin{eqnarray}
\hspace{-0.95in}&& \quad   
L_{[5]} \,\, \,\,  = \, \, \,\,\,   U_{5} \cdot \, L_{[4]} \, + \, L_{[3]}
 \,\,\, \,   = \, \, \,\,\, \, 
 U_{5} \cdot \, (U_{4} \cdot \, L_{[3]} \, + \, L_{[2]}) \, + \, L_{[3]}
  \, \, \, = \, \, \,\, \,\cdots 
\nonumber \\
\hspace{-0.95in}&& \quad   \quad \quad     \quad   \,  = \, \, \, 
  (U_{5} \cdot \,U_{4} \cdot \, U_{3} \cdot \, U_{2} \cdot \,  U_{1}
\,\,\, + U_{5} \cdot \, U_{4} \cdot \, U_{1} \,  \,
+ \, U_{5} \cdot \, U_{2} \cdot \,  U_{1} \, 
\nonumber \\
\label{last3}
\hspace{-0.95in}&&  \qquad \quad   \quad  \quad \quad    \quad 
 + \, U_{5} \cdot \, U_{4} \cdot \,  U_{3}
+ \, U_{3} \cdot \, U_{2} \cdot \, U_{1} 
\,\,\,\,  + U_{1} \, \, + \, U_{3}\,+ \, U_{5}) \cdot r(x),  
\end{eqnarray}
\begin{eqnarray}
\hspace{-0.95in}&& \quad    
L_{[6]} \,\,\,\,  = \, \, \, \,\, U_{6} \cdot \, L_{[5]} \, + \, L_{[4]}
 \, \,\, \, = \, \, \, \,\,\,
 U_{6} \cdot \, (U_{5} \cdot \, L_{[4]} \, + \, L_{[3]}) \, + \, L_{[4]}
 \, \, \, = \, \, \,\, \,\cdots 
\nonumber \\
\label{last4}
\hspace{-0.95in}&&   \quad  \quad   \quad  
\,  = \, \, \,
 (U_{6} \cdot \, U_{5} \cdot \,U_{4} \cdot \, U_{3} \cdot \, U_{2} \cdot \,  U_{1}
\,\,\,\,\, + U_{6} \cdot \, U_{5} \cdot \, U_{4} \cdot \, U_{1} \,  \,
+ \, U_{6} \cdot \, U_{5} \cdot \, U_{2} \cdot \,  U_{1} \, 
\nonumber \\
\hspace{-0.95in}&& \quad  \quad  \quad    \quad    \qquad  
\,\, + \,U_{6} \cdot \,  U_{5} \cdot \, U_{4} \cdot \,  U_{3}
+ \,U_{6} \cdot \,  U_{3} \cdot \, U_{2} \cdot \, U_{1}  \,
+ \, U_{4} \cdot \, U_{3} \cdot \, U_{2} \cdot \,  U_{1}
\nonumber \\
\hspace{-0.95in}&& \quad  \quad  \quad   \quad \quad \qquad   \qquad  
  \, \,   + U_{6} \cdot \, U_{1} \,
   \,+ \, U_{6} \cdot \, U_{3} \,+ \, U_{6} \cdot \, U_{5} \,
 + U_{4} \cdot \, U_{1}  \, + \, U_{2} \cdot \,  U_{1} \, 
\nonumber \\
\hspace{-0.95in}&& \quad  \quad  \quad  \quad \quad  \quad   \quad  \qquad    \quad  \qquad   
 \, \, + \, U_{4} \cdot \,  U_{3}\,  \, \,\, + \, 1) \cdot \, r(x), 
\end{eqnarray}
\begin{eqnarray}
\label{last5}
\hspace{-0.95in}&& \quad   
L_{[7]} \,\,\,\,  = \, \, \,\,\,  U_{7} \cdot \, L_{[6]} \, + \, L_{[5]}
 \,\,\, \,  = \, \, \,\,\,\, 
 U_{7} \cdot \, (U_{6} \cdot \, L_{[5]} \, + \, L_{[4]}) \, + \, L_{[5]} 
 \, \, \, = \, \, \,\, \,\cdots 
\nonumber \\
\label{last6}
\hspace{-0.95in}&&  \quad  \quad 
\,  = \, \, \,
 (U_{7} \cdot \, U_{6} \cdot \, U_{5} \cdot \,U_{4} \cdot \, U_{3} \cdot \, U_{2} \cdot \,  U_{1}
\,\,\,\,\, + U_{7} \cdot \, U_{6} \cdot \, U_{5} \cdot \, U_{4} \cdot \, U_{1} \,  \,
\nonumber \\
\hspace{-0.95in}&& \quad  \quad  \quad    \quad   
+ \,U_{7} \cdot \, U_{6} \cdot \,  U_{3} \cdot \, U_{2} \cdot \, U_{1}  \,
+ \, U_{7} \cdot \, U_{4} \cdot \, U_{3} \cdot \, U_{2} \cdot \,  U_{1} \,
+ \, U_{7} \cdot \, U_{6} \cdot \, U_{5} \cdot \, U_{2} \cdot \,  U_{1}
\nonumber \\
\hspace{-0.95in}&& \quad  \quad  \quad   \quad 
 + \, U_{7} \cdot \, U_{6} \cdot \,  U_{5} \cdot \, U_{4} \cdot \,  U_{3}
+\, U_{5} \cdot \,U_{4} \cdot \, U_{3} \cdot \, U_{2} \cdot \,  U_{1} 
\, + U_{7} \cdot \, U_{6} \cdot \, U_{1} \,
\nonumber \\
\hspace{-0.95in}&& \quad  \quad  \quad   \quad   
+ \, U_{7} \cdot \, U_{6} \cdot \, U_{3} \,
+ \, U_{7} \cdot \, U_{6} \cdot \, U_{5} \,
 + U_{7} \cdot \, U_{4} \cdot \, U_{1}  
\, + \,U_{7} \cdot \, U_{2} \cdot \,  U_{1} \, 
\nonumber \\
\hspace{-0.95in}&& \quad  \quad  \quad  \quad
  + \,U_{7} \cdot \,  U_{4} \cdot \,  U_{3}\,  \,
\,\, + U_{5} \cdot \, U_{4} \cdot \, U_{1} \,  \,
+ \, U_{5} \cdot \, U_{2} \cdot \,  U_{1} \,
+ \, U_{5} \cdot \, U_{4} \cdot \,  U_{3}
 \nonumber \\
\hspace{-0.95in}&& \quad  \quad  \quad  \quad 
+ \, U_{3} \cdot \, U_{2} \cdot \, U_{1} 
\,\,\,\,  + U_{1} \, \, + \, U_{3}\,+ \, U_{5}
\,\, + \, U_{7} ) \cdot \, r(x), 
\end{eqnarray}
and so on. The number of terms in the $\, L_{[N]}$'s, namely
$\, 1, \, 2, \, 3, \, 5, \, 8, \, 13, \, 21, \cdots $
 corresponds to the Fibonacci sequence, as a simple consequence 
of recursion (\ref{recurs}).

\vskip .1cm 

\vskip .1cm 

{\bf Remark 1:} In previous papers~\cite{bridged,unabridged} we reported on the
equivalence of two properties, the homomorphism of an irreducible operator 
with its adjoint, and the occurrence of a rational 
(possibly hyperexponential\footnote[3]{Hyperexponential 
solutions~\cite{SingUlm} are obtained with the command ``expsols'' in DEtools in Maple.})
solution for the exterior, or symmetric, square of that operator. If we assume that the 
exterior (resp. symmetric) square of the operator $\, L_{[N]}$ has a rational 
solution $\, r(x)$, one immediately deduces, from
the intertwining relation (\ref{tower1}), that  $\, r(x)$
is {\em also a solution} of the exterior (resp. symmetric) 
square of $\, L_{[N-1]}$. Using the
tower of intertwining relations (\ref{recurs2}), one deduces that 
all the exterior (resp. symmetric) squares of all the
 $\, L_{[N-p]}$ intertwiners {\em have the same rational solution} $\, r(x)$, 
especially the last one, namely  $\, L_{[1]} \,  = \, \,  U_{1} \cdot \, r(x)$.
It will be seen, in forthcoming sections, that the existence of 
a rational solution requires  the self-adjoint operator $\, U_{1}$
to be of {\em order-one} for symmetric squares, and  {\em order-two} for 
exterior squares. If the (irreducible) operator $\, U_{1}$ is of higher order,
one {\em does not} have a rational solution but a {\em drop of the order} 
of the symmetric, or exterior, square. 

\vskip .2cm 

{\bf Remark 2:} Generically the $\, L_{[N]}$'s, in the tower of intertwiners 
described in (\ref{recurs}) and (\ref{recurs2})  
in section (\ref{towersimpleright}),
are irreducible.  However, in the euclidean right-division 
recursion, it may happen that one or several $\, L_{[N]}$'s are reducible. Let us 
consider, for instance, (\ref{lastm1}) in the case 
where $\, U_{1} \,  = \, \,  - r(x) \cdot \, U_2 \cdot \, r(x)$. 
The operator $\, L_2$ thus reads: 
\begin{eqnarray}
\label{lastm1factor}
\hspace{-0.95in}&&  \, \, 
 - (U_{2} \cdot \, r(x) \cdot \, U_2 \cdot \, r(x) \,\, - 1) \cdot \, r(x) 
\,  \, = \, \, \,  
-  (U_{2}\cdot \, r(x) \, - 1) \cdot \, (U_{2}\cdot \, r(x) \, + 1). 
\end{eqnarray}
One should note that we {\em have not used this irreducibility
assumption}\footnote[1]{The euclidean right division can, of course, still
be performed with reducible operators.} in section (\ref{towersimpleright}).

\vskip .1cm 

\subsection{Parity constraint on the order of the $\, U_n$'s \newline}
\label{parity}

Let us now prove that these $\, U_n$'s have orders of 
the {\em same parity}. From (\ref{recurs2}) one easily deduces 
\begin{eqnarray}
\label{deduces} 
\hspace{-0.95in}&& \quad \quad \quad \quad \quad \quad 
(adjoint(L_{[N]}) \, - \, adjoint(L_{[N-2]})) \cdot  \, L_{[N-1]}
\nonumber \\
\hspace{-0.95in}&& \quad \quad \quad \quad \quad\quad \quad \quad \quad \quad 
 \,\, \,  \,  = \, \, \,\, \, 
  adjoint(L_{[N-1]})  \cdot  \, (L_{[N]} \, - \, L_{[N-2]}), 
\end{eqnarray}
or, recalling (\ref{recurs}), namely 
$\, L_{[N]} \,  = \, \,   U_N \cdot \,L_{[N-1]} \, + \,  L_{[N-2]}$:
\begin{eqnarray}
\label{deduces2} 
\hspace{-0.95in}&& \quad  \quad \quad \quad \quad \quad 
(adjoint(L_{[N]}) \, - \, adjoint(L_{[N-2]})) \cdot  \, L_{[N-1]}
\nonumber \\
\hspace{-0.95in}&& \quad \quad \quad \quad \quad \quad \quad \quad \quad \quad 
 \,\, \,  \,  = \, \, \,\, \, 
  adjoint(L_{[N-1]})  \cdot  \, U_N \cdot \,L_{[N-1]}. 
\end{eqnarray}
From (\ref{deduces2}) and (\ref{recurs}) one sees that $\, adjoint(L_{[N]})$
can be written alternatively: 
\begin{eqnarray}
\label{deduces3} 
\hspace{-0.95in}&& \, 
adjoint(L_{[N]}) \,  \,  = \, \, \,\,  
adjoint(L_{[N-2]}) \, + \, \, adjoint(L_{[N-1]}) \cdot  \, U_N   
 \\
\hspace{-0.95in}&& \, \,  \,   \,\,    = \, \, \,\,  
adjoint(L_{[N-2]}) \, + \, \, adjoint(U_N \cdot \, L_{[N-1]})
\,  = \, \, \,
  adjoint(L_{[N-2]} \, + \, U_N \cdot \, L_{[N-1]}).\nonumber
\end{eqnarray}
The equality (of the form $\, adjoint(A+B) = \,adjoint(A) \, +adjoint(B)$) 
between the two last terms in (\ref{deduces3}) can be fulfilled 
{\em only if} the parity of the order of $ \, L_{[N-2]}$ is equal to the
parity of the order of $ \,U_N \cdot \, L_{[N-1]}$ (the order of 
$ \,U_N \cdot \, L_{[N-1]}$ is, of course, the same as the order of $\, L_{[N]}$, see 
(\ref{euclid1})). 

Using, in the euclidean right divisions (\ref{recurs}),
 relation $\, L_{[N-1]} \,  = \, \,   U_{N-1} \cdot \,L_{[N-2]} \, + \,  L_{[N-3]}$,
one finds that the parity of the order of $ \, L_{[N-1]}$ is equal to the
parity of the order of $ \, U_{N-1} \cdot \,L_{[N-2]}$. Since 
 the parity of the order of $ \, L_{[N-2]}$ is equal to the
parity of the order of $ \,U_N \cdot \, L_{[N-1]}$, one straightforwardly 
deduces that  the {\em parity of the order of} $ \, U_{N-1}$
and $ \, U_{N}$ {\em are the same}. 

\vskip .1cm 

As a result, one finds, by recursion, that {\em all the} $\, U_n$'s  have 
orders of the {\em same parity}.

\vskip .2cm 

{\bf Remark:} The $\, U_n$'s, in these decompositions, have no reason 
to have the same order, they just need to have orders of 
the {\em same parity}. However, when one considers the successive
 intertwining relations (\ref{recurs2}), one knows that, {\em generically}, 
the intertwiner between two operators of the same order $\, q$ (like $\, L_{[M]}$  
and $\, adjoint(L_{[M]})$) is of order $\, q-1$. If one assumes this
``generic'' situation for all the intertwiners $\, L_{[M]}$ in the 
``tower of intertwiners'' (\ref{recurs}), one finds that
{\em all the} $\, U_n$'s in the decomposition 
are of {\em order one}. We will see in \ref{towards} 
that this corresponds to a differential Galois group $\, SO(q, \,  \mathbb{C})$.
The other case corresponds to the intertwiner 
between the  even order-$q$ operators $\, L_{[M]}$  
and $\, adjoint(L_{[M]})$, to be of order $\, q-2$. Again if one assumes this
``maximal even order'' situation for all the intertwiners $\, L_{[M]}$,  one finds that
 {\em all the} $\, U_n$'s being of {\em order two}. 
We will see, in \ref{towards}, that this corresponds to a 
differential Galois group $\, Sp(q, \,  \mathbb{C})$.

\vskip .1cm 
\vskip .1cm 

{\bf Definition:} We will call ``{\em generic}'' the decompositions  where  {\em all
the} $\, U_n$'s are of {\em order one} (orthogonal differential Galois groups), 
or the decompositions  where  {\em all the} $\, U_n$'s are of 
{\em order two} (symplectic differential Galois groups).

\vskip .1cm 

It turns out that the examples from physics are, most of the time, 
{\em not generic} in the above mathematical 
sense\footnote[1]{Operator $\, {\cal L}_6$, the first order-six example
of the paper, which has a ``generic'' decomposition, is a mathematical example,
it does not emerge from physics.} (see the Calabi-Yau 3-folds 
examples in section (\ref{lastminute}) below, and the order-six or eight 
lattice Green examples in~\cite{bridged,unabridged}), but most operators 
equivalent with these have such a ``generic'' decomposition.

\vskip .1cm 

\subsection{Canonical decomposition for the adjoint operator \newline}
\label{towerdual}

Since these structures, and decompositions, rely on the
homomorphisms between an operator $\, L_{[N]}$ and its adjoint, 
one can also consider the obvious viewpoint which amounts to seeing
 $\, adjoint(L_{[N]})$, the adjoint 
of an operator $\, L_{[N]}$, exactly on the {\em same footing} 
as $\, L_{[N]}$. 

Switching to the adjoint of the operator 
one can get the decomposition of this adjoint 
in two ways. These two decompositions are detailed
in \ref{towerdualapp}. 
One decomposition, described in \ref{towerdualleftsub}, 
 amounts to performing {\em euclidean 
left divisions} on the adjoints of the tower of 
intertwiners described in section (\ref{towersimpleright}). 
However, it is known that the euclidean left division
of a differential operator is more involved 
than the euclidean right division. As a consequence this decomposition is 
less efficient that the euclidean right divisions described 
in (\ref{towersimpleright}).

The other decomposition, described in \ref{towerdualright},
amounts to performing the 
euclidean right division described 
in (\ref{towersimpleright}), but, this time, {\em on the adjoint 
of the operator}. 
This adjoint operator is an operator of the same order, 
we can call $\, M_{[N]}$, for which the same euclidean right division 
calculations of section (\ref{towersimpleright}) can be performed, the first step 
corresponding to find the intertwiner $\, M_{[N-1]}$ in the intertwining 
relation:
\begin{eqnarray} 
\label{tower1M} 
\hspace{-0.8in}&& \quad \quad \quad \quad 
adjoint(M_{[N]}) \cdot  \, M_{[N-1]}
 \,\, \,  \,  = \, \, \,\, \,  \,  adjoint(M_{[N-1]})  \cdot  \, M_{[N]}. 
\end{eqnarray}
The command $\, Homomorphisms(M_{[N]}, \, adjoint(M_{[N]}))$ gives this first
intertwiner $\, M_{[N-1]}$ in the ``tower'' of intertwiners
(see section (\ref{towersimpleright})).
Since we have in mind that  $\, M_{[N]}$ is  $\, adjoint(L_{[N]})$, the 
intertwining relation (\ref{tower1M}) is in fact
\begin{eqnarray}
\label{tower1Mbis} 
\hspace{-0.8in}&& \quad \quad \quad \quad 
L_{[N]} \cdot  \, M_{[N-1]}
 \,\, \,  \,  = \, \, \,\, \,  \,  adjoint(M_{[N-1]})  \cdot  \, adjoint(L_{[N]}), 
\end{eqnarray}
which is different, from the intertwining relation (\ref{tower1}),
the DEtools Maple command, giving this first
intertwiner $\, M_{[N-1]}$, being, now,
$\, Homomorphisms(adjoint(L_{[N]}), \, L_{[N]})$. 
Performing the same calculations as in section (\ref{towersimpleright}) 
(see (\ref{lastm1}),  (\ref{last}),  (\ref{last2}), ...),
we will have another decomposition for these adjoints, deduced
 from successive right divisions. The relation between this decomposition
for $\, adjoint(L_{[N]})$, and the decomposition for $\, L_{[N]}$ described 
in (\ref{towersimpleright}), is detailed in \ref{towerdualright}.

\vskip .1cm

The fact that the (euclidean right division) decomposition of $\, adjoint(L_{[N]})$ 
will be more, or less, efficient than the (euclidean right division) 
decomposition of $\, L_{[N]}$, described in section (\ref{towersimpleright}),
will depend on the very nature of $\, L_{[N]}$. 

\vskip .1cm

\subsection{Getting the decomposition for very large differential operators}
\label{getting}

Before performing simple right (or left, see \ref{towerdualleftsub})
euclidean divisions, all these various decompositions, require to find,  a first 
intertwiner, for instance from the DEtools Maple command
$\, Homomorphisms(L_{[N]}, \, adjoint(L_{[N]}))$ or from the 
command $\, Homomorphisms(adjoint(L_{[N]}), L_{[N]})$.

Despite the fact that $\, L_{[N]}$ and $\, adjoint(L_{[N]})$ should be on the same footing, 
we have seen, experimentally, in a vast majority
of physical examples, that, curiously, the command $\, Homomorphisms(L_{[N]}, \, adjoint(L_{[N]}))$
requires {\em much more} computer time and resources to be performed, than the ``reverse'' 
command $\, Homomorphisms(adjoint(L_{[N]}), L_{[N]})$. This corresponds to the fact that 
$\, L_{[N]}$ and $\, adjoint(L_{[N]})$ are, most of the time, {\em not on the same footing 
for most of the} $\, L_{[N]}$'s {\em emerging in physics}. This is illustrated in 
section (\ref{lastminute}), where the rightmost operators ($U_1$ in (\ref{lastm1}),
(\ref{last}), (\ref{last2}), ..., (\ref{last5})) are of order {\em four}, when the 
leftmost operators ($U_N$ in the $\, L_{[N]}$'s of (\ref{lastm1}),
(\ref{last}), (\ref{last2}), ..., (\ref{last5})) are of order {\em two}: consequently,
for operators $\, L_{[N]}$ of order $q$, 
the intertwiner obtained from $\, Homomorphisms(adjoint(L_{[N]}), L_{[N]})$ is much simpler
than the one obtained from $\, Homomorphisms(L_{[N]}, \, adjoint(L_{[N]}))$, since they 
are respectively of order $\, q-4$ and $\, q-2$.

\vskip .1cm 

In the simple algorithm described in section (\ref{towersimpleright}), 
performing {\em euclidean right divisions} of differential operators is 
{\em almost instantaneous}, 
even for very large operators, {\em once the first intertwiner} 
 $\, Homomorphisms(L_{[N]}, \, adjoint(L_{[N]}))$ {\em has been obtained}. 
Unfortunately, in practice, for the very large differential operators
emerging in physics, this first intertwiner $\, Homomorphisms(L_{[N]}, \, adjoint(L_{[N]}))$ 
corresponds to calculations that require\footnote[5]{From the DEtools Maple command,
or, through a different algorithm which amounts to switching to linear differential 
theta systems~\cite{forthcoming}.} much larger computer resources than the
resources required for the other intertwiner, 
$\, Homomorphisms(adjoint(L_{[N]}), L_{[N]})$. Fortunately, 
the intertwiner $\, Homomorphisms(L_{[N]}, \, adjoint(L_{[N]}))$ 
can actually be simply obtained from the simpler intertwiner, 
$\, Homomorphisms(adjoint(L_{[N]}), L_{[N]})$. This  is a consequence of the
fact that they are {\em inverse of each other, modulo the operator} $\, L_{[N]}$.

A first set of such {\em inversion relations}
between these two intertwiners was already noticed in section (\ref{decompcalL6})
(see  (\ref{inversemodulo}), (\ref{inversemoduloadj})  
(\ref{inversemodulo2}), (\ref{inversemodulo2adj})). These inversion relations
were also noticed in section 2.1 of~\cite{bridged}. 

\ref{inverseapp} shows explicitly how these two intertwiners of the operator
with its adjoint, are {\em actually inverse of each other, modulo the operator}
and the adjoint operator. Recalling the intertwining relation (\ref{tower1})
and the intertwining relation (\ref{tower1M}) (or (\ref{tower1Mbis})), 
one actually has the two inversion relations: 
\begin{eqnarray}
\label{notappinotherwordsunity} 
\hspace{-0.8in}&& \quad \quad \quad \quad 
 M_{[N-1]} \cdot  \, L_{[N-1]} \,\, \,  \,  = \, \, \,\,\,\, \, 
\Omega_{ML} \cdot \,  L_{[N]}  \, \,\,\,  + \, C_{ML}, 
\\
\label{notappinotherwordsunity2} 
\hspace{-0.8in}&& \quad \quad \quad \quad 
 L_{[N-1]} \cdot  \, M_{[N-1]} \,\, \,  \,  = \, \, \,\,\,\,\, 
 \Omega_{LM}  \cdot \, adjoint(L_{[N]}) \,\,\,\,    + \,C_{LM}, 
\end{eqnarray}
where the constants $\, C_{ML}$ and $\,C_{LM}$ 
are equal,
and where $\,\Omega_{ML}$ and  $\,\Omega_{LM}$ are two operators
adjoint of each other: $\, \Omega_{LM}  \,  = \, \,   adjoint(\Omega_{ML})$.

\vskip .1cm 

Since getting the inverse of a given operator, modulo a given 
operator, is just a {\em linear} problem,
we will use these inversion relations, when studying quite massive 
linear differential operators, to get 
the (hard to get) intertwiner
we need for the (euclidean right division)
 decomposition, namely $\, Homomorphisms(L_{[N]}, \, adjoint(L_{[N]}))$, 
from the (easier to get) intertwiner $\, Homomorphisms(adjoint(L_{[N]}), \,L_{[N]})$.

\vskip .1cm 

For very large linear differential operators with selected differential 
Galois groups, using these inversion relations is, in practice, the 
simplest way to get the intertwiner corresponding to 
$\, Homomorphisms(L_{[N]}, \, adjoint(L_{[N]}))$. 
In a following section (\ref{lastminute}), these inversion relations will be
systematically\footnote[1]{The intertwiner corresponding to 
$\, Homomorphisms(L_{[N]}, \, adjoint(L_{[N]}))$ can be obtained with the ``standard'' 
Maple DEtools command, only for the simplest order-six operators
considered below in section (\ref{lastminute}), but this requires 
a lot of time and computer memory.
} used on quite ``massive'' linear differential
operators, recently obtained by P. Lairez~\cite{Lairez,Lairez2},
 annihilating periods arising from {\em mirror symmetries}
associated with reflexive 4-polytopes 
defining various selected Calabi-Yau 3-folds, in order to obtain 
$\, Homomorphisms(L_{[N]}, \, adjoint(L_{[N]}))$,
the intertwiner required to get the (euclidean right division)
decomposition of these operarors.

\vskip .2cm 

\section{Compatibility of the decompositions with the
equivalence of operators and generic decompositions.}
\label{equivdecomp}

Operators with the previously described decompositions, have necessarily 
{\em selected differential Galois groups}~\cite{bridged,unabridged}. Two 
equivalent operators necessarily 
have the same differential Galois groups~\cite{vdP}. Let us see what 
happens to these decompositions when the operator is changed 
into an {\em equivalent one}~\cite{vdP}. We follow, 
here, an {\em experimental mathematics approach}. We built a large number of 
linear differential operators corresponding to the previous decompositions 
(since building self-adjoint linear differential
operators of arbitrary order is quite easy), and performed, systematically, 
the algorithm described in section (\ref{towersimpleright}), to get 
the new decomposition.

\subsection{Equivalence of operators for generic decompositions.}
\label{equivdecompgen}

In section (\ref{decomp}) we have defined two kinds of ``generic decompositions'',  
namely the decompositions where {\em  all} the self-adjoint operators
$\, U_n$'s are of degree {\em one} (which will be seen 
to correspond to differential Galois
groups in $\, SO(q, \,  \mathbb{C})$), and the decompositions
 where {\em all} the self-adjoint operators
$\, U_M$'s are of degree {\em two} (which will be seen 
to correspond to differential Galois groups in $\, Sp(q, \,  \mathbb{C})$).

We have obtained the following experimental result: the form 
of a ``generic decomposition''
is (generically) stable by operator equivalence~\cite{vdP}. For instance, 
if one considers an operator $\, L_{5}$ given by (\ref{last3}), where 
the five $\, U_n$'s are {\em all order-one} 
(resp. {\em all order-two}) self-adjoint
operators, the equivalent operator $\, {\tilde L}_5$ defined
by\footnote[2]{Of course $\, D_x^n$ can be replaced by more involved 
operators.} (${\cal I}_n$ is an order-$n$ intertwiner)
\begin{eqnarray}
\label{equivL5}
\hspace{-0.95in}&& \quad  \quad \qquad \qquad  \qquad 
  {\cal I}_n \cdot \, L_5 \,\, \,  \,  = \,\, \, \,\, {\tilde L}_{5} \cdot \, D_x^n, 
\end{eqnarray}
has (generically) also a decomposition of the {\em same form} as (\ref{last3}):
\begin{eqnarray}
\label{last3tilde}
\hspace{-0.95in}&& \quad  \quad   
{\tilde L}_{5} \,\, \,  \,  = \, \, \,\, \,  \,
  ({\tilde U}_{5} \cdot \,{\tilde U}_{4} 
\cdot \, {\tilde U}_{3} \cdot \, {\tilde U}_{2} \cdot \,  {\tilde U}_{1}
\,\,\, + {\tilde U}_{5} \cdot \, {\tilde U}_{4} \cdot \, {\tilde U}_{1} \, 
+ \, {\tilde U}_{5} \cdot \, {\tilde U}_{2} \cdot \,  {\tilde U}_{1} \, 
\nonumber \\
\hspace{-0.95in}&&  \qquad \quad  \quad   \quad \quad    \quad 
 + \, {\tilde U}_{5} \cdot \, {\tilde U}_{4} \cdot \,  {\tilde U}_{3}
+ \, {\tilde U}_{3} \cdot \, {\tilde U}_{2} \cdot \, {\tilde U}_{1} 
\,\,\,\,  + {\tilde U}_{1} \,+ \, {\tilde U}_{3}\,+ \, {\tilde U}_{5}) \cdot \rho(x),  
\end{eqnarray}
where all the $\, {\tilde U}_{n}$'s are  {\em all order-one}
(resp. {\em all order-two}) self-adjoint
operators, and $\, \rho(x)$ is a function.

\vskip .1cm 

\subsection{Equivalence of operators for non-generic decompositions.}
\label{equivdecompnongen}

Let us now, consider,
also in an experimental mathematics viewpoint, 
 a few ``non-generic'' decompositions, where 
the order of {\em all} the $\, U_M$'s do not reduce to 
order one or order two. 

Let us first consider an {\em order-five} linear differential 
operator $\, {\cal L}_5$ of the form (\ref{last}) 
\begin{eqnarray}
\label{simplestform311} 
\hspace{-0.8in}&& \quad \quad \quad  \quad \quad \quad \quad 
{\cal L}_5 \, \,\,  \, = \, \, \, \,  \, 
M_3 \cdot \, N_1 \cdot \, P_1 \,\,\,    + \, M_3 \,\,\,   + \, P_1,
\end{eqnarray}
where $\, N_1$ and $\, P_1$ are self-adjoint operators of {\em order one}, 
but where the first self-adjoint operator $\, M_3$ is of {\em order three}.
If one changes $\, {\cal L}_5$ into an equivalent operator $\, {\tilde {\cal L}}_{5}$
(${\cal I}_n$ is an order-$n$ intertwiner)
\begin{eqnarray}
\label{equivL5bis}
\hspace{-0.95in}&& \quad  \quad \qquad \qquad  \qquad 
  {\cal I}_n \cdot \, {\cal L}_{5} 
\,\, \,  \,  = \, \,\, \,\, {\tilde {\cal L}}_{5} \cdot \, D_x^n, 
\qquad \quad \quad n \ge \, 2, 
\end{eqnarray}
one finds that the equivalent operator $\, {\tilde {\cal L}}_{5}$ has the
{\em generic}  decomposition (\ref{last3}), or (\ref{last3tilde}) where the $\, {\tilde U}_{M}$ 
are  {\em all order-one}  self-adjoint
operators, and $\, \rho(x)$ is a function.

Let us now consider
 an order-twelve linear differential operator $\, {\cal L}_{12}$ of the form (\ref{last}) 
\begin{eqnarray}
\label{simplestform311} 
\hspace{-0.8in}&& \quad \quad \quad  \quad \quad \quad \quad 
{\cal L}_{12} \, \,\,   = \, \, \, \,\,    
M_4 \cdot \, N_4 \cdot \, P_4 \,\,\,  \,  + \, M_4 \,\,\,   + \, P_4,
\end{eqnarray}
where $\, M_4$, $\, N_4$ and $\, P_4$ are self-adjoint operators of {\em order four}.
If one changes $\, {\cal L}_{12}$ into an equivalent operator $\, {\tilde {\cal L}}_{12}$
(${\cal I}_n$ is an order-$n$ intertwiner)
\begin{eqnarray}
\label{equivL12}
\hspace{-0.95in}&& \quad  \quad \qquad \qquad  \qquad 
  {\cal I}_n \cdot \, {\cal L}_{12}
 \,\, \, \, \,  = \, \,\, \,\, {\tilde {\cal L}}_{12} \cdot \, D_x^n, 
\qquad \quad \, \,  \,\, \,  n \ge \, 3, 
\end{eqnarray}
one finds that the equivalent operator $\, {\tilde {\cal L}}_{12}$ has the
 {\em generic} decomposition (\ref{last4}), where 
the {\em six} $\,  U_{M}$ are  {\em all  order-two} 
 self-adjoint operators, and $\, r(x)$ is a function.

Let us give another simple illustration of these results with the decomposition of 
an operator equivalent to an order-three self-adjoint operator. Let us 
consider an order-three self-adjoint operator 
\begin{eqnarray}
\label{calL3}
\hspace{-0.8in}&& \quad \quad \quad  \quad \, \, 
{\cal L}_3 \, \, \, = \, \, \,\, \,   a_3(x) \cdot \, D_x^3 \,  \,  \,  \, \,
+ \, {{3} \over {2}} \cdot \, {{d a_3(x)} \over {dx}}\cdot \, D_x^2 
\,\, \,   \, 
+ \, a_1(x) \cdot \, D_x \,  \, 
\nonumber \\
\hspace{-0.8in}&& \quad \quad \quad \quad \quad \quad \quad \quad \quad
 + \,  {{1} \over {2}} \cdot \, {{d a_1(x)} \over {dx}} \, \, \,  \, 
-{{1} \over {4}} \cdot \, {{d^3 a_3(x)} \over {dx^3}},
\end{eqnarray}
and its equivalent operator\footnote[1]{The $\, D_x^3$ interwiner
in (\ref{Dx3}), the equivalent between $\,{\cal L}_3$ 
and $\, {\tilde {\cal L}}_3 $ can be replaced,
by the order-two operator corresponding 
to $\, D_x^3$ mod. $\,{\cal L}_3$. } 
\begin{eqnarray}
\label{Dx3}
\hspace{-0.8in}&& \quad \quad \quad \quad \quad \quad \quad  \quad     
{\cal I}_3 \cdot \, {\cal L}_3 
\, \,\,\, = \, \, \,\,\, {\tilde {\cal L}}_3 \cdot \, D_x^3,
\end{eqnarray}
where $\, {\cal I}_3$ is an order-three intertwiner. Introducing the 
(order-two) intertwiner $\, {\tilde {\cal L}}_2$, obtained from 
the Maple DEtools command 
$Homomorphisms({\tilde {\cal L}}_3, adjoint({\tilde {\cal L}}_3)$,
 such that
\begin{eqnarray}
\label{homocalL3}
\hspace{-0.8in}&& \quad \quad \quad \quad 
adjoint({\tilde {\cal L}}_2)  \cdot \, {\tilde {\cal L}}_3
 \, \, \, = \, \,\, \, 
adjoint({\tilde {\cal L}}_3) \cdot \, {\tilde {\cal L}}_2,
\end{eqnarray}
and performing the successive euclidean right-divisions:
\begin{eqnarray}
\label{succcalL3}
\hspace{-0.7in}&&  \quad  \quad  \quad  \quad
{\tilde {\cal L}}_3 \, \, = \, \, \,
 U_3 \cdot {\tilde {\cal L}}_2 \, + \, {\tilde {\cal L}}_1, 
\quad   \quad  \quad  \qquad 
{\tilde {\cal L}}_2 \, \, = \, \, \, 
U_2 \cdot {\tilde {\cal L}}_1 \, + \, {\tilde {\cal L}}_0,
\nonumber \\
\hspace{-0.7in}&& \quad  \quad  \quad  \quad   \quad 
\quad {\tilde {\cal L}}_1 \, = \, \, U_1 \cdot r(x), 
\quad  \quad  \quad  \qquad \qquad  {\tilde {\cal L}}_0 \, = \, \, r(x), 
\end{eqnarray}
one finds that the $\, U_n$'s are actually self-adjoint 
{\em order-one} operators, $\, r(x)$ being a function. Thus, one gets 
that the order-three operator $\, {\tilde {\cal L}}_3$, 
equivalent to a self-adjoint order-three operator, has a decomposition (\ref{last}): 
\begin{eqnarray}
\label{lastL3}
\hspace{-0.95in}&& \quad  \quad  \quad  \quad  \quad \quad  \quad    \quad   
{\tilde {\cal L}}_3  \,\, \,  \,  = \, \, \,\, \,  \, 
 (U_{3} \cdot \, U_{2} \cdot \, U_{1} \,\, + U_{1} \,+ \, U_{3}) \cdot r(x), 
\end{eqnarray}
where the $\, U_n$'s are {\em order-one} self-adjoint operators.

\vskip .1cm 

More generally, the simplest example of non-generic decomposition 
corresponds to (\ref{lastm2}), namely an operator 
$  \,L_{[1]} \, = \, \,  U_{1} \cdot r(x)$, where the {\em unique} 
operator $\, U_{1}$ is a {\em self-adjoint} operator 
of order $\, q \ge 3$. This operator is such that its differential Galois group
is $\, SO(q, \,  \mathbb{C})$, for $\, q$ {\em odd}, where one finds a drop of order 
of the {\em symmetric square} of $\, U_{1}$,
and $\, Sp(q, \,  \mathbb{C})$,  for $\, q$ {\em even},
where one finds a drop of order of the {\em exterior square} of $\, U_{1}$.
Again, an equivalent operator (like (\ref{equivL5}), (\ref{equivL5bis})), 
will, for $\, n$ large enough\footnote[5]{For instance, for $\, n\, = \,1$, 
and $\, q \, = \,5$, an equivalent operator of $\, U_1$ 
has a (non-generic) decomposition 
$\, (M_1 \, N_1 \, P_3 \, + M_1 \, + \, P_3) \cdot \, r(x)$. However $\, n\, = \,2$
and $\, q \, = \,5$, yields a generic decomposition in five order-one
self-adjoint operators (see (\ref{last3})). For $\, n\, = \,1$, 
and $\, q \, = \,6$, an equivalent operator of $\, U_1$ 
has a (non-generic) decomposition
$(M_2 \, N_4\, +1)  \cdot \, r(x)$, when for $\, n\, = \,2$, 
and $\, q \, = \,6$, one has the generic decomposition 
$\, (M_2 \, N_2 \, P_2 \, + M_2 \, + \, P_2) \cdot \, r(x)$.}, 
change $\, L_{[1]}$ into an equivalent operator with a 
{\em generic decomposition}, namely
a decomposition where all the $\, U_{n}$'s are of order one, for $\, q$ odd,
and  a decomposition where all the $\, U_{n}$'s are 
of {\em order two}, for $\, q$ {\em even}.

\vskip .1cm 

More involved examples of equivalence of 
operators with {\em exceptional differential Galois groups}, 
already sketched in~\cite{bridged}, are detailed in \ref{except}.
Again, on these highly non trivial examples, it is shown that 
the equivalent of these operators have {\em quite involved} (generic or 
non-generic) decompositions, like  (\ref{last5}) 
(see also (\ref{last3screw}) below). 

\vskip .2cm 

{\bf To sum-up:} For {\em involved enough equivalence} of operators, the non-generic 
decomposition of an order-$q$ operator turns into a {\em generic decomposition}
for its equivalent operator. The order of {\em all} the self-adjoint operators 
$\,  U_{n}$'s, in the decomposition of the equivalent operator,
 is {\em one} (when the self-adjoint 
operators in the non-generic decomposition of the initial operator 
are of {\em odd} orders, which will be seen, in \ref{towardsredu},
to correspond to differential Galois groups in $\, SO(q, \,  \mathbb{C})$).
The order of all the self-adjoint operators $\,  U_{n}$'s, in the decomposition 
of the equivalent operator,
is {\em two} when the self-adjoint operators in the non-generic 
decomposition of the initial operator are of {\em even} orders (corresponding to
differential Galois groups in $\, Sp(q, \,  \mathbb{C})$, see 
\ref{towardsredusympl}).

\vskip .1cm 

\subsection{Towards the generic situation 
for selected differential Galois groups: reduced form 
for differential systems}
\label{towards}

In the previous sections we saw that the existence of a homomorphism 
between an operator and its adjoint (characteristic of 
selected differential Galois group~\cite{bridged,unabridged}) is the key 
ingredient to get the canonical decomposition of the operator. Another 
way to see that operators with selected differential Galois groups
{\em necessarily have} the decomposition, described in 
section (\ref{towersimpleright}), is sketched in \ref{towards}, analyzing, 
separately, in \ref{towardsredu} the operators 
with orthogonal differential Galois groups,
and in \ref{towardsredusympl} the operators 
with symplectic differential Galois groups. This approach amounts to introducing 
 the concept of {\em reduced form}~\cite{Symbolic,AparicioLast} for 
{\em linear differential systems} associated with these selected differential
operators. More specifically,  \ref{towards}  shows that the 
{\em most general operators with 
selected differential Galois group correspond to the} 
``{\em generic decompositions}''.

\vskip .1cm 

{\bf To sum-up:} Our different ``experimental mathematics'' approaches 
show that {\em all the linear differential operators, with
selected differential Galois groups, correspond to the decompositions
described in section} (\ref{towersimpleright}). In other words 
these decompositions can be seen as an {\em algebraic description} 
of the operators with selected differential Galois groups.

\vskip .1cm 

\section{Rational solutions of the exterior or symmetric squares of
the operators versus decompositions}
\label{ratiodecomp}

We have seen, in previous papers~\cite{bridged,unabridged}, 
that the existence of rational solutions for 
the exterior, or symmetric, square of an operator (or drop of the order of these squares)
was equivalent to the existence of a homomorphism between the operator
and its adjoint. Since the existence of a homomorphism between the operator
and its adjoint is the key ingredient to get the canonical decomposition of operators 
described in this paper, let us see what is the relation between these decompositions
and the rational solutions of the exterior, or symmetric, square of the operator.

Following, again, our experimental mathematics approach, we have considered 
for all our examples of operators with their decompositions
(see section (\ref{decomp}), namely (\ref{lastm1}), 
(\ref{last}), (\ref{last2}), (\ref{last3}), (\ref{last4}), ...), their symmetric  
and exterior squares, seeking for rational solutions of these squares. For simplicity, and
 without any loss of generality, we restrict the
decompositions (\ref{lastm1}), 
(\ref{last}), (\ref{last2}), (\ref{last3}), (\ref{last4}), ...
to $\, r(x) \, = \, \, 1$ (finding the symmetric, or exterior, square
of an operator $\, L \cdot r(x)$, from the symmetric, or exterior, square
of an operator $\, L$,  is straightforward).

For decompositions such that the self-adjoint operators $\,  U_{n}$'s 
are all of {\em odd order}, one gets, for the symmetric square of these order-$q$ operators,
either a rational solution of the symmetric square, or a drop of the order of
the symmetric square, the order being less than $\, q \, (q+1)/2$ (the two situations 
corresponding to differential Galois groups in $\, SO(q, \,  \mathbb{C})$). For decompositions 
such that the self-adjoint operators $\,  U_{n}$'s 
are all of {\em even order}, one gets, for the exterior square of these order-$q$ operators,
either a rational solution of the exterior square, or a drop of the order of
the exterior square, the order being less than $\, q \, (q-1)/2$ (the two situations 
corresponding to differential Galois groups in $\, Sp(q, \,  \mathbb{C})$). In both cases 
(odd or even order), one finds that the rational solution, or the drop of the order
of the symmetric, or exterior, square of the operator,
{\em depend  only on} $\,  U_{1}$, the rightmost  self-adjoint 
operator in the larger product in these 
decompositions. If the order of $\,  U_{1}$ is higher than one or two, one has a drop
of the order of the symmetric, or exterior, square of the operator. One has a rational 
solution for the symmetric, or exterior, square of the operator {\em only} when  
the self-adjoint operator is of {\em order one} 
\begin{eqnarray}
\label{formU1} 
\hspace{-0.95in}&&   \qquad \qquad \quad \quad  \, 
 U_1 \,\,\, = \, \,  \, \,  \, 
a_1(x) \cdot \, D_x \,\,   + \, {{1} \over {2}}  \cdot \, {{d a_1(x)} \over {dx}}, \, \quad 
\end{eqnarray}
and the rational solution of the {\em symmetric square} of the operator is $\, 1/ a_1(x)$, 
or when  the self-adjoint operator is of {\em order two}
\begin{eqnarray}
\label{formU1deux} 
\hspace{-0.95in}&&   \qquad \qquad \quad \quad  \, 
 U_1 \,\,\, = \, \, \,\,   \,  
a_2(x) \cdot \, D_x^2 \,\, \,  
+ \,  \, {{d a_2(x)} \over {dx}} \cdot \, D_x \,\, \,  + \, a_0(x), 
\end{eqnarray}
and the rational solution of the {\em exterior square} of the operator is $\, 1/ a_2(x)$.

\subsection{Drop of the order of the exterior or symmetric squares of the operators}
\label{drop}

We have considered a large number of examples
corresponding to the drop of the order of the squares
of the operators,
when the order $\, q$ of the self-adjoint operator 
$\,  U_{1}$ is higher than one or two:
\begin{eqnarray}
\label{selfadjN}
\hspace{-0.95in}&& \, \, \,\, \, U_1 
 \, \, \,\,  = \, \, \,   \,  \, \, \, 
 a_q(x) \cdot \, D_x^q \,  \,\,  \,  
 + \, {{q} \over {2}} \cdot \, {{d a_q(x)} \over {dx}}\cdot \, D_x^{q-1} 
 \, \,  \, \,+ \,  a_{q-2}(x) \cdot \, D_x^{q-2} 
\nonumber \\ 
\hspace{-0.95in}&&\quad    \, \,\, \,
 +   \, {{q-2} \over {2}}  \,\cdot \,  \Bigl(  {{d a_{q-2}(x)} \over {dx}}
 \,\, \, 
 -  \,  {{q \cdot \,  (q-1) } \over {12}}  
\, \cdot \, {{d^3 a_q(x)} \over {dx^3}}
   \Bigr)  \cdot \, D_x^{q-3}  \,\, + \, \,  a_{q-4}(x) \cdot \, D_x^{q-4}
\nonumber  
\end{eqnarray}
\begin{eqnarray}
\hspace{-0.95in}&&\quad  \, \, \, \,\, \,
 +   \, {{q-4} \over {2}}  \,\cdot \, \Bigl( {{d a_{q-4}(x)} \over {dx}}
 \, \, \,  -  \,  {{(q-2) \cdot  \, (q-3) } \over {12}}  
\, \cdot \, {{d^3 a_{q-2}(x)} \over {dx^3}}\, 
 \\ 
\hspace{-0.95in}&&\quad   \, \, \, \,\, \,
+ \,  {{ q \cdot  \,(q-1) \cdot  \, (q-2) 
\cdot  \, (q-3) } \over {120}}  
\, \cdot \, {{d^5 a_{q}(x)} \over {dx^5}}
   \Bigr)  \cdot \, D_x^{q-5} \, \,  \,  
+ \, \,  a_{q-6}(x) \cdot \, D_x^{q-6} \, \, \,  + \, \,   \,  \cdots 
 \nonumber 
\end{eqnarray}

We found the following result. If one changes the order-$q$ 
operator (\ref{selfadjN}) for an equivalent one 
(${\cal I}_n$ is an order-$n$ intertwiner)
\begin{eqnarray}
\label{equivL5ter}
\hspace{-0.95in}&& \quad  \quad  \quad \qquad \qquad \, 
  {\cal I}_n \cdot \, U_1 \,\, \,  \,  = \, \, \,  \,\, {\tilde L}_{q} \cdot \, D_x^n, 
\end{eqnarray}
one finds that the exterior, or symmetric, squares 
of the equivalent operator $\, {\tilde L}_{q}$
has, again, a rational solution\footnote[2]{This is in agreement with the previous 
section: for large enough $\, n$, non-generic decompositions reduce 
to generic decompositions (which have rational solutions for 
their exterior or symmetric squares).} for large enough $\,n$.

For instance, for odd orders $\, m$ of the self-adjoint operator (\ref{selfadjN}), 
one {\em recovers a rational solution} 
for the {\em symmetric square} of $\, {\tilde L}_{q}$, 
for $\, q \, = \, 3$ with $\, n\, = \, 1$, for $\, q \, = \, 5$ with $\, n\, = \, 2$, 
for $\, q \, = \, 7$ with $\, n\, = \, 3$, 
and more generally,  with $\, n\, = \, (q-1)/2$.
The rational solution of the symmetric square of that equivalent operator 
$\, {\tilde L}_{m}$ is found to be $\, 1/a_q(x)$, the inverse of 
the head coefficient of operator (\ref{selfadjN}).

For even orders $\, q$ of the self-adjoint  operator (\ref{selfadjN})
one {\em recovers a rational solution} 
for the {\em exterior square} of $ \, {\tilde L}_{q}$, 
for $\, q \, = \, 4$ with $\, n\, = \, 1$, for $\, q \, = \, 6$ with $\, n\, = \, 2$, 
for $\, q \, = \, 8$ with $\, n\, = \, 3$, and more generally,  
with $\, n\, = \, (q-2)/2$.
The rational solution of the exterior square of that equivalent operator 
$\, {\tilde L}_{q}$ is also found to be $\, 1/a_q(x)$, the inverse of the head coefficient
of operator (\ref{selfadjN}).

\vskip .1cm

{\bf Remark:} Of course, if one considers the {\em adjoint} of an operator
having one of the decompositions described in section (\ref{tower}), 
$\, L_{[N]} \, = \, \, (U_N \cdots U_1 \, + \,  \cdots) \cdot r(x)$, the 
rational solution of the 
symmetric, or exterior, squares of this  adjoint {\em depend only} 
on $\, U_N$ (instead of $\, U_1 \cdot \, r(x)$ for $\, L_{[N]}$).

\vskip .1cm
\vskip .1cm 

\section{Decomposition of large operators with selected differential Galois groups.}
\label{lastminute}

The euclidean right division of an operator with a particular intertwiner provides a simple
well-defined algorithm to get, very quickly, a canonical decomposition of operators with selected 
differential Galois groups, as well as a ``tower'' of intertwiners. However, most 
of the time in physics, or in the challenging problems of 
mathematical physics, the operators with selected differential Galois groups are quite ``massive''
operators (several Mega-octets, ...) of quite large 
order (12, 21, ... see~\cite{L12L21}), and one remarks, experimentally, 
that the intertwiner one needs to calculate in the first step of the algorithm, namely 
$\, Homomorphisms(Oper, \, adjoint(Oper))$, requires massive computer resources compared to 
$ \, Homomorphisms(adjoint(Oper), \,Oper)$. Furthermore the ``Homomorphisms''
 command, implemented in DEtools
in Maple, is not efficient enough for such ``massive'' operators.  

Let us sketch, or fully perform, in the next section, the study of some quite 
``massive'' operators that are important in physics, or mathematical physics.
This study is done using the ideas of section (\ref{getting}), 
together with a brand new algorithm that requires to
work on the linear {\em theta-system}~\cite{forthcoming} associated with the operators. 

\subsection{Operators annihilating periods arising from mirror symmetries.}
\label{lastminute}

P.~Lairez obtained\footnote[8]{We thank P. Lairez for generously 
sending us these explicit examples of selected operators before
public access on the web~\cite{Lairez2}. } recently, in a systematic analysis, 
a set of 210  explicit linear differential
operators annihilating periods arising from {\em mirror symmetries}\footnote[2]{Using 
a criterion of Namikawa, Batyrev and Kreuzer found~\cite{Kreuzer} 30241 
reflexive 4-polytopes such that the corresponding Calabi-Yau hypersurfaces 
are smoothable by a flat deformation. In particular, they found 210 reflexive 4-polytopes 
defining 68 topologically different Calabi-Yau 3-folds with $h_{11}= \, 1$.}
 (associated with reflexive 4-polytopes 
defining 68 topologically different 
Calabi-Yau 3-folds, see~\cite{Lairez2,Kreuzer,Lairez}). These periods are also
{\em diagonals of rational functions}~\cite{Christol,Short,Big}.

Among these 210 operators many correspond to the ``standard'' 
Calabi-Yau ODEs that have already been analyzed in various papers~\cite{CalabiYauIsing}.
They are {\em order-four} irreducible operators satisfying 
 the ``{\em Calabi-Yau condition}\footnote[1]{They are, up to a conjugation by a function, 
irreducible {\em order-four self-adjoint} operators~\cite{bridged}.}''~\cite{CalabiYauIsing} 
corresponding to say that the exterior square of these order-four operators
is of {\em order five}. However, remarkably, the other operators are {\em higher order}
operators of even orders $\, N \, = \, \, 6, \, 8, \, 10, \, 12, \, 14, \, 16,  \, \cdots ,  \, 24$. 

The study of such ``massive'' operators relies on the ideas of section (\ref{getting}).
The intertwiner $\, Homomorphisms(Oper, \, adjoint(Oper))$, that
one needs to calculate in the first step of the euclidean right division 
algorithm described in section (\ref{towersimpleright}), is, in fact, obtained from the 
(much easier to get) intertwiner $ \, Homomorphisms(adjoint(Oper), \,Oper)$,
since  these two intertwiners are {\em inverse of each other modulo the operator}
one considers: getting one intertwiner from the other one 
is essentially a {\em linear} problem. 
The complexity of the problem is reduced to calculating the 
(much easier to get) intertwiner $ \, Homomorphisms(adjoint(Oper), \,Oper)$. 
However, for most of the ``massive'' operators of this list~\cite{Lairez2}
of 210 operators, even obtaining these (simpler to get) intertwiners 
remains beyond our computer resources, using the ``Homomorphisms'' 
command of DEtools in Maple. In order to achieve this first
step, we have developed a brand new algorithm that requires to
work on the linear {\em theta-system} associated 
with the operators. The details of these calculations 
being slightly technical will be explained in a forthcoming publication~\cite{forthcoming}.
Let us just give the result of these (still massive\footnote[9]{Switching from
the operator approach to a linear theta-system approach~\cite{forthcoming}
yields drastic reduction of the computing time as well as the memory 
required to perform the calculations.}) calculations. 

Performing these calculations, we found\footnote[5]{These calculations
do not use the ``Homomorphisms'' command available in Maple in DEtools. 
The calculations are performed on the associated differential theta-systems, the 
Homomorphism of operator corresponding to ``gauge transformations'' 
on the system. This will be explained in a forthcoming 
paper~\cite{forthcoming}.} the following decompositions
for operators {\em up to order sixteen}:
\begin{eqnarray}
\label{order6}
\hspace{-0.95in}&& \quad    \quad   \quad \quad \quad \quad \,
{\cal L}_6 \, \,\,  \, = \, \, \,\, 
 (M_2 \cdot \, N_4  \, + \, 1) \cdot r(x),  
\end{eqnarray}
\begin{eqnarray}
\label{order8}
\hspace{-0.95in}&& \quad    \quad   \quad \quad \quad \quad \, 
{\cal L}_8 \, \,\,  \, = \, \, \,\, 
 (M_2 \cdot \, N_2 \cdot \, P_4 \, + \, M_2 \, + \, P_4) \cdot r(x), 
\end{eqnarray}
\begin{eqnarray}
\label{order10}
\hspace{-0.9in}&&\quad \quad 
{\cal L}_{10} \, \,\,  \, = \, \, \,\, 
(M_2 \cdot \, N_2 \cdot \, P_2  \cdot \, Q_4 
\, + \, M_2 \cdot \, Q_4  \, + \, P_2 \cdot \, Q_4 \, 
+ \, M_2 \cdot \, N_2 \, + \, 1) \cdot \, r(x),
\nonumber 
\end{eqnarray}
\begin{eqnarray}
\label{order12}
\hspace{-0.9in}&&\quad \quad 
{\cal L}_{12} \, \,\,  \, = \, \, \,\, 
(M_2 \cdot \, N_2 \cdot \, P_2  \cdot \, Q_2 \cdot \, R_4 \, 
+ \, M_2 \cdot \, N_2  \cdot \, R_4  \,
+ \, M_2 \cdot \,  Q_2 \cdot \, R_4  
\nonumber \\ 
\hspace{-0.9in}&&\quad \quad \quad \quad \quad 
+ \, M_2 \cdot \, N_2  \cdot \, P_2  \,
+ \,  P_2  \cdot \, Q_2 \cdot \, R_4 
+ \, M_2 + \,  P_2  + \, R_4) \cdot \, r(x),
\end{eqnarray}
\begin{eqnarray}
\label{last4forL14}
\hspace{-0.95in}&& \quad  \,  
{\cal L}_{14} \,\, \,  \,  = \, \, \,\, \,  \,
 (M_2 \cdot \, N_2 \cdot \, P_2  \cdot \, Q_2 \cdot \, R_2 \cdot \, S_4
\,\,\,\,\, + M_2 \cdot \, N_2 \cdot \, R_2  \cdot \, S_4 \,  \,
+ \, M_2 \cdot \, N_2 \cdot \, P_2  \cdot \, S_4  \, 
\nonumber \\
\hspace{-0.95in}&& \quad  \quad   \quad  \quad    \quad   
\,\, + \, M_2 \cdot \, N_2 \cdot \, P_2  \cdot \, Q_2 
+ \, M_2 \cdot \, Q_2 \cdot \, R_2 \cdot \, S_4
+ \, P_2  \cdot \, Q_2 \cdot \, R_2 \cdot \, S_4
\nonumber \\
\hspace{-0.95in}&& \quad  \quad  \quad \quad     \quad \quad \qquad  
  \, \,   +  \,M_2 \cdot \, S_4
   \,+ \, M_2 \cdot \, Q_2 \,+ \, M_2 \cdot \, N_2  \,
 + \, P_2 \cdot \, S_4  \, + \, R_2 \cdot \, S_4 \, 
\nonumber \\
\hspace{-0.95in}&& \quad  \quad  \quad  \quad   \quad \quad  \quad   \quad   \quad  \qquad   
 \, \, + \, P_2 \cdot \, Q_2\,  \, \,\, + \, 1) \cdot \, r(x), 
\end{eqnarray}
\begin{eqnarray}
\label{order16}
\hspace{-0.9in}&&\quad  \,  
{\cal L}_{16} \, \,\,  \, = \, \, \,\, 
(M_2 \cdot \, N_2 \cdot \, P_2 \cdot \,Q_2 \cdot \, R_2 \cdot \, S_2 \cdot \,  T_4
\,\,\,\,\, + M_2 \cdot \, N_2 \cdot \, P_2 \cdot \, Q_2 \cdot \, T_4 \,  \,
\nonumber \\
\hspace{-0.95in}&& \quad      \quad   
+ \,M_2 \cdot \, N_2 \cdot \,  R_2 \cdot \, S_2 \cdot \, T_4  \,
+ \, M_2 \cdot \, Q_2 \cdot \, R_2 \cdot \, S_2 \cdot \,  T_4 \,
+ \, M_2 \cdot \, N_2 \cdot \, P_2 \cdot \, S_2 \cdot \,  T_4
\nonumber \\
\hspace{-0.95in}&& \quad    \quad 
 + \, M_2 \cdot \, N_2 \cdot \,  P_2 \cdot \, Q_2 \cdot \,  R_2
+\, P_2 \cdot \,Q_2 \cdot \, R_2 \cdot \, S_2 \cdot \,  T_4 
\, + M_2 \cdot \, N_2 \cdot \, T_4 \,
\nonumber \\
\hspace{-0.95in}&& \quad   \quad   
+ \, M_2 \cdot \, N_2 \cdot \, R_2 \,
+ \, M_2 \cdot \, N_2 \cdot \, P_2 \,
 + M_2 \cdot \, Q_2 \cdot \, T_4  
\, + \,M_2 \cdot \, S_2 \cdot \,  T_4 \, 
\nonumber \\
\hspace{-0.95in}&& \quad   \quad
  + \,M_2 \cdot \,  Q_2 \cdot \,  R_2\,  \,
\,\, + P_2 \cdot \, Q_2 \cdot \, T_4 \,  \,
+ \, P_2 \cdot \, S_2 \cdot \,  T_4 \,
+ \, P_2 \cdot \, Q_2 \cdot \,  R_2
 \nonumber \\
\hspace{-0.95in}&& \quad    \quad 
+ \, R_2 \cdot \, S_2 \cdot \, T_4 
\,\,\,\,  + T_4 \, \, + \, R_2\,+ \, P_2
\,\, + \, M_2 ) \cdot \, r(x), 
\end{eqnarray}
where $\, r(x)$ is a rational function, and where the
$\, M_n$, $\, N_n$, $\, P_n$, $\, Q_n$, $\, R_n$, $\, S_n$ and $\, T_n$ 
operators are {\em self-adjoint}
 operators of order $\, n$. One notes that the ``rightmost'' self-adjoint
operator (in the first and largest term of the decomposition) 
is always of order {\em four}. 

For instance, for the order-twelve 
operator $\, {\cal L}_{12}$,
the intertwinners $\,{\cal L}_{10}$ and $\tilde{{\cal L}}_{8}$
\begin{eqnarray}
\label{intertwin1}
\hspace{-0.9in}&&\quad \quad \quad \quad \quad \quad 
{\cal L}_{12} \cdot \, \tilde{{\cal L}}_{8} \, \,\, = \, \,\, \, 
 adjoint(\tilde{{\cal L}}_{8}) \cdot \, adjoint({\cal L}_{12}), 
\qquad \\ 
\label{intertwin2}
\hspace{-0.9in}&&\quad \quad \quad \quad \quad \quad 
adjoint({\cal L}_{12}) \cdot \,  {\cal L}_{10}  \,\, \, = \, \,\, \,
 adjoint({\cal L}_{10}) \cdot \,{\cal L}_{12}, 
\end{eqnarray}
read respectively
\begin{eqnarray}
\label{order12interHOM1}
\hspace{-0.9in}&&\, \, \, \quad
{\cal L}_{10} \,\, = \,\,\, (N_2 \cdot \, P_2  \cdot \, Q_2 \cdot \, R_4 \, 
+ \, N_2  \cdot \, R_4  \, + \, Q_2 \cdot \, R_4  
+ \, N_2  \cdot \, P_2  \, + \, 1) \cdot \, r(x),
\end{eqnarray}
and 
\begin{eqnarray}
\label{order12interHOM}
\hspace{-0.9in}&&\,  \, \, \quad
\tilde{{\cal L}}_{8}\, \,= \, \, \,
 {{1} \over {r(x)}} \cdot \, (Q_2 \cdot \,P_2  \cdot \, N_2 \cdot \, M_2
+ \, Q_2 \cdot \, M_2  + \, Q_2 \cdot \,  P_2 
+ \, N_2  \cdot \, M_2  \, + \,  1). 
\end{eqnarray}

\vskip .1cm 

In this order-twelve operator $\, {\cal L}_{12}$ example, we first obtain (from
a theta-system approach~\cite{forthcoming}) the intertwiner 
$\, \tilde{{\cal L}}_{8}$. We then obtain the 
$\, {\cal L}_{10}$ intertwiner using the fact that $\, \tilde{{\cal L}}_{8}$ 
and $\, {\cal L}_{10}$ are inverse of each other modulo $\, {\cal L}_{12}$
(see section (\ref{getting})).  

\vskip .1cm 

\vskip .1cm 

{\bf Remark 1:} The form  (\ref{order12}) of the decomposition
of, for instance, $\, {\cal L}_{12}$ with the 
rightmost self-adjoint {\em order-four} operator $\, R_4$, yields that the intertwiner 
$\, \tilde{{\cal L}}_{8}$ is of {\em order eight}, when the other intertwiner 
$\, {\cal L}_{10}$ is of {\em order ten}.  This explains why, in all this
set of 210 operators (and, apparently, many other in physics), the intertwiner 
corresponding to $ \, Homomorphisms(adjoint(Oper), \,Oper)$ 
is much easier to obtain than the intertwiner corresponding 
to $\, Homomorphisms(Oper, \, adjoint(Oper))$. 

\vskip .1cm 

{\bf Remark 2:} From the fact that the ``rightmost'' self-adjoint
operator $\, U_1$ in these decompositions (\ref{order6}), ... (\ref{order12}), 
(\ref{last4forL14}), (\ref{order16}), is actually an {\em order-four} operator, 
 one immediately
deduces (see section (\ref{drop})),
that its exterior square has a drop of order (no rational solution), but
that the exterior square of its {\em adjoint} has a rational solution
(corresponding to  the exterior square of the left-most 
order-two operator $\, M_2$).

\vskip .1cm 

{\bf The results:} Switching not only to differential systems, but
differential {\em theta systems}~\cite{forthcoming}, to obtain 
$ \, Homomorphisms(adjoint(Oper), \,Oper)$,
and, then, using the inversion relation to obtain the intertwiner 
corresponding to $\, Homomorphisms(Oper, \, adjoint(Oper))$,
were two crucial steps to get these results for such very large operators.
Even so, the calculations still remain quite ``massive''. For instance, among 
the last {\em order-fourteen} operators, for which we have been able to perform
these calculations, for the order-fourteen (degree 185) operator, 
denoted v.23.592 in the list~\cite{Lairez2}, 
the first theta system step  required 381 CPU hours,
when the inversion relation step, which is essentially a linear 
calculation, took only 3 CPU hours. In constrast the last
euclidean right-division step takes less than two minutes.
The only\footnote[2]{At the present moment ...}
{\em order sixteen} operator, for which we have been 
able to perform these calculations, is denoted v.23.696 in the 
list~\cite{Lairez2}. This operator is the same as
operator v22.1476 in the list~\cite{Lairez2},  
associated with topology 13 (see~\cite{Lairez}). 
It is of degree 190. This operator 
annihilates a diagonal of rational function having the series expansion:
\begin{eqnarray}
\label{seriesL16}
\hspace{-0.9in}&&\,  \, \, \quad \, 
1  \, \,+ 18\cdot \, t^2  \,\, + 138\cdot \, t^3 \, \, + 2094\cdot \, t^4  \, \,
+ 29520\cdot \, t^5  \,\, + 465210\cdot \, t^6 \, \,
+ 7569240\cdot \, t^7  
\nonumber \\ 
\hspace{-0.9in}&&\,  \, \, \, \quad \quad 
\, + 128131710\cdot \, t^8 \, \, + 2225959680\cdot \, t^9 
 \,\, + 39546740268\cdot \, t^{10} \,\,  + \, \,\, \cdots  
\end{eqnarray}
For this  order sixteen operator the first theta system step  
required 683 CPU hours (i.e. one month) and 39 Gigas of Memory, when the inversion 
relation step, which is essentially a linear calculation,
took only 6 CPU hours. In constrast the last
euclidean right-division step takes less than ten minutes.

\vskip .1cm 

We found that all the thirty-two {\em order-six} operators were  of the form
(\ref{lastm1}), all the seven {\em order-eight} operators were  of the form
(\ref{last}),  all the sixteen {\em order-ten} operators were  of the form
(\ref{last2}), all the fifteen  {\em order-twelve} operators were  of the form
(\ref{last3}), all the {\em order-fourteen} operators\footnote[5]{With the notations
of~\cite{Lairez2}, the order-fourteen operators are v.26.354,
v.23.469,  v.23.473 and v.23.592, v.23.375 and v.23.585.
 The  operators v.23.473 and v.23.592 are, in fact the same operator. 
The  operators  v.23.375 and v.23.585 are, also the same operator.}
were  of the form (\ref{last4}),
and, finally, that one {\em order-sixteen} operator\footnote[3]{With the notations
of~\cite{Lairez2}, the other order-sixteen operators are the  degree 206 operators
v19.5882 and v21.845  which are in fact the same operator associated with topology 22,
the degree 221 operators v21.120, v21.2347, v22.1519 which are in fact 
the same operator associated with topology 10,
the  degree 236 operators v22.316, v22.357 and v23.42 which are in fact 
the same operator associated with topology 14.}, was of the form (\ref{last6}). For all
these operators all the $\, U_n$'s are {\em order-two} self-adjoint operators,
except the rightmost operator $\, U_1$ which is an {\em order-four}
self-adjoint operator. 

\vskip .1cm 

{\bf Conjecture:} We conjecture that {\em all the other}\footnote[1]{There
are eight (different) operators of order 18, two operators of order
20, three of order 22 and one operator of order 24.} 
{\em operators of higher orders} 
($\, 16$, $\, 18$, $\, 20$, $\, 22$, $\, 24$) also 
have decompositions generalizing the form (\ref{last4}), namely the 
rightmost self-adjoint operator  being of {\em order four}, 
all the other self-adjoint operators being of {\em order two}.  

\vskip .1cm 

{\bf By-product:} As a by-product these decompositions show that the differential Galois
groups of all these operators of even order $\, q$ are included in the symplectic 
groups $\, Sp(q, \, \mathbb{C})$. This is coherent with 
previous results of Bogner~\cite{BognerGood} on Calabi-Yau operators.

\vskip .1cm 

{\bf Remark 3:} Considering the successive euclidean right divisions 
described in section (\ref{towersimpleright}), starting with
the first euclidean right division
of the operators by the intertwiner corresponding to
$ \, Homomorphisms(adjoint(Oper), \,Oper)$, yields a set of  
operators $\, L_{[n]}$ ($n\, = \, \, 1, \, \cdots, \, N-1$, see (\ref{recurs})), 
and a set of self-adjoint operators $\, U_N$. One finds 
that the Wronskians of all these 
operators are rational functions. Furthermore, the critical exponents of all the
singularities of these operators (the $\, L_{[n]}$'s 
in the tower of intertwiners (\ref{recurs}) and the $\, U_n$'s)  
are {\em rational numbers} which are
half integers, and, in fact, {\em most of the time integers}. 
However, one actually finds that these underlying 
operators are {\em not globally nilpotent}\footnote[3]{We thank A. Bostan
for providing, here, several $\, p$-curvature calculations, which confirmed our
more limited $\, p$-curvature calculations.}. 
These results are in agreement with
the results in~\cite{bridged,unabridged}
where the decompositions of an order-six and an order-eight operator
yield self-adjoint operators that are, also, {\em not globally nilpotent} 
(see sections (3.6) and (3.7) in~\cite{bridged}).
Along a similar line, note that the series-solutions, 
analytic at $\, x=0$, of these ``underlying'' operators are, also, 
{\em not globally bounded}~\cite{Christol,Short,Big} in contrast with 
the 210 operators~\cite{Lairez2}
(which correspond to diagonal of rational functions). This excludes the 
possibility that these underlying operators could annihilate diagonals 
of rational functions. One finds, however, that the 
rightmost self-adjoint order-four operator $\, U_1$ corresponds 
to a Maximal Unipotent Monodromy 
(MUM) structure\footnote[5]{All the other $\, L_{[n]}$'s in the ``tower
of intertwiner'', of order higher than four, are {\em not} MUM.} 
(see section (6.2) in~\cite{CalabiYauIsing}).

In the decomposition 
of these operators~\cite{Lairez2}, corresponding to Calabi-Yau manifolds,
the rightmost self-adjoint operators $\, U_1$
 satisfy a large number of the properties defining the ``standard''
Calabi-Yau order-four ODEs~\cite{bridged,TablesCalabi}: they
are self-adjoint, they satisfy the ``Calabi-Yau condition'', 
see section (4) in~\cite{bridged}, they
have rational Wronskians, the critical exponents of all their singularities 
are rational numbers, however they are {\em not globally nilpotent},
and the series-solutions, analytic at $\, x=0$, of these operators are 
{\em not globally bounded}\footnote[2]{
And this is also the case for the corresponding nome or Yukawa 
couplings~\cite{bridged,unabridged}.}, and, probably, these
 series-solutions cannot be represented as $\,n$-fold integrals. Therefore,
these order-four self-adjoint operators $\, U_1$ are {\em not ``standard''
Calabi-Yau order-four ODEs}~\cite{bridged,TablesCalabi}: they 
correspond to some interesting generalization of Calabi-Yau order-four ODEs.

\vskip .1cm 

{\bf Remark 4:} For all the operators for which we have been able to get
the intertwiner $ \, Homomorphisms(adjoint(Oper), \,Oper)$
necessary to get the decompositions, we have remarked, with some surprise,
that this intertwiner is an operator with {\em polynomial coefficients}.
If one looks at the intertwining relation (\ref{intertwin1}), it is easy 
to see that a different normalisation of the operator, 
$\, {\cal L}_{12} \, \rightarrow \, F(x) \cdot \,  {\cal L}_{12}$,
yields an intertwiner 
$\, \tilde{{\cal L}}_{8} \, \rightarrow \,  \, \tilde{{\cal L}}_{8} \cdot \, F(x)$
that does not have this polynomial property anymore. This polynomial property 
corresponds to a particular normalization of the 210 operators in~\cite{Lairez2}: 
actually, the operators in~\cite{Lairez2} are all normalized so that they are operators with 
{\em polynomial coefficients}.
One can imagine to get, in an easier way, all these results, using the assumption that 
the intertwiners corresponding to $ \, Homomorphisms(adjoint(Oper), \,Oper)$ 
are operators with polynomial coefficients. 

\vskip .1cm 

\subsection{Revisiting the $\, L_{12}^{(left)}$ 
and $\, L_{21}$ operators of the Ising model.}
\label{revisting}

In a recent paper~\cite{L12L21}, we have shown that two quite large 
linear differential operators of order 12 and 21,
$\, L_{12}^{(left)}$  and $\, L_{21}$, which
 correspond to factors in the minimal order operators annihilating
 respectively the $\, \tilde{\chi}^{(5)}$ and $\, \tilde{\chi}^{(6)}$
components of the magnetic susceptibility of the Ising model, were 
actually such that the {\em exterior square} of $\, L_{12}^{(left)}$,
 and the {\em symmetric square} of $\, L_{21}$, have, both, a {\em rational 
function solution}. The differential Galois group of 
$\, L_{12}^{(left)}$ is in $\, Sp(12, \,  \mathbb{C})$,
and the differential Galois group of $\, L_{21}$
 is in $\,SO(21, \,  \mathbb{C})$. In order to get
the decomposition of these operators, one just needs to 
get the first intertwiner (see (\ref{tower1}) in section (\ref{towersimpleright})),
the other intertwiners being obtained, almost instantaneously, from 
euclidean right divisions. Unfortunately, that first step cannot
be performed: finding the intertwiners corresponding to homomorphisms 
of these very large linear differential operators
with their adjoints, is beyond our computer resources. Let us, however, 
sketch the various scenarii for the decompositions of these operators
using the new results of this paper. 

Since  the exterior square of $\, L_{12}^{(left)}$,
and the symmetric square of $\, L_{21}$, have, both, a rational 
function solution~\cite{L12L21}, we know that the ``rightmost'' 
self-adjoint operator $\, U_1$, in the corresponding decompositions, 
will be an operator of {\em order two} for $\, L_{12}^{(left)}$,
and an operator of {\em order one} for $\, L_{21}$.  However, since 
finding the rational solutions of the exterior square of the {\em adjoint}
of $\, L_{12}^{(left)}$, or of the symmetric square of the {\em adjoint}
of $\, L_{21}$, is also beyond our computer resources at the present moment,
we cannot find\footnote[5]{Knowing the leftmost self-adjoint operator
$\, U_N$ in the decompositions, one can get the adjoint of 
$\, L_{[N-1]}$ and  $\, L_{[N-2]}$, from a simple euclidean right division
of the adjoint of $\, L_{[N]}$ by $\, U_N$ (see below (\ref{recursapp})).} 
the order of the left-most self-adjoint operator 
$\, U_N$ in the decompositions of  $\, L_{12}^{(left)}$
 or $\, L_{21}$ (see (\ref{recurs}), see also 
section (\ref{decomp})), and, consequently, one cannot 
discard the ``generic'' scenario. 

If operator $\, L_{12}^{(left)}$ has a decomposition 
corresponding to the ``generic'' scenario, it would have 
a decomposition like $\, L_{[6]}$ in (\ref{last4}), the six
{\em self-adjoint operators being of order two}. Similarly,
if operator  $\, L_{21}$ has a decomposition  
corresponding to the ``generic'' scenario for $\,SO(21, \,  \mathbb{C})$,
it would have a decomposition with 21 {\em self-adjoint operators 
of order one}. However, it has been remarked, several times in this paper,
that the operators emerging in physics, are, often, 
{\em not of the generic type}.

Even if the decomposition of $\, L_{12}^{(left)}$ does not correspond to the
 ``generic'' scenario, for instance,
\begin{eqnarray}
\label{scenario2}
\hspace{-0.95in}&& \,   \quad   \quad   \quad   \qquad  
L_{12}^{(left)} \,  \, = \, \, \, \, 
(X_{2} \cdot \, Y_{8} \cdot \, Z_{2} \, +\, X_{2} \, +\,Z_{2}) \cdot \, r(x), 
\end{eqnarray}
we are, however, sure that an equivalent operator 
will be of the generic form (\ref{last4}), namely
a decomposition in terms of {\em six} self-adjoint operators. Similarly, even if 
the decomposition of $ \, L_{21}$ does not correspond to the
 ``generic'' scenario (\ref{last4}),  a general enough equivalent operator 
will be of that generic form (\ref{last4}).

\section{Speculations on diagonals of rational functions and selected 
differential Galois groups}
\label{Specul}

For the $\,\chi^{(n)}$'s components of the magnetic 
susceptibility of the Ising model~\cite{Landau,Landau2}, for $\,n \ge\, 7$,  
even obtaining just the
next operator annihilating $\,\chi^{(7)}$ is, and will certainly remain, 
out of reach for many years, possibly decades\footnote[2]{We have, 
however, been able to describe the singularities
of all the $\,\chi^{(n)}$'s from a Landau singularity 
approach~\cite{Landau,Landau2}.}. 

All the small factors of the minimal order operators 
annihilating  the $\,\chi^{(n)}$'s 
we have obtained and studied~\cite{L12L21}, correspond to 
{\em globally bounded}~\cite{Christol,Short,Big}
series solutions, in fact hypergeometric series with {\em integer coefficients}, 
having an elliptic function, or {\em modular function}, interpretation. These solutions are  
{\em diagonals of rational functions}. It was shown
in~\cite{Short,Big} that the $\,\chi^{(n)}$'s  are actually {\em diagonals of rational functions},
but we have seen that {\em all the factors} of the operators 
annihilating the $\,\chi^{(n)}$'s, themselves,
seem to have solutions that are {\em diagonals of rational functions}. Is this 
property always verified for any diagonal of rational function, 
or is it a consequence 
of our "physical framework" ?

Furthermore, we have seen~\cite{Short,Big} that all these factors 
correspond to {\em selected differential Galois groups,
these operators being homomorphic to their adjoints}. 
This raises the question to see whether the factors of the (minimal order) 
operators, {\em annihilating diagonals of  rational functions,
are, quite systematically, homomorphic to their adjoint}, thus corresponding to 
{\em selected differential Galois groups}, or, if this 
"{\em duality property}" is, on the contrary, 
a {\em consequence of our "physical framework"}. 
A first experimental examination of hundreds of (simple enough) diagonals
of three variables seem to {\em systematically yield}  such a 
{\em duality} (homomorphism of the factor to its adjoint), but one must 
be careful before generalizing too quickly these results.
In fact, the (minimal order) 
operators,  annihilating diagonals of a rational functions,
are {\em not systematically} homomorphic to their adjoint. 
Let us consider the series expansion of the simple hypergeometric 
function\footnote[9]{We thank A. Bostan for providing 
this simple hypergeometric example.} 
$\,_3F_2([1/3,1/3,1/3],[1,1],3^6 \, x)$. It is a series with {\em integer}\footnote[5]{Even 
$\,_3F_2([1/3,1/3,1/3],[1,1],3^5 \, x)$ has a series with integer coefficients.}
coefficients. One verifies easily that this
hypergeometric series is the {\em Hadamard cube} of the series (with integer
coefficients) of the algebraic function $\, (1\, -9\, x)^{-1/3}$, and is, thus, the diagonal
of a rational function~\cite{Short,Big}. On the other side, one can show that the 
order-three operator annihilating this hypergeometric series has a differential Galois group 
which is $\, SL(3, \, \mathbb{C})$, and thus it {\em cannot}\footnote[1]{Using 
the fact that the symmetric square of that irreducible
order-three operator has no rational solution and has logarithms 
in its series-solutions~\cite{Bouchet}, or using the algorithm in~\cite{Ragot}
and showing that there is no invariant of degree 2,3,4,6,8,9,12.}
 {\em be homomorphic to its adjoint}
(even in some involved algebraic extension). 

Accumulating more examples of operators annihilating diagonals 
of rational functions,  in order to find if they are
homomorphic to their adjoints or not,
should help to clarify the {\em relation between diagonals of 
rational functions and selected differential Galois groups} 
(and associated decompositions), in order to understand
 why the diagonals of rational functions
{\em emerging in physics} seem to have, systematically,
 this "{\em duality property}".

\vskip .1cm 

\section{Conclusion}
\label{conclu}

Trying to understand why an order-six operator $\, {\cal L}_6$ 
was homomorphic to its adjoint, thus having
a selected differential Galois group, we have discovered 
that this was, in fact, a consequence of a decomposition (\ref{decompL6})
of this order-six operator into three order-two {\em self-adjoint} operators.
This provides a first example of a {\em selected differential Galois group}
that does not emerge from a decomposition  like (\ref{genr1}) or 
(\ref{genr2}), but actually emerges from a more general {\em new type of 
decomposition} namely (\ref{decompL6}). 
A first set of {\em selected differential Galois groups}
actually emerges from a decomposition (\ref{decompL6}), where {\em one just needs
to impose that the orders of the three self-adjoint operators} 
have the {\em same parity} (not necessarily even). 
We, then, discovered in section (\ref{tower}),
a recursion, {\em based on euclidean right-division of operators},
on a sequence of linear differential operators 
where the {\em intertwiner in the homomorphism
with the adjoint of an operator $\, L_{[N]}$ is actually the next 
operator $\, L_{[N-1]}$ in the sequence}: we do have 
a ``{\em tower of intertwiners}''. These
canonical decompositions in an {\em arbitrary number} of self-adjoint
operators, provide an {\em infinite number} of linear differential operators 
which have, automatically, {\em selected 
differential Galois groups}. We defined the ``generic'' decompositions as the one
where the self-adjoint operators $\, U_n$, in the decompositions,
are {\em all of order one}, or {\em all of order two}. It was 
found\footnote[2]{Using Kolchin's result~\cite{Kolchin1,Kolchin2}
 that the most general operators with 
selected differential Galois groups can be reduced, up to operator equivalence, 
to operators associated with some simple 
``reduced form''~\cite{Symbolic,AparicioLast} for 
the associated linear differential systems.}  ``experimentally'' 
that the {\em most general operators with selected differential 
Galois groups do correspond to what we have 
called the ``generic decompositions''}. To some extent, this provides 
an {\em algebraic approach of  differential Galois groups}: the existence 
of such simple algebraic decompositions of the operators
is the ``deus ex machina'' for selected differential Galois groups.
 
According to the {\em parity} order of the 
underlying self-adjoint operators required to build them,
the {\em exterior, or symmetric,} squares of the
selected differential operators, or equivalent 
operators, described in this paper, 
 have {\em rational solutions}.
We have seen that the ``generic'' character of the decomposition
is (generically) {\em preserved by the operator equivalence}. In contrast, 
operators equivalent to operators 
with ``non-generic'' decomposition eventually have a ``generic'' decomposition
for involved enough operator equivalence. Non-genericity\footnote[5]{An apparently 
quite frequent situation in physics, or enumerative combinatorics~\cite{bridged,unabridged}.} 
is, thus, not preserved by the operator 
equivalence. We have also found the remarkable result 
that the rational solutions (or drop of 
order) of the symmetric, or exterior, squares of the operators with these 
decompositions, {\em depend only on}  $\, U_1 \cdot \, r(x)$, where $\, U_1$
is the {\em rightmost 
self-adjoint operator}. Rational solutions always emerge 
for equivalent operators, for an involved enough equivalence.

\vskip .1cm 

Since we have a well-defined algorithm to get
these canonical decompositions of operators with
selected differential Galois groups, we used it to 
obtain the decompositions of various remarkable
(and quite massive) operators. 
For instance, in a quite systematic analysis of a set, obtained 
recently by P. Lairez~\cite{Lairez2,Lairez}, of 210 
 explicit linear differential
operators associated with reflexive 4-polytopes 
defining 68 topologically different Calabi-Yau 3-folds,
we found, with quite large calculations, that all the 
 order-six, order-eight, order-ten, order-twelve,
 order-fourteen operators, and a first
{\em order-sixteen} operator, 
had the decompositions detailed in this paper,
all the $\, U_n$'s being {\em order-two} self-adjoint operators
except the rightmost operator $\, U_1$ which is an {\em order-four}
self-adjoint operator. We conjecture that all these operators 
of higher order ($14$, $\, 16$, $\, 18$, $\, 20$, $\, 22$, $\, 24$)
also have this kind of particular decomposition ($\, U_N$ of order-two
and  the rightmost operator $\, U_1$ of {\em order-four}).

\vskip .1cm 

All the structures, decompositions, 
discovered in this paper, can be seen as a simple {\em algebraic description}
of the linear differential operators with selected differential Galois groups.
We have seen that the various linear differential operators 
emerging in the Ising model, in a large number
of integrable models of lattice statistical mechanics, or enumerative 
combinatorics (lattice Green 
functions~\cite{Short,Big,bridged,unabridged,Guttmann,GlasserGuttmann,Kou}), 
correspond to selected linear differential operators. They 
are globally nilpotent~\cite{bo-bo-ha-ma-we-ze-09}
operators, or operators associated with reflexive polytopes (and Calabi-Yau 3-folds),
all associated with {\em diagonals of rational functions}~\cite{Short,Big}),
which also correspond to selected differential Galois groups. This paper provides simple, 
and computationally efficient,
{\em algebraic tools} to study, and describe, these selected operators
and their selected differential Galois groups.

\vskip .3cm 

\vskip .3cm 

{\bf Acknowledgments:} We thank A. Bostan, G. Christol
 and P. Lairez for fruitful
 discussions on diagonals of rational functions.
We thank P. Lairez for generously sending us explicit examples 
of selected operators. 
 We thank D. Bertrand for fruitful discussions on
 differential Galois groups and (self-adjoint) dualities
in geometry. S. B. would like to thank the CNRS and the LPTMC 
for kind support. This work has been performed without
 any support of the ANR, the ERC or the MAE. 

 \vskip .5cm 

 \vskip .5cm 

\appendix

\section{Analysis of the order-six operator $\, {\cal L}_6$ }
\label{analyL6}

This order-six linear differential operator $\, {\cal L}_6$ 
in~\cite{bridged,unabridged}, analyzed in section (\ref{revisit}) annihilates 
the diagonal of the (three variables) rational function
\begin{eqnarray}
\label{genericappend}
\hspace{-0.95in}&&
R(x,\, y, \, z)\, \,  = \, \,   {{1} \over { 1\, -P(x,\,y, \, z)}}
\,  = \, \,  
 {{1} \over { 
1 \, -3\, x \, -5\, y \, -7\, z \,+x\, y \,+2\,y\,z^2\, +3\,x^2\,z^2}}.
\end{eqnarray}

The expansion (\ref{genericdiag}) of the rational function (\ref{genericappend}) 
can also be obtained from an expansion using {\em multinomial coefficients}:
\begin{eqnarray}
\label{multi}
\hspace{-0.95in}&& \, \, \qquad   \, R(x,\, y, \, z)
\, \, = \, \, \,  \,\, \sum_{n=0}^{\infty} \, P(x,\,y, \, z)^n 
\nonumber  \\
\hspace{-0.95in}&& \quad \quad \qquad 
\, \, = \, \, \,  \,\,  
 \,   \sum_{m_1 \, \cdots \, m_6} \, 
 {\,m_1 +  \ldots + m_6  \choose m_1,\,  m_2,\,   \ldots, \,  m_6}  
\cdot \, A_{m_i}(x, \, y, \, z), \quad \quad \quad  \quad  \, \, \,
 \hbox{where:}
\nonumber  \\
\hspace{-0.95in}&& \quad A_{m_i}(x, \, y, \, z)
\, \, = \, \, \,   
\, (3\, x)^{m_1}  \, (5\, y)^{m_2}  \, (7\, z)^{m_3} 
 \, (-x\, y)^{m_4}  \, (-2\,y\,z^2)^{m_5} 
\, (-3\,x^2\,z^2)^{m_6}. \nonumber 
\end{eqnarray}

Then the expansion (\ref{genericdiag})
of the diagonal of this rational function (\ref{genericappend}) reads
\begin{eqnarray}
\label{dmulti}
\hspace{-0.95in}&&  \, \quad  \quad   \quad 
Diag(R(x,\, y, \, z)) 
\, \, = \, \, \,  \,\, 
\, \sum_{M=0}^{\infty} \, \sum_{N,m_1,m_5} \, A_{N,m_1,m_5} \cdot \, x^M, 
\end{eqnarray}
\begin{eqnarray}
\label{dmulti2}
\hspace{-0.95in}&&  \hbox{where:} \quad  \quad  \quad \, \, 
A_{N,m_1,m_5} \, = \,  \
\, {N \choose m_1, m_5, p_1, p_2,p_3,p_4}  
\cdot \, 2^{m_5} \cdot \,3^{q_1} \cdot \, \,5^{q_2}
 \cdot \, \,7^{q_3}   \cdot \, (-1)^{q_4},
\nonumber 
\end{eqnarray}
with 
\begin{eqnarray}
\label{dmulti3}
\hspace{-0.95in}&& \quad \,
 p_1\, = \, \, 3m_1\, -2N\, -4M\, -5m_5, 
\quad  \quad \quad 
p_2 \, = \, \,2N\, -2m_1\, +2m_5\, -3M, 
\nonumber  \\
\hspace{-0.95in}&& \quad \qquad \, \, \, 
 p_3 \, = \, \,2N\, -3m_1\, -3M\, +4m_5, 
 \quad  \quad \, \, 
p_4 \, = \, \,2M\, -N\, +m_1\, -2m_5, 
\end{eqnarray}
and:
\begin{eqnarray}
\label{dmulti4}
\hspace{-0.95in}&& \qquad q_1\, = \, \, 2M -N +2m_1 -2m_5,
 \quad  \quad \quad 
q_2 \, = \, \,3m_1 -5m_5 +4M -2N, \quad 
\nonumber  \\
\hspace{-0.95in}&& \quad \quad \quad \quad 
q_3 \, = \, \,2N -2m_1 +2m_5 -3M, 
\quad \quad \, \,  q_4 \, = \, \,N -2m_1 -M +3m_5,  
\end{eqnarray}
where the summation in (\ref{dmulti}) is taken over all the 
integers $\, m_1$, $\, m_5$ and $\, N$, provided the $\, p_i$'s 
in (\ref{dmulti3}) are not negative.

\vskip .1cm 

\subsection{Decomposition of the order-four operator $\, {\cal L}_4$}
\label{decompL4app}

The order-four linear differential operator 
 $\, {\cal L}_4$, of section (\ref{revisit}), reads
\begin{eqnarray}
\label{decompP4app}
\hspace{-0.9in}&& \quad  \,  \quad  \quad  \quad \quad \quad  
{\cal L}_4  \,  \, \, = \, \, \, \,\, 
 (N \cdot \, P \, \, + \, 1) \cdot \, r(x), 
\end{eqnarray}
where  $\, N $ and $\,P $ are two order-two {\em self-adjoint operators}, 
and where $\, r(x)$ is the rational function
\begin{eqnarray}
\label{defadex}
\hspace{-0.9in}&&  \quad  \quad 
r(x) \,  \, \, \, = \, \, \, \,\, c_r \cdot \,
x \cdot \, \Bigl({{ p_{28}} \over {p_{43} \cdot \, p_{10}}}\Bigr)^2,  \\
\hspace{-0.95in}&& \quad  \quad 
\hbox{with:}\qquad \qquad 
c_r \, \, = \, \,\, -175746210353375850313251934961664000000,  
\nonumber
\end{eqnarray}
where $\, p_{43}$ is an {\em apparent polynomial}
 of degree 43 given below in (\ref{defp43one}),  
where $\, p_{10}$ is given below in (\ref{p10}), and where $\, p_{28}$
is a polynomial of degree 28 given in(\ref{defp28}).
The two  self-adjoint operators $\, N $ and $\,P $ 
are quite large order-two operators.
The order-two self-adjoint operator $\, N$ reads:
\begin{eqnarray}
\label{defsN}
\hspace{-0.9in}&& \quad  \,  \,
 {{1} \over {c_N}} \cdot \,  N \, \,\, \, = \, \, \,  \, \,\,
{{x^2 \cdot \, p_{28}^3} \over {p_{10} \cdot \,p_{43}^4}} \cdot \, D_x^2
 \, \, \,  \,  
+ \,\, {{x \cdot \, p_{28}^2  \cdot \, p_{81}} 
\over {p_{10}^2 \cdot \,p_{43}^5}} \cdot \, D_x\, \,  \, + \, 
{{x \cdot \, p_{28}^2  \cdot \, p_{123}} 
\over {p_{10}^2 \cdot \,p_{43}^6}},  \\
\hspace{-0.95in}&& \quad  \quad 
\hbox{with:}\qquad \qquad \quad 
 c_N \, = \, \,-738664786498877076270961404149760000000, 
\nonumber 
\end{eqnarray}
where $\,p_{81}$ is a  polynomial of degree 81 in $\, x$, with integer coefficients, 
and where $\,p_{123}$ is a  polynomial of degree 123, also with integer coefficients.

The order-two self-adjoint operator $\, P$ reads:
\begin{eqnarray}
\label{defP}
\hspace{-0.95in}&&  \,  \, 
c_P \cdot \, P \, \, \,  = \, \,  \,  \, 
 \, \, {{ p_{10}^3 \cdot \, p_{12} \cdot \, p_{43}^4 } \over { 
x \cdot \, p_{28}^4}} \cdot \, D_x^2 \,
\, \, +  \, {{ p_{10}^2 \cdot \, p_{93} \cdot \, p_{43}^3 } \over { 
x^2 \cdot \, p_{28}^5}} \cdot \, D_x  
  \, \,  \, + \, {{ p_{10}^2 \cdot \,p_{164} \cdot \, p_{43}^2 } \over {
 x^3 \cdot \, p_{28}^6}}, \\
\hspace{-0.95in}&& \quad  \quad 
\hbox{with:}\qquad \qquad  c_P \, = \, \, 305114948530166406793840164864000000,  
\nonumber 
\end{eqnarray}
where $\, p_{93}$ and $\, p_{164}$ are polynomials with integer coefficients
of degree 93 and 164 respectively. The 
order-two operator $\, P$ can be redefined 
in a simpler way introducing another 
self-adjoint operator $\, r(x) \cdot \,P \cdot \, r(x)$: 
\begin{eqnarray}
\label{defPother}
\hspace{-0.95in}&&  \, \quad \quad  
 {{1} \over {c}} \cdot \, r(x) \cdot \, P \cdot \, r(x)
 \, \, \,  = \, \,  \,  \,  \, 
{{  p_{12} } \over { p_{10} }} \cdot \, x \cdot \, D_x^2
\, \,  + \,  {{  p_{22} } \over { p_{10}^2 }} \cdot \, D_x \, 
\, \,  + \, 2 \, {{  p_{21} } \over { p_{10}^2 }}, \\
\hspace{-0.95in}&&  \, \qquad \quad  \hbox{with:} \quad \quad  \quad \quad 
c  \, \, = \, \,  \,  \, 101229817163544489780433114537918464000000, 
\nonumber 
\end{eqnarray}
where $\, p_{22}$ and $\, p_{21}$ are  degree 22 and 21 polynomials 
with integer coefficients (see (\ref{defp22}), (\ref{defp21}) given 
below).

\subsection{Euclidean right division of  $\, {\cal L}_6$ by  $\, {\cal L}_4$ \newline }
\label{decompL4appsub}

Recalling the euclidean right division (\ref{rightdivL6L4}) of  the order-six
operator $\, {\cal L}_6$ by the order-four operator $\, {\cal L}_4$, 
namely $\, {\cal L}_6 \, = \, M \cdot \, {\cal L}_4 \, + \, {\cal L}_2$, 
the order-two self-adjoint operator $\, M$ reads:
\begin{eqnarray}
\label{defM}
\hspace{-0.9in}&& \,  \,  \,  \,  \quad   \quad  
c_M \cdot \,  M 
\, \, \,\,  = \, \, \, \,  \, \, {{p_{43}^3} \over {p_{28}}} \, \cdot \, D_x^2 
 \, \, \, \,   + \, {{p_{70} \cdot \,p_{43}^2} \over { p_{28}^2}} \, \cdot \, D_x 
\, \,  \,  +  \, {{q_{70} \cdot \,p_{43}^2} \over { x \cdot \, p_{28}^2}}, 
\\
\hspace{-0.95in}&& \quad  \quad  \quad  \quad  
\hbox{with:}\qquad \qquad  c_M \, = \, \, 697405596640380358385920376832000000,
\nonumber 
\end{eqnarray}
where $\, p_{70}$ and  $\, q_{70}$ are two polynomials of 
degree 70 in $\, x$, with integer coefficients, 

\subsection{The polynomials occurring in  the analysis of $\, {\cal L}_6$ }
\label{polynomial}

The two polynomials $\, p_{10}$ and  $\, p_{12}$ 
occurring in section (\ref{revisit}), in
the rational function of the exterior square of $\, {\cal L}_6$, 
are two polynomials of degree
ten and twelve with integer coefficients, which read respectively:
\begin{eqnarray}
\label{p10}
\hspace{-0.95in}&&  \, \quad 
p_{10} \, \, = \, \, \, 145212480\,{x}^{10} \, 
-1851804864\,{x}^{9}\,-471355865712\,{x}^{8}\,
-2127618407544\,{x}^{7} 
\nonumber \\
\hspace{-0.95in}&&  \, \quad \quad \quad \,
-2504188513576\,{x}^{6}\, -4687345201826\,{x}^{5}\,
-9012222844732\,{x}^{4}\,
\nonumber \\
\hspace{-0.95in}&&  \, \quad \quad \quad \quad 
 \,\, \, -18285528253377\,{x}^{3}\, 
-22232025838680\,{x}^{2}\,-10741640355390\,x\,
\nonumber \\
\hspace{-0.95in}&&  \, \quad \quad \quad \quad \quad 
\,\, \, -1453000612770, 
\end{eqnarray}
and 
\begin{eqnarray}
\label{p10}
\hspace{-0.95in}&& \quad \, 
p_{12} \, \, = \, \, \, 474360768\,{x}^{12} \,  \, 
-21346234560\,{x}^{11} \, \,  -9830736566352\,{x}^{10}
\nonumber \\
\hspace{-0.95in}&&  \, \quad \,  \quad \quad 
\,\, \, -49730104754400\,{x}^{9}
+332161583716500\,{x}^{8}\, +2281890913038548\,{x}^{7}\, 
\nonumber \\
\hspace{-0.95in}&&  \, \quad \, \quad \quad \,\, \,
+4259219378255537\,{x}^{6}\, 
+2726995508245092\,{x}^{5}\, +266148400806530\,{x}^{4}\,
\nonumber \\
\hspace{-0.95in}&&  \, \quad \, \quad\quad 
 \,\, \, +1339537706508092\,{x}^{3}
+2659961825724861\,{x}^{2}\, +1339804492447785\,x\,
\nonumber \\
\hspace{-0.95in}&&  \, \quad \, \quad \quad \quad
 \,\, \,  -484333537590.
\end{eqnarray}

The degree-28 polynomial $\, p_{28}$, occurring in the rational function 
(\ref{defadex}) emerging in the decomposition of the order-four intertwiner 
(\ref{decompP4})
and also in the decomposition of $\, {\cal L}_6$  (see (\ref{decompL6})),
reads:
\begin{eqnarray}
\label{defp28}
\hspace{-0.95in}&&  
p_{28} \, \, = \, \, \, 222000402253116211200\,{x}^{28} \, 
+16671177334581067210752000\,{x}^{27} 
\nonumber \\
\hspace{-0.95in}&& \quad  \, \,
 +997686473094955341487964160\,{x}^{26} \, \,  
-424558055468758430724178378752\,{x}^{25}\, 
\nonumber \\
\hspace{-0.95in}&& \quad  
-5352263005372177429702409969664\,{x}^{24}
\,+28532549906147363004230907949056\,{x}^{23}\,
 \nonumber \\
\hspace{-0.95in}&& \quad  \, 
-817011419352436691241775865192448\,{x}^{22}\, 
-16274396213266656792629292697691136\,{x}^{21}\, 
\nonumber \\
\hspace{-0.95in}&& \quad  \, 
-90232371132900789295067378226024960\,{x}^{20}\,
-500637681004648369297528874000878080\,{x}^{19}\,
 \nonumber \\
\hspace{-0.95in}&& \quad  \,
 -4993606665623240703449658338613645312\,{x}^{18}\, 
-30245133133694295061972050348437464512\,{x}^{17}\,
\nonumber \\
\hspace{-0.95in}&& \quad  \, 
 -185743134083879510307905273106717264704\,{x}^{16}\, 
-976930467381467189083133231381070485120\,{x}^{15}\, 
\nonumber \\
\hspace{-0.95in}&& \quad  \, 
-3367039413961560591653749004449857274584\,{x}^{14}\,
 -7294585923067316376931832664546378582832\,{x}^{13}\,
\nonumber \\
\hspace{-0.95in}&& \quad  \,  
 -9938986858255476817819640602774955842078\,{x}^{12}\,
-8869602400042451664718223608577965685292\,{x}^{11}\,
\nonumber \\
\hspace{-0.95in}&& \quad  \, 
-7915734277920977075124503067364891962671\,{x}^{10}\,
-13323343876410843499695772267774078620818\,{x}^{9}\,
\nonumber \\
\hspace{-0.95in}&& \quad  \,  
-23035143510457266975680110399567225817550\,{x}^{8}\,
-27995703147600523231627072322667435627270\,{x}^{7}\,
\nonumber \\
\hspace{-0.95in}&& \quad  \,   
-23240989943800867264930403245305575019984\,{x}^{6}\,
-12663707457789877548806711493284323118250\,{x}^{5}\,
\nonumber \\
\hspace{-0.95in}&& \quad  \, 
-4207201352793289661461971125476855152780\,{x}^{4}\,
-802705633170266051006530606597781060580\,{x}^{3}\,
\nonumber \\
\hspace{-0.95in}&& \quad  \,  
-154800451744689329012558083417852734900\,{x}^{2}\,
-66136165509758890646569575954920184000\,x\,
\nonumber \\
\hspace{-0.95in}&& \quad  \,  
-17095610140484667152552076876139296000. 
\end{eqnarray}
The polynomials $\, p_{22}$ and $\, p_{21}$, occurring as 
coefficients of the order-two self-adjoint 
operator $\,\, r(x) \cdot \, P \cdot \, r(x)\, $ (see (\ref{defPother})), read:
\begin{eqnarray}
\hspace{-0.95in}&& 
p_{22} \, \, \, = \, \, \, \,
206649310607953920\,{x}^{22}-9713173628131319808\,{x}^{21} \, 
-2426922106370025707520\,{x}^{20}
\nonumber \\
\hspace{-0.95in}&& \quad  \,  +70600350852546385296384\,{x}^{19} \, 
+14163950463400033391748096\,{x}^{18}  
\nonumber \\
\hspace{-0.95in}&& \quad  \,   \, +130185601423920337305765888\,{x}^{17}\,
+286975902003933399495977088\,{x}^{16} \,
\nonumber \\
\hspace{-0.95in}&& \quad 
 -623168847910615654838222592\,{x}^{15} 
\,-3538729092035340386109171216\,{x}^{14}\, 
 \nonumber \\
\hspace{-0.95in}&& \quad  \,  
-10952980116620522700303057376\,{x}^{13}\, 
-43198517288449764379831927088\,{x}^{12}
\nonumber \\
 \hspace{-0.95in}&& \quad  \, 
 -145043720261824787740194197788\,{x}^{11}\, 
-367790107127544232638909982720\,{x}^{10} \,
\nonumber \\
\hspace{-0.95in}&& \quad  
-659752751521470013936611019364\,{x}^{9} \,
 -760856686133164797192096477412\,{x}^{8} \, 
\nonumber \\
\hspace{-0.95in}&& \quad  \, 
-514919966483376918743532932868\,{x}^{7} \,
 -189221289620448844441499010432\,{x}^{6} \,
\nonumber \\
\hspace{-0.95in}&& \quad  \,  
 -70630559928767816259500053080\,{x}^{5} \,
 -79750467708757827084054819315\,{x}^{4} \,
 \nonumber \\
\hspace{-0.95in}&& \quad  \, 
-64947815604716490637198723800\,{x}^{3} \,
   -25997244208301276956743517260\,{x}^{2} \, 
\nonumber \\
\hspace{-0.95in}&& \quad  \, 
-3893473497037260884458428900\,x \,
 +703736926903331829024300,
\end{eqnarray}
and
\begin{eqnarray}
\label{defp21}
\hspace{-0.95in}&&
 p_{21} \, \, \, = \, \, \, \,
33738662956400640\,{x}^{21} \,
-1065211258519296000\,{x}^{20} \,
\nonumber \\
\hspace{-0.95in}&& \quad  \, \,
+137751133024444133376\,{x}^{19}\,
+13971600975964488256512\,{x}^{18} 
\nonumber \\
\hspace{-0.95in}&& \quad  \, \,
 \,+2072616787347233631183360\,{x}^{17} \,
+19143502759621498372119936\,{x}^{16} 
\nonumber \\
\hspace{-0.95in}&& \quad  \,\,
+111056760563865193378425600\,{x}^{15}\, 
+609147860679604401066131808\,{x}^{14}
\nonumber \\
\hspace{-0.95in}&& \quad  \, \,
+2284440127728770464121151168\,{x}^{13}\, 
+3816824280036865582719277232\,{x}^{12} \,
\nonumber \\
\hspace{-0.95in}&& \quad  \, \,
-916281024394298349389193376\,{x}^{11} \,
-17325258089488435492938995946\,{x}^{10} \,
\nonumber \\
\hspace{-0.95in}&& \quad  \,
-47923688835171521983291837594\,{x}^{9} \,
-85220533413397361142814764127\,{x}^{8} \,
\nonumber \\
\hspace{-0.95in}&& \quad  \, \,
-95547159126942042724808925384\,{x}^{7} \,
-57905642741824134044096538894\,{x}^{6} \,
\nonumber \\
\hspace{-0.95in}&& \quad  \,
-6830662192589519324624321892\,{x}^{5} \,
+16304405412896867488602623040\,{x}^{4} \,
\nonumber \\
\hspace{-0.95in}&& \quad  \,
+12975784425095916109623440580\,{x}^{3} \,
+3377129346163911842811657630\,{x}^{2} \,
\nonumber \\
\hspace{-0.95in}&& \quad  \,
-415617758773209736393116600\,x \,
-216750973486226203339484400.
\end{eqnarray}
The polynomials $\, p_{43}$, occurring 
in the head coefficient of  $\, {\cal L}_6$ , reads:
\begin{eqnarray}
\label{defp43one}
\hspace{-0.95in}&&
 p_{43} \, \, \, = \, \, \, \,
697115132002046172480720076800000\,{x}^{43} 
\nonumber \\
\hspace{-0.95in}&& \quad  \,
+46966788589019958554494032194568192000\,{x}^{42}
\nonumber \\
\hspace{-0.95in}&& \quad  \,
-21099793949237885955113953175859206553600\,{x}^{41}
\nonumber \\
\hspace{-0.95in}&& \quad  \,
-12218070507034966285297837769765669044224000\,{x}^{40}
\nonumber \\
\hspace{-0.95in}&& \quad  \,
-14194373444333574120149253605673357064273920\,{x}^{39}
\nonumber \\
\hspace{-0.95in}&& \quad  \,
+95681633992983148850231449432522337810723635200\,{x}^{38}
\nonumber \\
\hspace{-0.95in}&& \quad  \,
+5520522767992453240508293629010426228437862318080\,{x}^{37}
\nonumber \\
\hspace{-0.95in}&& \quad  \,
+114296137092097802339644622301518644478404747591680\,{x}^{36}
\nonumber \\
\hspace{-0.95in}&& \quad  \,
+417910689150890521997273968958589501995213946224640\,{x}^{35}
\nonumber \\
\hspace{-0.95in}&& \quad  \,
-31508493936457860589092556909074176639521503684624384\,{x}^{34}
\nonumber \\
\hspace{-0.95in}&& \quad  \,
-903936764481753141424148779858885101259602207109447680\,{x}^{33}
\nonumber \\
\hspace{-0.95in}&& \quad  \,
-14332265098518509709934412539054897619963416778500300800\,{x}^{32}
\nonumber \\
\hspace{-0.95in}&& \quad  \,
-149481107350715305817082579283208875899765288515401691136\,{x}^{31}
\nonumber
\end{eqnarray}
\begin{eqnarray}
\hspace{-0.95in}&& \quad  \,
-1033538497267096011790975433433897490651064449893529853952\,{x}^{30}
\nonumber \\
\hspace{-0.95in}&& \quad  \,
-4821442010547424238037959567906394473464364860695632913408\,{x}^{29}
\nonumber \\
\hspace{-0.95in}&& \quad  \,
-19978165382552239749293130530244544553933980930279985998848\,{x}^{28}
\nonumber \\
\hspace{-0.95in}&& \quad  \,
-140867377223273230659981893431123146189730880776359488689664\,{x}^{27}
\nonumber 
\\ 
\hspace{-0.95in}&& \quad  \,-1159867839188205256138512414882994988836222652247822650381184\,{x}^{26}
\nonumber  \\
\hspace{-0.95in}&& \quad  \,
-6351896417634861870867164399785963779858852813259236265509632\,{x}^{25}
\nonumber \\
\hspace{-0.95in}&& \quad  \,
-19971763593409793575488427179831444424654145339690894395028704\,{x}^{24}
\nonumber \\
\hspace{-0.95in}&& \quad  \,
-29276418982909322276170630931225213840998278949306320170659232\,{x}^{23}
\nonumber \\
\hspace{-0.95in}&& \quad  \,
+9654630434461854336187543726596136785120347219785702648099920\,{x}^{22}
\nonumber \\
\hspace{-0.95in}&& \quad  \,
+85853277293687339742074678932453004464244583743701538472162264\,{x}^{21}
\nonumber \\
\hspace{-0.95in}&& \quad  \,
-160054706451279429494171232085742639976407036877062557712330264\,{x}^{20}
\nonumber \\
\hspace{-0.95in}&& \quad  \,
-2258895857748459378897561858150946902130393238862424109243558684\,{x}^{19}
\nonumber \\
\hspace{-0.95in}&& \quad  \,
-10587987305051901240209836269848805589728925726684541835278857721\,{x}^{18}
\nonumber \\
\hspace{-0.95in}&& \quad  \,
-31197739833767694054120129704437013088350311917897807218379021982\,{x}^{17}
\nonumber \\
\hspace{-0.95in}&& \quad  \,
-61702779900612867731214872614165179314709755885273509788298565787\,{x}^{16}
\nonumber \\
\hspace{-0.95in}&& \quad  \,
-84666583576312950514740894147953872788530635983760466811973453234\,{x}^{15}
\nonumber \\
\hspace{-0.95in}&& \quad  \,
-84581321256078449348114201458605757961462588640328430113579371469\,{x}^{14}
\nonumber \\
\hspace{-0.95in}&& \quad  \,
-66579501385417743645896831995666623321786452789989601988868930220\,{x}^{13}
\nonumber \\
\hspace{-0.95in}&& \quad  \,
-46422144734306012640614115287296842756324723788172904199227424928\,{x}^{12}
\nonumber \\
\hspace{-0.95in}&& \quad  \,
-34109048367984268267671082613328375922384379783373944543955842292\,{x}^{11}
\nonumber \\
\hspace{-0.95in}&& \quad  \,
-31579571071937009738017989004416659378464666295456745499499451123\,{x}^{10}
\nonumber \\
\hspace{-0.95in}&& \quad  \,
-31383477409339726943053162665868124737503098706615744587744129055\,{x}^{9}
\nonumber \\
\hspace{-0.95in}&& \quad  \,
-25192805778861042021357196910841906492061035149750125700349112500\,{x}^{8}
\nonumber \\
\hspace{-0.95in}&& \quad  \,
-15484947468128477413170938540582845959272279364669181133069873400\,{x}^{7}
\nonumber \\
\hspace{-0.95in}&& \quad  \,
-8825647465199759499489079986639230436941230964864953414638740550\,{x}^{6}
\nonumber \\
\hspace{-0.95in}&& \quad  \,
-5592482889305405022888968560846164620653256382698605479016168500\,{x}^{5}
\nonumber \\
\hspace{-0.95in}&& \quad  \,
-3175821728174664086144593028022649626932243816098004497271215500\,{x}^{4}
\nonumber \\
\hspace{-0.95in}&& \quad  \,
-1197312010623323136288515293659836786239749091267282912166901000\,{x}^{3}
\nonumber \\
\hspace{-0.95in}&& \quad  \,
-261689835566995733872747571036555312536936557291652102921551000\,{x}^{2}
\nonumber \\
\hspace{-0.95in}&& \quad  \,
-27051165864672155225865153613729609826904997195126025314800000\,x
\nonumber \\
\hspace{-0.95in}&& \quad  \,
-3137634679643321707303313577336229275827023029827551200000.
\end{eqnarray}

\vskip .1cm 

\section{Canonical decomposition for the adjoint\newline}
\label{towerdualapp}

Switching to the adjoint of the operator 
one can get the decomposition of this adjoint 
in two ways. One amounts to performing {\em euclidean 
left division} on the adjoints of the tower of 
intertwiners described in (\ref{towersimpleright}). 
The other one corresponds to performing the 
euclidean right division described 
in (\ref{towersimpleright}) but, this time, on the adjoint 
of the operator. 

\subsection{Canonical decomposition for the adjoint: euclidean left division \newline}
\label{towerdualleftsub}

If one takes the adjoint of the relations (\ref{recurs}) of section (\ref{towersimpleright}), 
one has (since the $\, U_n$'s are self-adjoint):
\begin{eqnarray}
\label{recursapp}
\hspace{-0.95in}&& \,   \, \, \,
adjoint(L_{[N]}) \,\, \,  \,  = \, \, \,\, \,  
 adjoint(L_{[N-1]}) \cdot \,U_N \, + \,  adjoint(L_{[N-2]}), 
\nonumber \\
\hspace{-0.95in}&& \,  \, \, \,
adjoint(L_{[N-1]}) \,\, \,  \,  = \, \, \,\, \,
  adjoint(L_{[N-2]}) \cdot \,U_{N-1}  \, + \,  adjoint(L_{[N-3]}), 
\nonumber \\
\hspace{-0.95in}&& \, \, \, \,
adjoint(L_{[N-2]}) \,\, \,  \,  = \, \, \,\, \, 
 adjoint(L_{[N-3]})  \cdot \,  U_{N-2} \, + \,  adjoint(L_{[N-4]}), 
\quad  \quad \cdots 
 \\
\hspace{-0.95in}&& \, \, \, \,
adjoint(L_{[N-p]}) \,\, \,  \,  = \, \, \,\, \, 
 adjoint(L_{[N-p-1]}) \cdot \,U_{N-p}  \, + \,  adjoint(L_{[N-p-2]}),
\quad  \quad \cdots \nonumber 
\end{eqnarray}
the intertwining relations (\ref{recurs2}) remaining the same 
(they are globally self-adjoint).
Of course, $\, L_{[N]}$ and $\, adjoint(L_{[N]})$ are on the same 
footing. A similar ``tower'' of intertwiners can be built on 
the  $\, adjoint(L_{[N]})$ and on the adjoint of the successive intertwiners 
$\, adjoint(L_{[N-p]})$, the only difference being that one must consider 
the {\em euclidean left division} of these successive adjoints. 
The first decompositions of these adjoints read:
\begin{eqnarray}
\label{adjlastm2}
\hspace{-0.95in}&& \quad 
adjoint(L_{[0]}) \,\, \,  \,  = \, \, \,\, \, r(x), \qquad \quad \, \,  \quad  
adjoint(L_{[1]}) \,\, \,  \,  = \, \, \,\, \,  r(x)  \cdot \,  U_{1}, 
\nonumber 
\end{eqnarray}
\begin{eqnarray}
\label{adjlastm1}
\hspace{-0.95in}&& \quad  
adjoint(L_{[2]}) \,\, \,  \,  = \, \, \,\, \,  \, 
  r(x)  \cdot \, (U_{1} \cdot \, U_{2} \,\, + 1), 
\end{eqnarray}
\begin{eqnarray}
\label{adjlast}
\hspace{-0.95in}&& \quad 
adjoint(L_{[3]}) \,\, \,  \,  = \, \, \,\, \,  \, 
 r(x)  \cdot \,  (U_{1} \cdot \, U_{2} \cdot \, U_{3} \,\, + U_{1} \,+ \, U_{3}), 
\nonumber 
\end{eqnarray}
\begin{eqnarray}
\label{adjlast2}
\hspace{-0.95in}&& \quad 
adjoint(L_{[4]}) \,\, \,  \,  = \, \, \,\, \,  \,
  r(x)  \cdot \, (U_{1} \cdot \, U_{2} \cdot \, U_{3} \cdot \,  U_{4}
\, + U_{1} \cdot \, U_{4} \,  
+ \, U_{3} \cdot \,  U_{4} + \, U_{1} \cdot \,  U_{2} \, + \, \, 1). 
\nonumber 
\end{eqnarray}

\vskip .1cm 

\subsection{Canonical decomposition for the adjoint: euclidean right division \newline}
\label{towerdualright}

The adjoint operator $\, adjoint(L_{[N]})$ is an operator, 
we can call $\, M_{[N]}$, for which 
the same euclidean right division 
calculations of section (\ref{towersimpleright}) can be performed, the first step 
corresponding to find the intertwiner $\, M_{[N-1]}$ in the intertwining 
relation:
\begin{eqnarray}
\label{tower1Mapp} 
\hspace{-0.8in}&& \quad \quad \quad \quad 
adjoint(M_{[N]}) \cdot  \, M_{[N-1]}
 \,\, \,  \,  = \, \, \,\, \,  \,  adjoint(M_{[N-1]})  \cdot  \, M_{[N]}. 
\end{eqnarray}
The command $\, Homomorphisms(M_{[N]}, \, adjoint(M_{[N]}))$ gives this first
intertwiner $\, M_{[N-1]}$ in the tower of intertwiners
(see section (\ref{towersimpleright})).
Since we have in mind that  $\, M_{[N]}$ is  $\, adjoint(L_{[N]})$, the 
intertwining relation (\ref{tower1Mapp}) is in fact
\begin{eqnarray}
\label{tower1Mbisapp} 
\hspace{-0.8in}&& \quad \quad \quad \quad 
L_{[N]} \cdot  \, M_{[N-1]}
 \,\, \,  \,  = \, \, \,\, \,  \,  adjoint(M_{[N-1]})  \cdot  \, adjoint(L_{[N]}), 
\end{eqnarray}
which is different from the intertwining relation (\ref{tower1}),
the DEtools Maple command giving this first
intertwiner $\, M_{[N-1]}$ being now
$\, Homomorphisms(adjoint(L_{[N]}), \, L_{[N]})$. 
Performing the same calculations as in section (\ref{towersimpleright}) 
(see (\ref{lastm2}), (\ref{lastm1}),  (\ref{last}),  (\ref{last2}), ...),
we will have another decomposition for these adjoints, deduced
 from successive right-divisions:
\begin{eqnarray}
\label{lastm2app}
\hspace{-0.95in}&& \,   \quad \, \, \,
M_{[0]} \,\, \,  \,  = \, \, \,\, \, \rho(x), 
\qquad \quad  \quad  \quad  \quad      \quad      
M_{[1]} \,\,   \,  = \, \, \,\,    V_{1} \cdot \rho(x),  \nonumber 
\end{eqnarray}
\begin{eqnarray}
\label{lastm1V}
\hspace{-0.95in}&& \,   \quad   \, \, \, 
M_{[2]} \,\, \,  \,  = \, \, \,\, 
 (V_{2} \cdot \, V_{1} \,\, + 1) \cdot \rho(x), 
\end{eqnarray}
\begin{eqnarray}
\label{lastV}
\hspace{-0.95in}&& \,    \quad \, \, \,
M_{[3]} \,\, \,  \,  = \, \, \,\,  
 (V_{3} \cdot \, V_{2} \cdot \, V_{1} \,\, + V_{1} \,+ \, V_{3}) \cdot \rho(x), 
\nonumber 
\end{eqnarray}
\begin{eqnarray}
\label{last2V}
\hspace{-0.95in}&& \,  \quad   \, \, \, 
M_{[4]} \, \,  \,  = \,  \,\, 
  (V_{4} \cdot \, V_{3} \cdot \, V_{2} \cdot \,  V_{1}
\, + V_{4} \cdot \, V_{1} \, 
+ \, V_{2} \cdot \,  V_{1} \, 
+ \, V_{4} \cdot \,  V_{3} \, + \, \, 1) \cdot \rho(x), 
 \quad \, \, \, \cdots \nonumber 
\end{eqnarray}

If one compares this decomposition for $\, M_{[4]} \, = \,  adjoint(L_{[4]})$
with the one in \ref{towerdualleftsub}, which is 
nothing but the adjoint of the decomposition
 in section (\ref{towersimpleright}), 
one finds that they are, actually, the same decompositions
provided $\, \rho(x) \, = \, \, r(x)$ for $\, N$ even,
 and  $\, \rho(x) \, = \, \, 1/r(x)$ for $\, N$ odd,
with the following change of operators:
\begin{eqnarray}
\label{change}
\hspace{-0.95in}&& 
V_{N} \, = \, \, r(x)  \cdot \, U_{1} \cdot \,r(x),  \, \,    \quad  
V_{N-1} \, = \, \, {{1} \over {r(x)}}  \cdot \, U_{2} \cdot \, {{1} \over {r(x)}},
  \, \,   \quad V_{N-2} \, = \, \, r(x)  \cdot \, U_{3} \cdot \,r(x), 
\nonumber \\
\hspace{-0.95in}&& \,   \quad   \, \, \, \, \,
V_{N-3} \, = \, \,\, {{1} \over {r(x)}}  \cdot \, U_{4} \cdot \, {{1} \over {r(x)}}, 
\quad  \, \,  \,  \, \,
 V_{N-4} \, = \, \,\, r(x) \cdot \, U_{5} \cdot \, r(x),
\quad  \, \,  \, \,   \cdots 
\end{eqnarray}
Therefore the $\, M_{[N]} \, = \, \, adjoint(L_{[N]})$  operator, and its successive 
intertwiners, read 
\begin{eqnarray}
\label{recap}
\hspace{-0.95in}&&  \quad \quad
M_{[N]}^{(N)} \, \,  \,  = \,  \,\, \, 
  (V_{N} \cdot \, V_{N-1} \cdots \,  V_{2} \cdot \,  V_{1}
\, +  \, \, \cdots ) \cdot \rho(x) 
\nonumber \\
\hspace{-0.95in}&& 
\qquad \qquad \quad \quad \,  \,  = \,  \,\,  \,\, 
r(x) \cdot \, 
(U_1 \,\cdot \, U_2 \, \cdots \,  U_{N-1} \cdots \,U_{N} \, +  \, \, \cdots ),
\nonumber \\
\hspace{-0.95in}&&  \quad \quad
M_{[N-1]}^{(N)} \, \,  \,  = \,  \,\, \, 
(V_{N-1} \cdot \, V_{N-2} \cdots \,  V_{2} \cdot \,  V_{1}
\, +  \, \, \cdots ) \cdot \rho(x)
 \nonumber \\
\hspace{-0.95in}&& 
\qquad \qquad \quad \quad \,  \,  = \,  \,\,  \,\, 
{{1} \over {r(x)}} \cdot \, 
(U_2 \,\cdot \, U_3 \, \cdots \,  U_{N-1} \cdots \,U_{N} \, +  \, \, \cdots ),
\nonumber \\
\hspace{-0.95in}&&  \quad \quad
M_{[N-2]}^{(N)} \, \,  \,  = \,  \,\, \, 
(V_{N-2} \cdot \, V_{N-3} \cdots \,  V_{2} \cdot \,  V_{1}
\, +  \, \, \cdots ) \cdot \rho(x)
 \nonumber \\
\hspace{-0.95in}&& 
\qquad \qquad  \quad \quad \,  \,  = \,  \,\,  \,\, \, 
r(x) \cdot \, 
(U_3 \,\cdot \, U_4 \, \cdots \,  U_{N-1} \cdots \,U_{N} \, +  \, \, \cdots ),
 \quad \, \, \cdots 
\end{eqnarray}

\vskip .1cm
\vskip .1cm
 
\section{The two intertwiners of an operator with its adjoint
are inverse of each other modulo the operator}
\label{inverseapp}

From the two intertwining relations (\ref{tower1})
\begin{eqnarray}
\label{tower1repeat} 
\hspace{-0.8in}&& \quad \quad \quad \quad \qquad
adjoint(L_{[N]}) \cdot  \, L_{[N-1]}
 \,\, \,  \,  = \, \, \,\, \,  \,  adjoint(L_{[N-1]})  \cdot  \, L_{[N]}. 
\end{eqnarray}
and (\ref{tower1M}), (\ref{tower1Mbis})
\begin{eqnarray}
\label{tower1Mapp} 
\hspace{-0.8in}&& \quad \quad \quad \quad \qquad
adjoint(M_{[N]}) \cdot  \, M_{[N-1]}
 \,\, \,  \,  = \, \, \,\, \,  \,  adjoint(M_{[N-1]})  \cdot  \, M_{[N]}, 
\end{eqnarray}
with $\, M_{[N]} \, = \, \, adjoint(L_{[N]})$, namely 
\begin{eqnarray}
\label{tower1Mbisrepeat} 
\hspace{-0.8in}&& \quad \quad \quad \quad 
L_{[N]} \cdot  \, M_{[N-1]}
 \,\, \,  \,  = \, \, \,\, \,  \,  adjoint(M_{[N-1]})  \cdot  \, adjoint(L_{[N]}), 
\end{eqnarray}
one gets
\begin{eqnarray}
\label{gets} 
\hspace{-0.95in}&& \quad  \qquad
L_{[N]} \cdot  \, M_{[N-1]} \cdot  \, L_{[N-1]} \,\, \,  \,  = \, \, \,\, \,  \, 
adjoint(M_{[N-1]}) \cdot  \, adjoint( L_{[N-1]}) \cdot  \, L_{[N]}
\nonumber \\
\hspace{-0.95in}&& \quad \quad \quad \quad \quad \qquad
\,  \, \,  = \, \, \, adjoint(L_{[N-1]} \cdot M_{[N-1]})  \cdot  \, L_{[N]},
\end{eqnarray}
and 
\begin{eqnarray}
\label{gets2} 
\hspace{-0.95in}&&  
 adjoint(L_{[N]}) \cdot  \,  L_{[N-1]} \cdot  \, M_{[N-1]}
 \,   = \, \,  
adjoint(L_{[N-1]}) \cdot  \, adjoint( M_{[N-1]}) \cdot  \, adjoint(L_{[N]}) 
\nonumber \\
\hspace{-0.95in}&& \quad \quad \quad \quad \quad \quad 
\,  \,  = \, \, \, adjoint(M_{[N-1]} \cdot L_{[N-1]})  \cdot  \, adjoint(L_{[N]}),
\end{eqnarray}
As $\, L_{[N]}$ is irreducible, 
relation (\ref{gets}) implies that the right division of 
$\,\, M_{[N-1]} \cdot  \, L_{[N-1]}\,$ by $\, L_{[N]}$ is 
a constant. Relation (\ref{gets2}) means that the right division 
of $\,\, L_{[N-1]} \cdot  \, M_{[N-1]}\,$ by $\, adjoint(L_{[N]})$ is 
a constant (see section 2.1 of~\cite{bridged}). This yields: 
\begin{eqnarray}
\label{inotherwordsunity} 
\hspace{-0.8in}&& \quad \quad \quad \quad 
 M_{[N-1]} \cdot  \, L_{[N-1]} \,\, \,  \,  = \, \, \,\,\,\, \, 
\Omega_{ML} \cdot \,  L_{[N]}  \, \,\,\,  + \, C_{ML}, 
\\
\label{inotherwordsunity2} 
\hspace{-0.8in}&& \quad \quad \quad \quad 
 L_{[N-1]} \cdot  \, M_{[N-1]} \,\, \,  \,  = \, \, \,\,\,\,\, 
 \Omega_{LM}  \cdot \, adjoint(L_{[N]}) \,\,\,\,    + \,C_{LM}, 
\end{eqnarray}
where $\,\Omega_{ML}$ and  $\,\Omega_{LM}$ are two operators of appropriate orders,
and where the constants $\, C_{ML}$ and $\,C_{LM}$ 
can be shown to be equal: $\, C_{ML} \,  = \, \,C_{LM}$.
As  $\, L_{[N]}$ is defined modulo a constant, we may choose 
 $\, C_{ML} \,  = \, \, 1$.

Using $\, adjoint(A+B) \, = \, adjoint(A) \, +adjoint(B)$, 
when $\, A$ and $\, B$ are of the same
parity order, and the fact that $\, \Omega_{LM}  \cdot \, adjoint(L_{[N]})$ 
is of the same order as $\, L_{[N-1]} \cdot  \, M_{[N-1]}$, and is thus of 
even parity, one finds, reinjecting (\ref{inotherwordsunity}) 
and (\ref{inotherwordsunity2})
\begin{eqnarray}
\label{getsmore} 
\hspace{-0.95in}&& \quad \quad   \quad  \, \,  
L_{[N]} \cdot  \, (\Omega_{ML} \cdot \,  L_{[N]}  \, \,\,\,  + \, 1)
 \,\, \,  \,  = \, \, \,\, \,  \, L_{[N]} \cdot  \, \Omega_{ML} \cdot \,  L_{[N]} 
 \, \,\,\,  + \,  \, L_{[N]}
\nonumber \\
\hspace{-0.95in}&& \quad \quad \quad \quad  \quad  \quad  
\,  \,  = \, \, \, 
adjoint(\Omega_{LM}  \cdot \, adjoint(L_{[N]}) \,\,\,\,    + \, 1)  \cdot  \, L_{[N]}
\nonumber \\
\hspace{-0.95in}&& \quad \quad \quad \quad  \quad  \quad  \,  \,  = \, \, \,
adjoint(\Omega_{LM}  \cdot \, adjoint(L_{[N]})) \cdot  \, L_{[N]} 
 \,\,    + \, \, L_{[N]}
 \nonumber \\
\hspace{-0.95in}&& \quad \quad \quad \quad  \quad  \quad  \,  \,  = \, \, \,
\,  \,   L_{[N]} \cdot \,adjoint(\Omega_{LM}) \cdot  \, L_{[N]} \,\,  
  + \, \, L_{[N]}, 
\end{eqnarray}
yielding 
\begin{eqnarray}
\label{endup}
\hspace{-0.8in}&& \quad \quad \quad \quad  \quad 
L_{[N]} \cdot  \, \Omega_{ML} \cdot \,  L_{[N]}  \, \, \,   = \, \, \, \, \, 
 L_{[N]} \cdot \,adjoint(\Omega_{LM}) \cdot  \, L_{[N]}, 
\end{eqnarray}
and thus:
\begin{eqnarray}
\label{enduponehas}
\hspace{-0.8in}&& \quad \quad \quad \quad  \quad \quad \quad 
\Omega_{LM}  \,\, \,  \,  = \, \, \,\,\,   adjoint(\Omega_{ML}). 
\end{eqnarray}

In fact the operator $\, \Omega_{ML}$ in these inversion 
relations modulo $\,  L_{[N]}$,
is exactly  $\,M_{[N-2]}^{(N-1)}$ of (\ref{recap}). For 
instance for $\,  L_{[6]}$, using the notation of (\ref{recap}),
one gets the following inversion relations,
 modulo $\,  L_{[6]}$, on $\, M_{[5]}^{(6)}$ and $\, L_{[5]}$:
\begin{eqnarray}
\label{inversionL6}
\hspace{-0.9in}&& \, \, \, 
M_{[5]}^{(6)} \cdot \, L_{[5]} 
  \,\, \,  \,  = \, \, \,\, \, \,
\Omega_{ML} \cdot \, L_{[6]} \,\, \,\, + \, 1
 \,  \, \,  = \, \, \,\, \, 
adjoint(\Omega_{LM}) \cdot \, L_6 \,\, \,+ \, 1, \\
\hspace{-0.9in}&& \, \, \, 
 L_{[5]} \cdot  \, M_{[5]}^{(6)} \,\, \,  \,  = \, \, \,\, \,\,
\Omega_{LM}  \cdot \, adjoint(L_{[6]}) \,\, \,\,  + \, 1, 
\end{eqnarray}
where the operator $\, \Omega_{ML}$ reads 
\begin{eqnarray}
\label{omegaLM}
\hspace{-0.95in}&& \, 
\Omega_{ML} \,  \,  = \, \, \, {{1} \over {r(x)}} \cdot \, 
  (U_{2} \cdot \, U_{3} \cdot \, U_{4} \cdot \,  U_{5}
\,\, + U_{2} \cdot \, U_{5} \, \,
+ \, U_{2} \cdot \,  U_{3} \, \,
+ \, U_{4} \cdot \,  U_{5} \,\, + \, \, 1),  
\end{eqnarray}
which is nothing but the adjoint of $\, M_{[4]}^{(5)}$ 
(see (\ref{recap}) in \ref{towerdualright}).  
The two relations (\ref{inversionL6}) read:
\begin{eqnarray}
\label{inversionL6other}
\hspace{-0.8in}&& \quad \quad \, 
M_{[5]}^{(6)} \cdot \, L_{[5]} 
  \,\, \,  \,  = \, \, \,\, \, \,
M_{[4]}^{(5)} \cdot \, L_{[6]} \,\, \,\, + \, 1
 \, \, \,  = \, \, \,\, \, 
M_{[4]}^{(5)} \cdot \, L_{[6]} \,\, \, -1, \\
\hspace{-0.8in}&& \quad \quad \, 
 L_{[5]} \cdot  \, M_{[5]}^{(6)} \,\, \,  \,  = \, \, \,\, 
adjoint(M_{[4]}^{(5)})  \cdot \, adjoint(L_{[6]}) \,\, \,\, -1,
\end{eqnarray}
The same calculations for $\,  L_{[7]}$ give 
an operator $\, \Omega_{ML}$ which reads
\begin{eqnarray}
\label{last3omega}
\hspace{-0.95in}&& \quad  \, \,
\Omega_{ML} \,\, \,  \,  = \, \, \,\, \,  \,
  (U_{2} \cdot \,U_{3} \cdot \, U_{4} \cdot \, U_{5} \cdot \,  U_{6}
\,\,\, + U_{2} \cdot \, U_{3} \cdot \, U_{6} \,  \,
+ \, U_{2} \cdot \, U_{5} \cdot \,  U_{6} \, 
\nonumber \\
\hspace{-0.95in}&&  \qquad \quad  \quad \quad  \,   \quad 
 + \, U_{2} \cdot \, U_{3} \cdot \,  U_{4}
+ \, U_{4} \cdot \, U_{5} \cdot \, U_{6} 
\,\,\,\,  + U_{2} \, \, + \, U_{4}\,+ \, U_{6}) \cdot r(x),  
\end{eqnarray}
which is nothing but $\, M_{[5]}^{(6)}$, 
the two inversion relations reading:
\begin{eqnarray}
\label{inversionL6otherbis}
\hspace{-0.8in}&& \quad \quad \quad 
M_{[6]}^{(7)} \cdot \, L_{[6]} 
  \,\, \,  \,  = \, \, \,\, \, \,
M_{[5]}^{(6)} \cdot \, L_{[7]} \,\,  + \, 1
\, \,  \, \, = \, \, \,\,\, 
M_{[5]}^{(6)} \cdot \, L_{[7]} \,\, \, + \,1, \\
\hspace{-0.8in}&& \quad \quad \quad 
 L_{[6]} \cdot  \, M_{[6]}^{(7)} \,\, \,  \,  = \, \, \,\, 
adjoint(M_{[5]}^{(6)})  \cdot \, adjoint(L_{[7]}) \,\, \,  + \, 1. 
\end{eqnarray}
More generally one has the two relations (with 
the notations of (\ref{recap})):
\begin{eqnarray}
\label{inversionL6othergeneral}
\hspace{-0.9in}&& \quad \, 
M_{[N-1]}^{(N)} \cdot \, L_{[N-1]} 
  \,\, \,  \,  = \, \, \,\, \, \,
M_{[N-2]}^{(N-1)} \cdot \, L_{[N]} \,\, \,\, \, - \, (-1)^N, \\
\label{inversionL6othergeneral2}
\hspace{-0.9in}&& \quad \, 
 L_{[N-1]} \cdot  \, M_{[N-1]}^{(N)} \,\, \,  \,  = \, \, \,\, \,\,
adjoint(M_{[N-2]}^{(N-1)})  \cdot \, adjoint(L_{[N]}) \,\, \,\,   - \, (-1)^N. 
\end{eqnarray}
Seeing (\ref{inversionL6othergeneral}) and (\ref{inversionL6othergeneral2})
as identities on the (self-adjoint same parity-order) $\, U_n$'s,
identity (\ref{inversionL6othergeneral2}) is nothing but 
identity (\ref{inversionL6othergeneral}), if the $\, U_n$'s
are changed according to the involution 
$\, U_n \, \leftrightarrow \, U_{N+1-n}$.

\vskip .3cm

\section{Decomposition for operators equivalent to 
operators with exceptional differential Galois groups}
\label{except}

Let us recall, for instance\footnote[2]{There is nothing particular 
with this example, the results are similar
for the other  order-seven linear differential operators with 
an exceptional differential 
Galois group $\, G_2$ in~\cite{bridged}.}, one of the six 
order-seven linear differential operators 
(called $\, E_2$ in~\cite{bridged}), which has the {\em exceptional differential 
Galois group} $\, G_2$:
\begin{eqnarray}
\label{except2}
 \hspace{-0.95in}&& \quad \quad \quad    \quad    
{\cal L}_7 \, \,= \, \, \,      \, \, \theta^7 \, \,  -128 \cdot \, x \cdot \, 
(8 \, \theta^4\, +16 \,\theta^3 \, + 20 \, \theta^2 \, +12 \, \theta \, +3) 
\, (2 \, \theta \, +1)^3 \nonumber \\ 
 \hspace{-0.95in}&& \quad \quad \quad \quad  \,      \quad \qquad 
+1048576 \, x^2 \cdot \, (2  \, \theta \, +1)^2 \cdot \, 
(2  \, \theta \, +3)^2 \cdot \, (\theta \, +1)^3.
\end{eqnarray}
where $\, \theta\, = \, \, x \cdot D_x$.

\subsection{A first equivalent operator \newline}
\label{firstequiv}

We introduce an operator $\, {\tilde {\cal L}}_7^{(3)}$ equivalent to $\, {\cal L}_7$:
\begin{eqnarray}
\label{except2equiv}
\hspace{-0.95in}&& \quad \quad \quad   \quad \quad  \,  \,  \quad   \quad   \quad    
 {\cal I}_3 \cdot \,  {\cal L}_7 \, \,\,\, \,= \,\, \, \,  \, \, 
{\tilde {\cal L}}_7^{(3)}  \cdot \, D_x^3, 
\end{eqnarray}
where $\, {\cal I}_3$ is an order-three linear differential operator. The  equivalent 
operator $\, {\tilde {\cal L}}_7^{(3)}$ is such that its {\em  symmetric square} has a 
(quite simple~\cite{bridged}) rational solution, namely 
$\,\rho(x) = \,  1/(1\, -4096\,x)^2/x^6$, when symmetric square
of $\, {\cal L}_7$ has a drop of order (order 27 instead 
of the generic order 28,  see~\cite{bridged}). 

Performing  the DEtools Maple command
``Homomorphisms($\, {\tilde {\cal L}}_7^{(3)}$, $adjoint({\tilde {\cal L}}_7^{(3)})$)'',  
we obtained an {\em order-six} operator that we will 
denote $\, {\tilde {\cal L}}_6^{(3)}$, such that
\begin{eqnarray}
\label{defP4G2}
\hspace{-0.9in}&& \quad  \quad \quad \quad \quad 
adjoint({\tilde {\cal L}}_6^{(3)})  \cdot \, {\tilde {\cal L}}_7^{(3)} 
\,  \, \, \, = \, \, \, \,\, 
  adjoint({\tilde {\cal L}}_7^{(3)}) \cdot \, {\tilde {\cal L}}_6^{(3)}. 
\end{eqnarray}
The successive euclidean right-divisions described in section (\ref{towersimpleright}),
enable to deduce, for $\, {\tilde {\cal L}}_7^{(3)}$, 
the (generic\footnote[9]{In contrast, the order-seven operator $\, {\cal L}_7$
has the most extreme non-generic decomposition, since it is of the 
form $\, U_1 \cdot \, \rho(x)$ where $\,\rho(x)$ is a function 
and $\, U_1$ is an order-seven self-adjoint operator.}) 
decomposition  (\ref{last5}), where the seven operators
$\, U_n$'s are {\em all order-one self-adjoint} operators, $\, r(x)$ being a quite 
involved rational function. The rational solution
$\,\rho(x)$ is also the rational solution of the symmetric square of 
the order-one operator $\, U_1 \cdot r(x)$ in this decomposition  (\ref{last5}).

\vskip .1cm 

{\bf Remark:} The order-one operator $\, U_1$ is quite involved, as well 
as the rational solution $\, r(x)$. In contrast the {\em self-adjoint} operator 
$\, r(x) \cdot  \, U_1 \cdot \, r(x) $ has a remarkably simple expression in terms of 
the rational solution $\,\rho(x)$:
\begin{eqnarray}
\label{verysimple}
\hspace{-0.9in}&& \quad  
 r(x) \cdot  \, U_1 \cdot \, r(x)  \, \, = \, \, \, \, \, 
- \, {{99225} \over {122825998336}} \cdot \, 
{{1} \over {\rho(x)}} \cdot \, 
\Bigl( D_x \, \, - {{1} \over {2}} \cdot \,  {{d \ln(\rho(x))} \over {dx}}   \Bigr).
\end{eqnarray}

\subsection{A second equivalent operator \newline}
\label{secondequiv}

We now introduce an operator $\, {\tilde {\cal L}}_7^{(2)}$ 
equivalent to $\, {\cal L}_7$:
\begin{eqnarray}
\label{except2equiv}
\hspace{-0.95in}&& \quad \quad \quad   \quad   \quad   \quad   \quad    
 {\cal M}_2 \cdot \,  {\cal L}_7 \,\,\, \, = \, \, \, \,\, 
 {\tilde {\cal L}}_7^{(2)}  \cdot \, D_x^2, 
\end{eqnarray}
where $\, {\cal M}_2$ is an order-two linear differential operator. The  equivalent 
operator $\, {\tilde {\cal L}}_7^{(2)}$ is such that its exterior square has a drop 
of order, but its order-35
 {\em exterior cube}~\cite{bridged} has a 
(quite simple)  {\em rational solution}, namely $\, 1/(1\, -4096\,x)^3/x^9$. 
Again, we follow the algorithm described in section (\ref{towersimpleright}),
to get the decomposition of this order-seven operator $\, {\tilde {\cal L}}_7^{(2)}$.
 Performing  the DEtools Maple command
``Homomorphisms($\, {\tilde {\cal L}}_7^{(2)}$, $adjoint({\tilde {\cal L}}_7^{(2)})$)'',  
we obtained an {\em order-six} operator that we will 
denote $\, {\tilde {\cal L}}_6^{(2)}$, such that
\begin{eqnarray}
\label{defP4G2app}
\hspace{-0.9in}&& \quad  \quad \quad \quad \quad \, \,
adjoint({\tilde {\cal L}}_6^{(2)})  \cdot \, {\tilde {\cal L}}_7^{(2)}
 \,  \, \, \, = \, \, \, \,\, 
  adjoint({\tilde {\cal L}}_7^{(2)}) \cdot \, {\tilde {\cal L}}_6^{(2)}. 
\end{eqnarray}
Performing the successive euclidean right-divisions described 
in section (\ref{towersimpleright}), we obtain the following decomposition, 
corresponding to decomposition
(\ref{last3}), {\em but in the non-generic case}
\begin{eqnarray}
\label{last3screw}
\hspace{-0.95in}&& \quad  \quad  \, \,   
{\tilde {\cal L}}_7^{(2)} \,\, \,  \,  = \, \, \,\, \,  \,
  (U_{5} \cdot \,U_{4} \cdot \, U_{3} \cdot \, U_{2} \cdot \,  U_{1}
\,\,\, + U_{5} \cdot \, U_{4} \cdot \, U_{1} \,  \,
+ \, U_{5} \cdot \, U_{2} \cdot \,  U_{1} \, 
\nonumber \\
\hspace{-0.95in}&&  \qquad \quad  \quad \,\,\, \quad    \quad 
 + \, U_{5} \cdot \, U_{4} \cdot \,  U_{3}
+ \, U_{3} \cdot \, U_{2} \cdot \, U_{1} 
\,\,\,\,  + U_{1} \, \, + \, U_{3}\,+ \, U_{5}) \cdot r(x),  
\end{eqnarray}
where $\, r(x)$ is a rational function, where 
$\, U_{5}$,  $\, U_{4}$, $\, U_{3}$,  $\, U_{2}$ are four 
{\em order-one self-adjoint} operators, but  $\, U_{1}$ is an 
{\em order-three self-adjoint} operator.

From this non-generic decomposition (\ref{last3screw}), it is clear, from
the results of the  section (\ref{ratiodecomp}), that the  
symmetric square of $\, {\tilde {\cal L}}_6^{(2)}$ has a drop of order 
(as a consequence of the order-three self-adjoint operator $\, U_{1}$),
but that the symmetric square of the
adjoint of $\, {\tilde {\cal L}}_6^{(2)}$ has a rational solution, 
associated with the order-one self-adjoint operator $\, U_{5}$.

\vskip .1cm 

 One notes that the rational solution 
of the exterior cube of $\, {\tilde {\cal L}}_7^{(2)}$, namely $\, 1/(1\, -4096\,x)^3/x^9$, 
is {\em the same} as the rational solution 
of the exterior cube of the order-three operator $\, U_{1}\cdot \, r(x)$,
or of the self-adjoint operator $\, r(x) \cdot \, U_{1}\cdot \, r(x)$ (see (\ref{verysimple})). 

\vskip .1cm 

\section{Towards the generic situation 
for selected differential Galois groups: reduced form 
for differential systems}
\label{towards}

Let us show (experimentally) in an alternative way, that a
 linear differential operator with a {\em selected  differential Galois group}
{\em necessarily has} a decomposition as described in section (\ref{decomp}).
In order to get some hint on this very general question 
we need to recall the concept of {\em reduced form}~\cite{Symbolic,AparicioLast} for 
{\em linear differential systems}.

\subsection{Reduced form of differential systems for orthogonal groups}
\label{towardsredu}

If one considers a linear differential system corresponding to an 
{\em antisymmetric} $\, q \times q$ matrix $\, A(x)$, whose
entries are rational functions of $\, x$
\begin{eqnarray}
\label{Ant}
\hspace{-0.9in}&& \qquad \quad  \quad  \quad  \quad  \quad  \quad 
 Y' \, \, \, \,  = \, \, \,  \, \, \, \, A(x) \cdot Y,  
\end{eqnarray}
one is sure that the
differential Galois group of this system will correspond to the
orthogonal group (this is a result by 
E. R. Kolchin~\cite{Kolchin1,Kolchin2,Kolchin3,AparicioLast}). Less 
obvious is the result by Kovacic and Kolchin~\cite{Symbolic,Kovacic} 
that {\em any} linear differential system
with an {\em orthogonal} group for its differential Galois group,
can be reduced to this canonical form (\ref{Ant}).  
Studying linear differential systems like (\ref{Ant}) is, thus a way, 
to get some hint of ``generic'' differential operators with
differential Galois groups which correspond to orthogonal groups.

 Following our experimental mathematics approach, we have 
built a large number of examples of  order-$q$ 
(mostly  $\, q\, = \, 4$, but also  
$\, q\, = \,\, 5, \, 6, \, 7,\, 8$) linear 
differential operators $\, {\tilde L}_N$ associated with
such a linear differential system (\ref{Ant}). Using the 
algorithm described in section (\ref{towersimpleright}), 
we obtained the decomposition of $\, {\tilde L}_N$.
We found that all our (numerous) examples have a 
{\em generic decomposition}, the order of {\em all}
 the self-adjoint operators $\,  U_{n}$'s
 being {\em one} (corresponding to
$\, SO(q, \,  \mathbb{C})$ differential Galois groups).

\subsection{Reduced form of differential systems for symplectic 
 groups}
\label{towardsredusympl} 

The (generic) symplectic case can be sketched, in a similar way,
introducing  linear differential systems like
 $\, Y' \,   = \, \, J \cdot A(x) \cdot Y$,  where $\, A(x)$
is a {\em symmetric} $\, 2p \times 2p$ matrix with rational functions entries,
and where $\, J$ is the {\em symplectic} matrix 
\begin{eqnarray}
\label{defJ}
\hspace{-0.95in}&& \qquad \qquad \qquad  \quad  \quad 
 J \, \, = \, \, \, 
\left[ \begin {array}{cc} 
0&{\it Id}\\ 
\noalign{\medskip}-{\it Id}&0
\end {array} 
\right],
\end{eqnarray}
where $\, {\it Id}$ denotes the $\, p \times p$ identity matrix,
and $\, 0$ denotes the  $\, p \times p$ null matrix. 
Again, we have 
built a large number of examples of  order-$q$ 
(mostly  $\, q\, = \, 6$, but also  $\, q\, = \, 8$) 
linear differential operators 
$\, {\tilde L}_q$ associated with
such a ``symplectic'' linear differential system. Using the 
algorithm described in section (\ref{towersimpleright}), 
we have obtained the decomposition of these $\, {\tilde L}_q$.
We have found that all our (numerous) examples actually have a 
{\em generic decomposition}, the order of {\em all}
 the self-adjoint operators $\,  U_{n}$'s
 being {\em two} (corresponding to
$\, Sp(q, \,  \mathbb{C})$ differential Galois groups).

\end{document}